# IMPACT IN INFORMETRICS AND BEYOND

## Leo Egghe, Hasselt University, Belgium

leo.egghe@uhasselt.be

ORCID: 0000-0001-8419-2932

## Ronald Rousseau, KU Leuven, Belgium

ronald.rousseau@kuleuven.be

&

## University of Antwerp, Belgium

ronald.rousseau@uantwerpen.be

ORCID: 0000-0002-3252-2538

The concept of impact is one of the most important concepts in the field of informetrics in the last two decades. It studies to what degree scientists have influence in their field, e.g., by their publications (mostly via citation analysis). This concept has been formally developed in a mathematical way by the present authors in 5 publications, collected in this text. To lead the reader more easily to these publications and to show what is – intuitively – behind the concept of impact, we present this introduction. A more advanced introduction can be found in the paper I which can be read after the present text.

We first have to fix a topic for which we want to find the influential "objects": authors, journals, … w.r.t. their "production" (e.g., publications generating citations, … - in short: their sources and items they generate (produce)). These objects are then said to have a certain degree of impact (to be defined in an exact way in the sequel).

There are three levels on which we can perform this. On the first level of investigating impact, we need a measure of these objects, i.e., a function of these objects, represented by their rank-frequency function, describing the number of items in the source on rank r (where sources are arranged in decreasing order of their number of items): a so-called impact measure. Of course, we need measures with specific properties: they must focus on the production of the most productive sources of an object. Examples: the h-index (Hirsch (2005)) or the g-index (Egghe (2006)) of an author calculated on the basis of his/her publications and the citations of these publications. In paper II we study a formal definition of impact measures based on these "left-hand" sides of the source-item rank-frequency functions that represent these objects (left in the sense of small ranks, i.e., the places of sources with the highest production, hence on the left part of the abscissa axis).

The second level of impact investigation is using impact bundles (or sheaves) as is done in paper III. To show the difference between an impact measure and an impact bundle we present a simple and classical example: the h-index and the h-bundle. The h-index of a function Z (representing object Z), denoted h(Z), is defined as x for which $Z(x) = x$, i.e., the abscissa of the intersection of the graph of Z and the straight line $y = x$ (cf. Hirsch (2005)). The h-bundle is defined in the same way but now the line $y = x$ is replaced by any increasing straight line through the origin: $y = \theta.x$, $\theta > 0$. Hence this abscissa is denoted $h_\theta(Z)$ (and hence $h_1(Z) = h(Z)$). So, we have a bundle of impact measures $h_\theta$ which is, of course, more powerful to measure the impact of an object Z. Impact bundles in paper III are characterized as increasing functions on Z for an order on Z that focuses on the left part of Z (i.e., the sources with the highest production in the sense of having the most items).

A third (and highest) level of impact investigations involves a variant of the Lorenz order in econometrics in the sense that we use the non-normalized form of the Lorenz order. In paper IV we study global impact measures being measures that respect the non-normalized Lorenz order between the rank-frequency functions that represent the objects Z, yielding a restricted class of impact measures, but, as shown in paper V, when using bundles, most known impact bundles (such as the h- and g-bundle) respect the non-normalized Lorenz order. For this reason, the non-normalized Lorenz order is defined as the measure- or bundle-independent notion of impact: we say that object Z has more impact than object Y if Y is smaller than Z in the sense of the non-normalized Lorenz order, i.e., if the non-normalized Lorenz curve of Y is below the one of Z. Hence this is a higher level of the treatment of impact: a mathematical definition of the concept itself, which is superior to defining impact measures or impact bundles (but who have their own merits). In papers IV and Egghe and Rousseau (2022), relations between the normalized and non-normalized Lorenz order are given.

We note that papers I, II, III, IV and V have never been uploaded in arXiv before as separate papers. We opted for this "combined" upload because of the added value of presenting a unified mathematical theory of impact.

This submission contains the following papers:

I. Egghe, L., & Rousseau, R. The logic behind the mathematical treatment of impact

II. Egghe, L. Impact measures. What are they?

III. Egghe, L., & Rousseau, R. Rank-frequency data and impact in a continuous model: Introducing impact sheaves

IV. Egghe, L., & Rousseau, R. Global impact measures

V. Egghe, L., & Rousseau, R. Global informetric impact: a description and definition

# The Logic Behind the Mathematical Treatment of Impact

## Leo Egghe, Hasselt University, Belgium

leo.egghe@uhasselt.be

ORCID: 0000-0001-8419-2932

## Ronald Rousseau, KU Leuven, Belgium

ronald.rousseau@kuleuven.be

&

## University of Antwerp, Belgium

ronald.rousseau@uantwerpen.be

ORCID: 0000-0002-3252-2538

## Abstract

We propose a definition for the fundamental notion of impact in informetrics.

## Introduction

There are many possible interpretations of the intuitive notion of impact. Impact occurs when a car collides with another object, or when a person's acts or decisions influence the life of another person (e.g., by criminal behavior, and for the perpetrator, facing the consequences in court). In these examples, we notice an influence of one "object" on another "object" through an action. It is also possible not to specify the "receiving object" and just focus on an action of one object (e.g., a person) for which we can describe impact without specifying on which object. A typical example is the impact of a scientific publication as measured by the received number of citations. Here the "receiving object" is not specified since it can be another publication, a researcher, a scientific community, or even the whole world (in case of an important invention with widespread practical consequences). In this informetric example (and we will continue henceforth in the field of informetrics) we can speak of one-dimensional impact in the sense that impact is measured by one number (e.g., the number of citations). Similarly, but broader, in the UK Research Excellence Framework (REF) the outside impact of research was defined as 'an effect on, change or benefit to the economy, society, culture, public policy or services, health, the environment or quality of life, beyond academia'.

It is much more interesting to consider impact in two dimensions through the connection of publications and received citations or - in general terminology - sources and the corresponding number of items (that they possess or have produced). The basic function in this framework (Egghe, 2005) is the rank-frequency function Z describing, for every rank r, the number $Z(r)$ of items in the source on rank $r = 1,2,…,T$, where sources are ranked in decreasing order of the number of items (ties are solved in a certain - here not specified - way) and where T (fixed) denotes the total number of sources. A typical, discrete, rank-frequency function occurs when a set of authors is ranked according to the number of publications, or when a scientist's publications are ranked according to the number of received citations. In this framework, the classical "impact measures" such as the h-index, $h=h(Z)$ (Hirsch, 2005), the g-index $g=g(Z)$ (Egghe, 2006)) and many variants, see e.g., (Egghe, 2010) can be applied and one says that a situation Z (another rank-frequency function) has (strictly) less impact than situation Y if $m(Z) < m(Y)$ where m is the used impact measure (e.g., m=h or m=g). Note that in the previous lines the terms, "impact" and "impact measure" are used in a heuristic way. It is the purpose of this note to show how to select from these classical measures those mathematical aspects of impact and impact measures that are essential and hence come to rigorous definitions of these notions.

**Our framework**

We will highlight now how we came to the definitions we were aiming at (Egghe, 2022, Egghe & Rousseau, 2022a,b,c) and what the logic is behind their introduction. For comparative reasons and further use, we will also use the measure $\mu(Z)$ (the arithmetic average (or mean) of Z). To work in a more comfortable framework, we will henceforth assume that the rank-frequency functions Z, Y are continuous functions with domain the interval [0,T], replacing the discrete set $\{1,…,T\}$, with T fixed.

What we learned from studying the classical measures such as $h(Z)$ or $g(Z)$ for a specific situation Z is the following: these measures focus on the sources with the higher number of items, those placed on the lower ranks r, and on their number of items. In other words: they focus on

the production (A) of the most productive sources (B)    ($\mathcal{I}$)

For aspect B this means that the left-hand side of the graph of Z, where ranks are low, is the part that really matters. These measures are not influenced by the production of the low productive sources, those with high ranks r. From this observation, we can already conclude that $\mu(Z)$ does not satisfy this principle since it depends on the total number of items in all sources and hence does not focus on (or is determined by) the most productive sources. This is where the notion of "measure bundle" ("**_bundle_**" for short) enters this story. For all measures m we can introduce an ad hoc parameter version $m_\theta$ with the purpose of "scanning" the rank-frequency function Z. Concretely, we consider the following two examples. For the bundle m=h we define (for θ a positive number): $x=h_\theta(Z) \Leftrightarrow Z(x) = \theta x$ (Egghe, 2021). Note that the classical h-index $h(Z)$ is equal to $h_1(Z)$. All these measures $h_\theta$ have the property to focus on the production of the most productive sources. An analogous definition can be given for the generalized g-index $g_\theta$ (Egghe, 2021). This technique also shows the way to use µ (in its "bundle version") as a tool for measuring impact: for all θ in [0,T], we define $\mu_\theta(Z)$ as the average number of items in the sources on ranks r ≤ θ. Note that $\mu_T(Z) = \mu(Z)$ but that all other $\mu_\theta(Z)$ values focus on the "left-hand part" of the graph of Z, here the interval [0,θ] (while for these measures also the production is taken into account) and hence satisfy the requirement (𝐽) for measuring impact.

We observe that (𝐽) consists of two parts A and B. It is clear that A deals with 'pure' production while B deals with the notion of "concentration" or "inequality", well-known in other fields such as econometrics. So (𝐽) is a combination of aspects A and B and hence the exact notion of "impact" must also be a combination of these two aspects. So far, we still used the word "impact" in a heuristic way but from now on we will replace this heuristic notion with a mathematically exact formulation.

## Impact measures and impact bundles

We noted already above that the B-part in (𝐽) focuses on the left-hand side of the graph of Z. That is why we proposed in Egghe (2022) the following basic form of the definition of an ***impact measure*** m. A function m is an impact measure if

> for every rank-frequency function Z, there exists a number $a_Z$ in ]0,T[ such that the condition Z < Y on $[0,a_Z]$ implies that m(Z) < m(Y)

We note that the definition in Egghe (2022) slightly differs from the above definition (for technical reasons) but - essentially - it is the same definition and, for reasons of simplicity, we work with the one above. It is easy to see that h and g are impact measures and that for m=h, $a_Z=h(Z)$ and for m=g, $a_Z=g(Z)$ while for m=µ, $a_Z$ does not exist and hence is not an impact measure for the reason mentioned above: µ also depends on the production of the least productive sources so that the value µ(Z) is not determined by the values of Z on a "left-hand part of the graph of Z".

It is intuitively clear that, when we work with bundles $m_\theta$, all these different measures (assumed to be impact measures) generate a range of values $a_Z$ in ]0,T[, so that we have:

for all values a in ]0,T[ and all rank-frequency functions Z, Y the condition Z < Y on [0,a] implies $m_\theta(Z) < m_\theta(Y)$, for all θ in a certain interval.

This is the condition for an ***impact bundle*,** see (Egghe & Rousseau, 2022a). It is easy to see that the bundles $h_\theta$ and $g_\theta$ are impact bundles but also the bundle $µ_\theta$ satisfies this condition: indeed, here we can take - given any value a > 0 - all θ in [0,a]. So $µ_\theta$ is an impact bundle while µ (the overall average, such as the journal impact factor) is not an impact measure in the sense explained above.

## Global impact bundles

The above definition of impact bundle is fine to define objects (bundles) that measure impact but is not suited to define "impact" itself. Yet it can be used in the following reasoning to define impact, based on ($J$). We may delete the bundle $m_\theta$ in the above definition of impact bundle but then we end up with the condition "Z < Y" (since the number a is not defined). This is not a wrong assumption for the notion of impact but it is too strict since it requires Z(x) < Y(x) for all x in [0,T] and it is clear that we want to have a wider range of situations Z, Y where Y has more impact than Z (or vice-versa) as suggested by condition B in ($J$). This condition is related to the classical notion of concentration where more or less concentration is defined via a relation between two functions Z and Y known as the dominance relation (Hardy et al., 1934) based on the classical Lorenz curve (Lorenz, 1905; Rousseau et al., 2018). We recall that the classical Lorenz curve is described within the unit square, i.e., two normalizations have been applied. Now, because of condition A in ($\mathcal{I}$) - we will define the easier non-normalized version of the Lorenz curve and show that this is a good decision. For a rank-frequency function Z, we define

$$I(Z)(r) = \int_0^r Z(x)dx$$

for r in [0,T] (or in the discrete setting: $I(Z)(r) = \sum_{i=1}^r Z(i)$ ), the cumulative number of items in the first r (most productive) sources. It is the normalized version of this function that is used in concentration theory (the Lorenz curve) but, as indicated above (because of A and B in ($\mathcal{I}$)), it is this simple non-normalized version that we need here. Indeed:

> For all rank-frequency functions Z, Y we define the impact order -< as:
>
> Z -< Y iff I(Z) ≤ I(Y) on [0,T]

Moreover, we define Z -<$_\neq$ as Z -< Y with Z≠Y (i.e., there exists an x such that Z(x) ≠ Y(x)). With this powerful tool we can formulate the following definition of a **_global impact bundle_** m$_\theta$:

> For all rank-frequency functions Z, Y:
>
> the condition Z -<$_\neq$ Y implies that m$_\theta$(Z) and m$_\theta$(Y) respect the impact order -<$_\neq$ order of Z and Y for all θ in a certain interval.

What does this definition mean? Take two rank-frequency functions Z and Y. We consider their difference Y - Z and suppose that this function does not switch between zero and non-zero an infinite number of times. Such a situation, infinitely many transitions, can exist in a purely mathematical sense but does not occur in practical cases. So, we can suppose that Y - Z has only a finite number of transitions (FNT) between 0 and a non-zero number. In that case, it is easy to show (Egghe & Rousseau,

2022b)) that the condition "$Z \prec_{\neq} Y$" implies one of the following two properties :

(i) $Z < Y$ on $[0,a]$ for a certain $a > 0$

(ii) $Z = Y$ on $[0,a]$ and $Z < Y$ on $]a,b]$ for certain numbers a and b such that $0 \leq a < b$.

Now we can make the definition of global impact bundle $m_\theta$ above more concrete by requiring that the same relations $<$ (in case (i)) or $=$ followed by $<$ (in case (ii)) are valid for $m_\theta$ for all $\theta$ in a certain interval. It follows immediately from this definition and the one of impact bundle that every global impact bundle is an impact bundle (since for the latter only (i) applies). It is also intuitively clear that situations as in (ii) ("equal left-hand parts") are allowed in measuring (different) impact as long as such an equal part is followed by an unequal ($<$) part. It is now easy to prove that the impact bundles $h_\theta$, $g_\theta$ and $\mu_\theta$ (and others) are global impact bundles.

## Impact

Based on the above results we now define impact (independent of the measure m or the bundle $m_\theta$) in the following sense: for two rank-frequency functions Z and Y, we say that

Z has less impact than Y if and only if $Z \prec_{\neq} Y$

In Egghe and Rousseau (2022c) we present results on the relation between the order relation $\prec$ and its normalized analog as used in concentration theory or its opposite, diversity theory. Heuristically (but exact results are in Egghe and Rousseau (2022c)) we can say that A and B in ($\jmath$) represent (respectively) production and the normalized $\prec$ and that they together, i.e. ($\jmath$)) represent the non-normalized $\prec$ hereby presenting the link between inequality and impact, the latter being "productivity + inequality" (in a heuristic sense).

## An analogy with the classical Lorenz curve

If the concave Lorenz curve $C_1$ is situated above the concave Lorenz curve $C_2$, different from $C_1$, then any acceptable concentration measure must lead to a strictly larger value for the data associated with $C_1$, than for the data associated with $C_2$. Similarly, with fixed T, if I(Y), the integral function of the rank-frequency function Y is situated above I(Z), the integral function of the rank-frequency function Z, $Z \neq Y$, then any impact measure or impact bundle, must give a value that is strictly larger for Y than for Z. Yet, if Lorenz curves intersect, then the relation between the concentration values of the two cases depends on the used – valid! - concentration measure. Similarly, if the graphs of I(Z) and I(Y) intersect, then one may have a higher impact than the other, or vice versa, depending on the impact measure or bundle one uses. For bundles, this relation is determined by $m_\theta$ where $m_\theta(Z)$ and $m_\theta(Y)$ respect the impact order $-<_{\neq}$ of Z and Y for all θ in a certain interval.

Figure 1, illustrates the difference between the notions of concentration and impact. Y has a higher impact than Z (obvious because Y> Z), but Z is more concentrated. Moreover, $Y_1$ with equation y= (2T-2x) has a higher impact than Y, with equation y = (T-x/T) because $I(Y_1) > I(Y)$. Indeed $I(Y_1)(x) = x(2T-x)$, while $I(Y)(x) = Tx - x^2/(2T)$ and for all $0 < x \leq T$, $x(2T-x) > Tx - x^2/(2T)$.

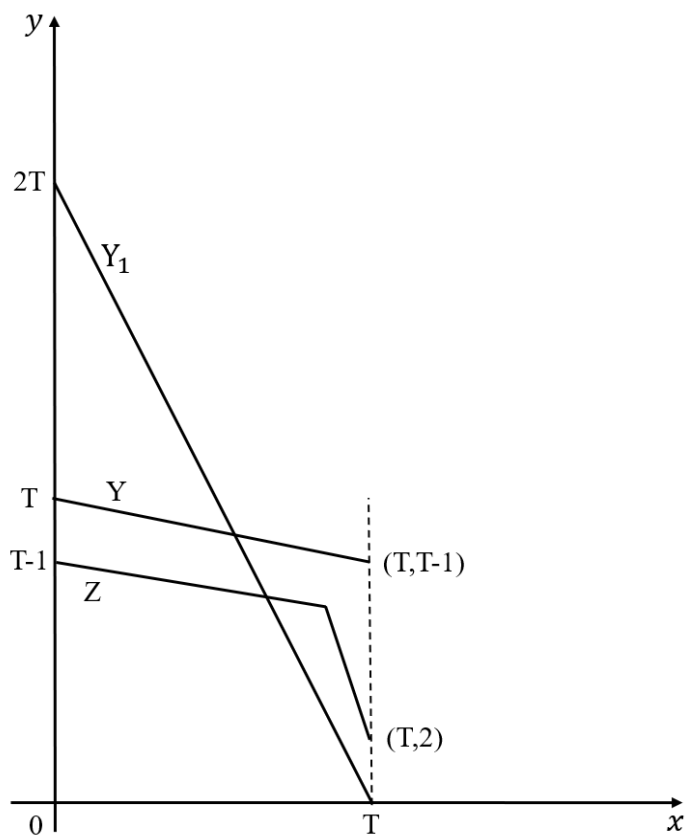

Figure 1. Impact (Y) vs. concentration (Z); $Y_1$ has a higher impact than Y.

We close this short note by remarking that, before we wrote our articles, Egghe (2022) and Egghe and Rousseau (2022a,b,c), many articles in informetrics already dealt with impact and impact measures but this only in what we consider as a heuristic sense. We ourselves (Egghe, 2021; Egghe & Rousseau, 2021) began with a study of impact-related measures and bundles but without studying impact itself (at least not in a mathematical way). We hope that our approach to the fundamental notion of impact will prove to be essential for the field of informetrics.

# Impact measures: what are they?

Leo Egghe

Hasselt University, Belgium

leo.egghe@uhasselt.be

Abstract

We propose a mathematical, axiomatic definition for the hitherto vague notion of "impact measure". For this we consider four axioms defined on a given set of (rank-frequency) functions (of which we want to measure impact). The most typical axiom explains how an impact measure should behave on the most productive sources appearing in this function (i.e. in the left side of this rank-frequency function).

We give an overview of "classical" impact measures and prove (in most but not in all cases) that they satisfy these axioms of impact measures.

This approach can be compared with (but is different from) the approach in econometrics where one defines what concentration is for a (rank-frequency) function. In this sense we are convinced that our approach is important in informetrics in order to understand the notion of "impact" and to further develop informetrics into the direction of an exact science.

Keywords: impact; axiomatic approach; h-index; g-index; pointwise defined measures; truncated average; truncated number of items; A-index

**Introduction**

Most scientists have some idea about what is meant by "impact" and "impact measures" but few can give a precise definition. Harter and Nisonger (1997) published one of the few attempts to define an impactful journal, namely one that publishes many articles and receives many citations. This idea was made concrete as follows: the journal impact factor (1994) is equal to the number of citations received by the journal in 1994 to articles published in 1992 and 1993. They keep the same publication window as the standard Garfield-Sher impact factor (JIF), not to give an undue advantage to older, established journals. This definition was mainly given to argue that the famous JIF (Journal impact factor) was a misnomer as it is a relative indicator, which 'punishes' a journal for publishing many articles.

In our opinion, bibliometricians, preferring a strictly logical and mathematical definition of impact and, of an impact measure, have not yet come up with a suitable definition. That is precisely what we will try to do.

In this contribution, we define a measure m as a function defined on a set of functions (its domain) with values in the positive real numbers $\mathbf{R}^+$ (its codomain). These functions represent source-item relations (Egghe, 2005) such as publications as sources and number of received citations over a given period as items. Which type of functions m can be said to measure impact?

As a way to a solution, we consider a somewhat similar idea, namely that of concentration or inequality, leading to concentration measures. In the discrete case, both theories consider arrays $X = (x_1, x_2, \ldots)$ ranked in decreasing order. Here $x_j$ is a non-negative number representing the number of items 'produced' by the source at rank j. In the continuous case, one considered continuous functions $y = Z(x)$, with $x \in [0,T]$ or $x \in \mathbf{R}^+$. In the two cases, impact and concentration, one assumes that the functions X or Z are monotone. We will moreover assume that they are decreasing. In this way, the most productive sources come first.

Now a function m is a (basic) concentration measure if it meets the following three requirements (Allison, 1978).

(i) $m(C) = 0$, with C a constant array or a constant function;

(ii) $m(aX) = m(X)$, with $a > 0$;

(iii) If Y is derived from X via an elementary transfer (see further), then $m(Y) > m(X)$.

Condition (i) is evident as a constant has no inequality. Condition (ii), known as scale invariance, expresses that only the distribution function matters. Finally, condition (iii) states that if $X = (x_1, x_2, \ldots, x_N)$ and $Y = (x_1, \ldots, x_{i-1}, x_i+a, x_{i+1}, \ldots, x_{j-1}, x_j-a, x_{j+1}, \ldots, x_N)$ with $a > 0$, then $m(Y) > m(X)$. Expressed in terms of monetary units, it states that if one takes from a poorer one and gives to an already richer one, then concentration (here of wealth) increases. We note that array Y must be decreasing and hence some rankings may change. We added the word "basic" as these requirements do not take the distance between sources (cells) into account.

What can we learn from this example? When studying impact measures condition (i) becomes

(I) $m(X) = 0$ if and only if $X=0$.

This already is a difference with concentration measures as a strictly positive, constant situation must have a positive impact.

Condition (ii) makes no sense for impact measures. If $a > 1$ then, at least intuitively, $aX$ must have a higher impact than X. More generally we will require for an impact measure m that

(II) $Y \geq X \Rightarrow m(Y) \geq m(X)$ and $Y = X \Rightarrow m(Y) = m(X)$.

Clearly, conditions (I) and (II) are necessary conditions for impact, but these conditions are certainly not sufficient, otherwise, the constant function zero, $m(X) = 0$ would be a valid impact measure.

In concentration theory the third condition provides sufficiency. We will formulate a third condition for a measure to be an impact measure in the next section. We will then check if well-known indicators such as the h-index, the g-index, the R-index, the average, and the total number of items (Rousseau et al., 2018) meet these three conditions and hence can be considered to be bona fide impact measures.

**Development and statement of the third condition for impact measures.**

Point (iii) from concentration provides a start in the direction of what we want to hold for an impact measure. Point (iii) considers sources with a high number of items. These receive even more items at the expense of a lower-ranked source. Impact measures should, in our opinion, concentrate on sources with a high number of items, and may neglect sources with a low number. The exact meaning of the words 'high' and 'low' will depend on the used measure. Representing mentally a source-item relation as a graph the sources with a high number of items are situated on the left-hand side (because arrays X or functions Z are decreasing).

In the following, we will only use continuous functions and omit the discrete case (arrays). This provides nice and pure arguments omitting possible discrete aberrations. For the moment, we keep the number of sources fixed as T. Hence functions are defined on the interval [0, T]. We note that the theory equally works for functions defined on ]0.T].

Let $U_T$ = {Z ‖ Z: [0,T] $\rightarrow$ $R^+$, continuous and decreasing} and we will consider different subsets $Z_T \subset U_T$.

Inequality between functions Y, Z in $U_T$ is expressed in the following two ways:

(α) Y ≥ Z, if for all x: Y(x) ≥ Z(x) $\in$ [0,T]:

(β) For a given set B $\subset$ [0,T] we say that Y $\gg$ Z on B if Y(x) > Z(x) on B.

Note that in the context of inequality (α), Y > Z means that Y ≥ Z and there exists a point $x_0$ such that $Y(x_0) > Z(x_0)$. Inequality (β) is different and much more demanding since we require a strict inequality in every element of B. For this reason, we denote it differently and with a double inequality sign, >>. Finally, we use the standard notations ≤ and < for the inequality between numbers, trusting the reader to make a distinction between inequality of functions and inequality of numbers.

From now on in this section we keep T fixed and hence simply write $Z$ for $Z_T$. We propose now the following requirement for impact measures:

(III.1) $\forall Z \in Z$, $\exists a_Z \in ]0,T[$ such that (Y $\in$ $Z$ and Y >> Z on [0, $a_Z$] implies that m(Y) > m(Z) ) .

We can also formulate the following requirement:

(III.2) $\forall Z \in Z$, $\exists b_Z \in ]0,T[$ such that (Y $\in$ $Z$ and Y << Z on [0, $b_Z$] implies that m(Y) < m(Z) ) .

We already note that the h-index and the g-index meet requirements (I), (II), (III.1), and (III.2), and this with $a_Z = b_Z = h_Z$ for the h-index and $a_Z = b_Z = g_Z$ for the g-index. This follows from their definitions and the continuity of Z. We return to this point later and will give complete proofs of these statements.

We know that the coefficient of variation V, defined as the standard deviation divided by the mean is a concentration measure, but it is not an impact measure as it does not meet the second requirement (II). For a constant function $Y = K > 0$ , defined on an interval [0,T], $V = 0$, but any non-constant continuous, decreasing function X for which $0 < X(x) < Y(x)$ on [0,T] has a V-value strictly larger than 0. Conversely, the h-index meets the requirements (I), (II), (III.1), and (III.2) but it is not a concentration measure because adding an item to the richest source never changes the h-index (except for the null case), see also (Egghe, 2009).

We may say that (III.1) and (III.2) express the distinguishing power of the functions m with respect to the left-hand sides of the functions in **Z**.

Proposition 1. If $Z \in$ **Z** meets requirement (III.1) or (III.2) then

$Y >> Z$ on $B = [0,T] \Rightarrow m(Y) > m(Z)$.

Proof. This is clear as $Y >> Z$ on [0,T] implies that $Y >> Z$ on [0,$a_Z$] and similarly on [0,$b_Y$].

Although (III.1) and (III.2) seem to be equivalent, this is actually not the case. Concretely, it is not true that for any **Z** and any m, meeting (III.1) implies meeting (III.2) or meeting (III.2) implies meeting (III.1). We will provide an example of a function m which meets (III.1) and not (III.2), and similarly, another function, on another **Z**, which meets (III.2) and not (III.1).

A. An example of a function m that meets (I), (II), and (III.1) but not (III.2)

Let $T > 0$ be fixed and consider the function $y = Z(x) \Leftrightarrow \frac{x}{T} + \frac{y}{T} = 1$, for $x \in$ [0,T].

For $0 < R < S < T$ we define the functions $y = Y_{R,S} \Leftrightarrow$

$$\begin{cases} \dfrac{x}{S} + \dfrac{y}{S} = 1, & \text{if } x \in [0, R] \\ y = S - R, & \text{if } x \in \,]R, T] \end{cases}$$

Function Z and functions $Y_{R,S}$ are illustrated in Fig.1.

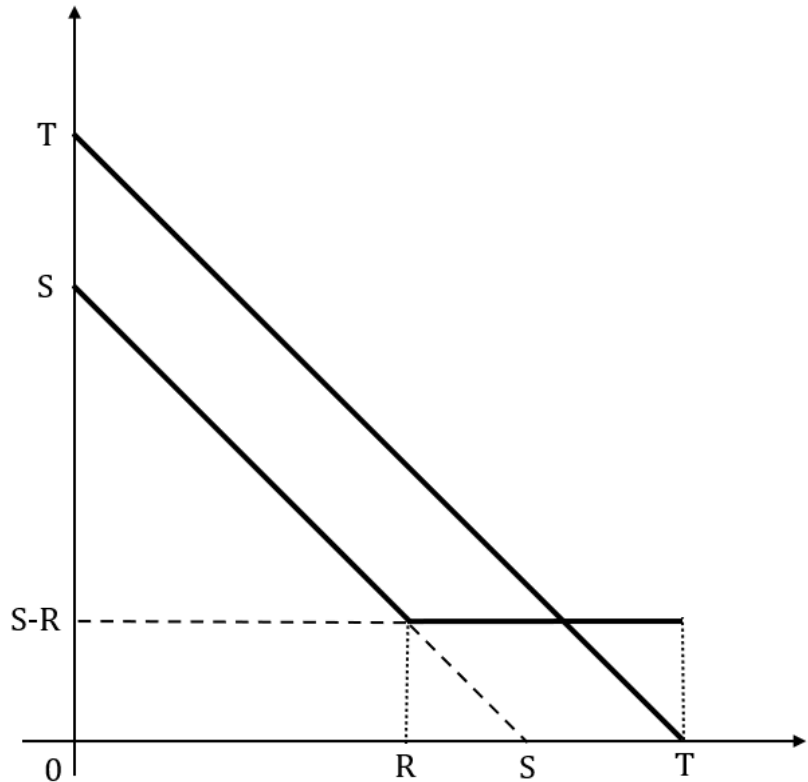

Fig. 1

We define **Z** as $\{Z, 0\} \cup \{Y_{R,S}\}$, where the second set refers to all functions $Y_{R,S}$ as defined above. We consider the measure m defined on **Z** as

$$m(Y) = \int_0^T Y(s)\,ds$$

It is clear that this function m meets requirements (I) and (II) on **Z**. We first determine R such that $m(Y_{R,S}) > m(Z) = \frac{T^2}{2}$ (this property will be needed further on). We see that $m(Y_{R,S})$ =(T-R)(S-R) + R(S+S-R)/2=TS-TR+$R^2$ − $R^2$/2 = TS-TR+$R^2$/2, which we require to be larger than $T^2$/2. Defining the quadratic function $f_S(R)$ as $R^2$ -2 TR + 2TS - $T^2$, we have to determine for which R this function is larger than or equal to zero. This function is zero if $R = \frac{2T \pm \sqrt{4T^2 - 4(2TS - T^2)}}{2}$ . Only the minus sign is meaningful as R < T. Taking the minus sign yields: $R = T - \sqrt{2T(T - S)}$, where we note that the expression under the square root sign is strictly positive. In order to have a strict inequality, we take R somewhat smaller, namely: $R = \left(T - \sqrt{2T(T - S)}\right)\frac{S}{T}$. This is the R-value we will use further on. Note though that we have to check if 0 < R < S because every valid R must meet this inequality.

a) R > 0 if $T - \sqrt{2T(T - S)} > 0 \Leftrightarrow T^2 > 2T(T-S) \Leftrightarrow S > T/2$.

b) R < S if $\left(T - \sqrt{2T(T - S)}\right)\frac{S}{T} < S$ which is clearly correct if S < T.

Now we consider a subset of **Z**, denoted **Z$_\#$** and defined as: **Z$_\#$** = $\{Z, 0\} \cup \left\{Y_S; \frac{T}{2} < S < T\right\}$, where $Y_S$ is short for $Y_{S,R}$, with R as determined above. We prove now that m meets the requirement (III.1) on **Z$_\#$**.

We will check (III.1) for the functions 0, Z and $Y_S$.

For the null function 0 we take, for example, $a_0$ = T/2 and from Y >> 0 on [0,T/2] we find that m(Y) > m(0) = 0.

For the function Z we see that there never exists a function Y and a point a such that Y >> Z on [0,a] (because $Y_S(0)$ = S < T = Z(0)). Hence for Z requirement (III.1) is void and hence logically always correct.

Now we take a fixed value S, denoted as $S_1$ and hence the corresponding $R_1$. We define $a_1 = T - \frac{S_1 - R_1}{2} < T$. Then, cf. Fig.1, the abscissa of the intersection of y = $S_1 - R_1$ with Z

= the abscissa of the intersection of $Y_{S_1}$ with Z

= T − (S$_1$-R$_1$) < a$_1$ = $T - \frac{S_1 - R_1}{2}$.

The functions 0 and Z are not larger than $Y_{S_1}$ on [0,a$_1$] so we do not have to check anything for them.

Consider now a function $Y_{S_2} \gg Y_{S_1}$ on [0,a$_1$]. Then, by the construction of the functions $Y_S$, $S_2 > S_1$. As a$_1$ is strictly larger than the abscissa of the intersection point of $Y_{S_1}$ with Z, $Y_{S_2} \gg Y_{S_1}$ in that point. As $S_2 > S_1$ and $S_2$

< T, we see that the function $Y_{S_2}$ is horizontal in that intersection point. From this, we derive that $S_2 - R_2 > S_1 - R_1$ and hence $Y_{S_2} \gg Y_{S_1}$ on [0,T]. We conclude that $m(Y_{S_2}) > m(Y_{S_1})$ proving that m meets condition (III.1) on $\mathbf{Z_\#}$.

Next, we prove that m does not meet the requirement (III.2).

Consider again the function $Z \in \mathbf{Z_\#}$ . Then for all a in ]0,T[ there exists a function $Y_S \ll Z$ on [0,a]. Indeed, it suffices to take $T - S + R > a$, which is possible as $a < T$ and $\lim_{S \to T}(T - S + R) = \lim_{S \to T}\left(T - S + \left(T - \sqrt{2T(T - S)}\right)\frac{S}{T}\right) = T$. Now, for this function $Y_S$, we have $m(Y_S) > T^2/2 = m(Z)$. This shows that m on $\mathbf{Z_\#}$ does not meet the requirement (III.2).

B. An example of a function m which meets (I), (II) and (III.2) but not (III.1).

Fix T > 0 and let $0 < R < S < T$ , $0 < P < Q \le \frac{2P+T}{3} < T$. Then we take $\mathbf{Z} = \{Z_S, 0\} \cup \{Z_{P,Q,R,S};\ \text{for all } P, Q, R, S \text{ as defined below}\}$.

For x ∈ [0,T]: we set $Z_S(x) = S$, and

$$Z_{P,Q,R,S}(x) = \begin{cases} S, & if\ x\ \in [0, P] \\ \left(\dfrac{S - R}{P - Q}x + \dfrac{RP - SQ}{P - Q}\right), & if\ x\ \in [P, Q] \\ R, & if\ x\ \in [Q, T] \end{cases}$$

The function $Z_{P,Q,R,S}$ is illustrated in Fig.2.

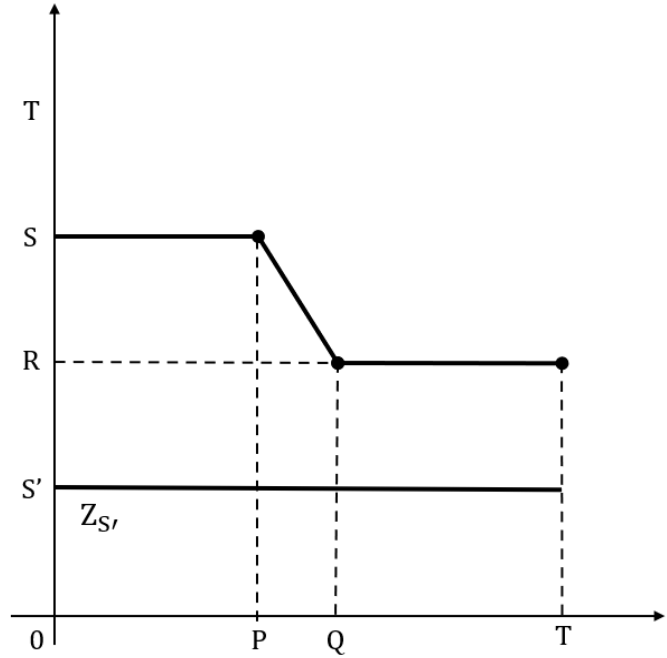

Fig. 2. An illustration of a function $Z_{S'}$ and a function $Z_{P,Q,R,S}$

The measure m is again defined as m(Z) = $\int_0^T Z(t)dt$, with Z ∈ **Z**.

It is obvious that m meets requirements (I) and (II). We show that m on **Z** meets the requirement (III.2).

For Z = $Z_{P,Q,R,S}$ we take 0 < $b_Z$ = (P+T)/2 < T. Then we have for all functions Y ∈ **Z**, with Y << Z on [0, $b_Z$] that Y << Z on [0,T], as (2P+T)/3 < $b_Z$, and hence m(Y) < m(Z).

For Z = $Z_S$ we take $b_Z$ = T/2 < T and note that for Z = 0 the condition is void.

Next, we show that m on **Z** does not meet the requirement (III.1).

Let Z = $Z_S$ and for every a ∈ ]0,T[: consider $Y_a = Z_{P',Q',R',S'}$ with P' = a, Q'= P'+ δ, S'= S+δ, R'=δ, with δ > 0 (to be determined). It is clear that $Y_a$ >> $Z_S$ on [0,a]. In order to obtain m($Y_a$) < m($Z_S$) we need: (S+δ).a + Sδ/2+δ² + (T-a-δ).δ < ST, or: aS+ δ(T+S/2) < ST, or 0 < δ < $\frac{2S(T-a)}{S+2T}$, what is possible as a < T.

Proposition 2. (i) If $sup_{Z \in \mathbf{Z}}(a_Z) < T$ , and if a measure m on **Z** meets requirement (III.1) then it also meets requirement (III.2) on **Z**.

(ii) If $sup_{Z \in \mathbf{Z}}(b_Z) < T$ , and if a measure m on **Z** meets requirement (III.2) then it also meets requirement (III.1) on **Z**.

Proof. (i) Let a = $sup_{Z \in \mathbf{Z}}(a_Z)$, with 0 < a < T. Then we take, for all Z in **Z**: $b_Z$ = a. If then Y, Z $\in \mathbf{Z}$, with Y << Z on $[0,b_Z]$ = [0,a] ⊃ [0,$a_Y$], we conclude by property (III.1) that m(Y) < m(Z), proving property (III.2). The proof of part (ii) is similar.

We note that the requirements in Proposition 2 are always met if **Z** is a finite set.

Proposition 3.

If for a measure m, requirements (III.1) and (III.2) hold, then we can take for every Z in **Z**: $a_Z = b_Z$.

Proof. It is obvious that if requirement (III.1) holds on $[0,a_Z]$ and if $a_Z$ ≤ a < T, then requirement (III.1) also holds on [0, a]. Similarly, if requirement (III.2) holds on $[0,b_Z]$ and if $b_Z$ ≤ b < T, then requirement (III.1) also holds on [0,b]. Hence taking, for every Z in **Z**: $c_Z$ = max($a_Z,b_Z$) we see that requirements (III.1) and (III.2) both hold on [0, $c_Z$].

As we consider (III.1) as well as (III.2) as very natural properties for an impact measure we are now looking for a property that includes both.

We use the following properties:

(III) $\forall X \in \mathbf{Z}$, $\exists a_X \in \,]0,T[$ such that: for all Y, Z in **Z**, (Y >> Z on [0, min($a_Y,a_Z$)] ) implies that m(Y) > m(Z).

This expression is trivially equivalent with:

(IIIa) $\forall X \in \mathbf{Z}$, $\exists a_X \in \,]0,T[$ such that: for all Y, Z in **Z**, (Y << Z on [0, min($a_Y,a_Z$)] ) implies that m(Y) < m(Z).

Besides (III) we also consider (III'):

(III') $\forall X \in \mathbf{Z}$, $\exists a_X \in \,]0,T[$ such that: for all Y, Z in **Z**, (Y >> Z on [0, max($a_Y,a_Z$)] ) implies that m(Y) > m(Z).

The idea is that we want to find out which of the two, (III) or (III'), is the most appropriate to characterize the notion of impact. The answer is given in the next theorem.

Theorem 1.

(III) ⇔ (III.1) ∧ (III.2) ⇒ (III.1) ∨ (III.2) ⇒ (III'), and (III') ⇏(III.1) ∨ (III.2)

Proof.

(a) (III) ⇒ (III.1) ∧ (III.2)

This follows immediately as, by (III) and (IIIa): $[0,a_Z] \supset [0, \min(a_Y,a_Z)]$ and $[0,a_Y] \supset [0, \min(a_Y,a_Z)]$.

(b) (III.1) ∧ (III.2) ⇒ (III)

We know already that, for all Z in **Z** we may take $a_Z = b_Z$ (in (III.1) and (III.2)) let now Y >> Z on $[0, \min(a_Y,a_Z)]$. Then we have either Y>>Z on $[0, a_Z]$ or Y>>Z on $[0,a_Y]$. In the first case (III.1) leads to m(Y) > m(Z), in the second case (III.2) yields m(Y) > m(Z). This proves part (b) and hence the equivalence of the first two expressions.

(c) (III.1) ∧ (III.2) ⇒ (III.1) ∨ (III.2) is just a logical implication.

(d) (III.1) ∨ (III.2) ⇒ (III')

Take $c_Z = a_Z$ or $b_Z$ depending on which of the two requirements holds. Now, $\forall Y, Z \in Z$, with Y>> Z on $[0, \max(a_Y,a_Z)]$ we have that Y >> Z on $[0,a_Y]$ as well as on $[0, a_Z]$. In case (III.1) holds we conclude from the first case that m(Y) > m(Z), if (III.2) holds the second case leads to the same conclusion.

(e) (III') ⇏ (III.1) ∨ (III.2)

We provide a counterexample. The impact function is still m(Z) = $\int_0^T Z(s)ds$. We already considered a case for which (III.1) holds and (III.2) does not hold. This case is defined in the square [0,T] x [0,T]. We also have an example where (III.2) holds and (III.1) does not, this time defined in [0,T] x ]0,T[.

Let **Y** be the set of functions used for the first case, and **Z** the set of functions used for the second case. We add the value T to each function

Z in **Z**, leading to **Z**$_{(+T)}$ = {Z+T: Z ∈ **Z** }. Then m still meets (III.2) while it does not meet requirement (III.1) on **Z**$_{(+T)}$. Put **Z**\* = **Y** ∪ **Z**$_{(+T)}$. We note that every function in **Z**$_{(+T)}$ is situated above every function in **Y**. Then m meets (III.1) hence (III') on **Y** and m meets (III.2) hence (III') on **Z**$_{(+T)}$. Hence m meets (III') on **Z**\*.

Now, m does not meet (III.2) on **Y**, hence also not on **Z**\* and m does not meet (III.1) on **Z**$_T$, hence also not on **Z**\*. This ends the proof of part (e), and hence of Theorem 1. □

From Theorem 1 we conclude that (III) is the property we are after. Hence we conclude that an impact measure m on the set **U** = {Z ‖ Z: [0,T] → **R**$^+$, continuous and decreasing}, or a subset **Z** of **U** is a function

m: **Z** ⊂ **U** → **R**$^+$: Z → m(Z)

which meets the following three requirements:

(I) m(Z) = 0 if and only if Z = 0.

(II) Y ≥ X ⇒ m(Y) ≥ m(X) and  Y = X ⇒ m(Y) = m(X).

(III) ∀ $X$ ∈ **Z**, ∃ $a_X$ ∈ ]0,$T$[  such that: for all Y, Z in **Z**, (Y >> Z on [0, min($a_Y$,$a_Z$)] implies that m(Y) > m(Z)).

A simple example: the continuous equivalent of the number of items, e.g. citations, of the largest source, e.g. article, namely m(Z) = Z(0).

We check the three requirements:

(I) m(Z) = 0 if and only if Z = 0, obviously;

(II) Y ≥ X ⇒ m(Y) = Y(0) ≥ m(X)=X(0)  and  Y = X  ⇒ m(Y) = Y(0) = X(0) = m(X).

(III) ∀ $X$ ∈ **Z**, ∃ $a_X$ ∈ ]0,$T$[  such that: for all Y, Z in **Z** (Y >> Z on [0, min($a_Y$,$a_Z$)] implies that m(Y) > m(Z)). Taking all $a_X$ = T/2, we see that Y >> Z on [0, T/2] implies that Y(0) > Z(0), or m(Y) > m(Z). Obviously, here we may take $a_X$ equal to any number strictly between 0 and T. Then inf($a_X$) = 0.

Proposition 4. If **Z** consists of functions ending with an interval where the function is zero, then requirements (III) and (III.1) are equivalent.

Proof. We have to show that these situations always meet requirement (III.2). Consider a function Z in **Z** such that Z = 0 on $[n_Z,T]$ ($0 \leq n_Z < T$). Now we take $b_Z = (T+n_Z)/2$ and consider a function Y in **Z** such that Y << Z on $[0, b_Z]$ . As such a function Y does not exist (III.2) is valid.

A similar observation can be made related to (III.1). Let M be a fixed strict positive number. If **Z** consists of functions Z for which Z(0) = M then all these functions Z meet requirement (III.1). Indeed, given Z, take any $a_Z$ with $0 < a_Z < T$. Then we need a function Y >> Z on $[0, a_Z]$, but such functions do not exist as Y(0) = Z(0) = M.

These observations show that (III.1) as well as (III.2) are important.

We stated earlier that we want to focus on the left-hand side of the function graph. That is what requirement (III) does. In this context, we note that if two different functions Y and Z in **Z** have the same value in zero, Y(0) = Z(0) = n (let us assume that **Z** = {Y,Z}), then condition (III) becomes empty and hence such functions always meet the requirement (III). Yet, such functions do not necessarily meet requirement (II). Assume that $0 < c_Z < c_Y < T_Z < T_Y < T$ and define the functions Y and Z as follows:

$$Y(x) = \begin{cases} n & x \in [0, c_Y] \\ -\dfrac{n}{(T_Y - c_Y)}(x - T_Y) & x \in [c_Y, T_Y] \\ 0 & x \in [T_Y, T] \end{cases}$$

and

$$Z(x) = \begin{cases} n & x \in [0, c_Z] \\ -\dfrac{n}{(T_Z - c_Z)}(x - T_Z) & x \in [c_Z, T_Z] \\ 0 & x \in [T_Z, T] \end{cases}$$

Then clearly Y ≥ Z on [0,T]. Yet, with $m(X) = \dfrac{X(0)}{T_X}$ , X in {Y,Z} we have $m(Y) = \dfrac{n}{T_Y} < \dfrac{n}{T_Z} = m(Z)$, contradicting requirement (II). Of course, (II) does not imply (III). To see this it suffices to take m(X), X in {Y,Z}, equal to a constant C > 0.

**Condition (IV) for a variable number of sources**

Until now we have kept T > 0 fixed, working on subsets of $\boldsymbol{U_T}$. We now consider the union $\boldsymbol{U}$ = $\cup_{T>0} \boldsymbol{U_T}$ = $\cup_{T>0} \{Z \parallel Z : [0, T] \rightarrow \boldsymbol{R^+}, \text{continuous and decreasing}\}$.

Assume that we take W > T and we extend functions Z defined on [0,T] and with Z(T) = 0, on ]T, W] by taking the value zero on this interval. Now we will require that a measure m does not increase strictly. More precisely: consider $\boldsymbol{U}$ and let a measure m be defined on $\boldsymbol{Z} \subset \boldsymbol{U}$. Then we formulate requirement (IV):

(IV). Let Z ∈ $\boldsymbol{U}$, with dom(Z) = [0,T] and Z(T) = 0. Then we take W > T, and define the function $Y_Z$ as

$Y_Z(x) = \begin{cases} Z(x), \text{for } x \in [0, T] \\ 0, \quad \text{for } x \in [T, W] \end{cases}$

It is clear that $Y_Z \in \boldsymbol{U}$. Then we require that $m(Y_Z) \leq m(Z)$.

We first give an example of a measure m that does not meet requirement (IV). Let m(Z) = T.Z(0), defined for any function Z in $\boldsymbol{U}$. For fixed T this is an impact measure, as it meets requirements (I), (II), and (III). Yet, with $Y_Z$ as defined above we have:

$$m(Y_Z) = W.Y_Z(0) = W.Z(0) = \frac{W}{T} m(Z) > m(Z).$$

Condition (IV) leads to a simple classification of impact measures. We can make a distinction between:

Type (IV.1): functions that meet the requirement $m(Y_Z) = m(Z)$, for all Z in $\boldsymbol{U}$

Type (IV.2): functions that meet the requirement $m(Y_Z) \leq m(Z)$ and for which $m(Y_Z) < m(Z)$ for at least one Z in $\boldsymbol{U}$.

Note that for Z=0, we always have $m(Z) = m(Y_Z) = 0$

Intuitively (details follow later) we may say that the h-index is of type IV.1, while an average is of type IV.2.

We conclude this investigation by recalling the definition of an impact measure m defined on $\boldsymbol{U}$ = $\cup_{T>0} \boldsymbol{U_T}$ = $\cup_{T>0} \{Z \parallel Z: [0,T] \to \boldsymbol{R}^+$, continuous and decreasing$\}$

The function m: $\boldsymbol{U} \to \boldsymbol{R}^+$ is an impact measure if it meets the following four requirements.

Restricted to any $\boldsymbol{Z_T} \subset \boldsymbol{U}_T$ is meets the following three requirements:

(I) m(Z) = 0 if and only if Z = 0.

(II) Y ≥ X ⇒ m(Y) ≥ m(X) and Y = X ⇒ m(Y) = m(X)

(III) $\forall\, X \in \boldsymbol{Z_T} \subset \boldsymbol{U_T}$, $\exists\, a_X \in\, ]0,T[$ such that: for all Y, Z in $\boldsymbol{Z_T}$, (Y >> Z on [0, min($a_Y$,$a_Z$)] ) implies that m(Y) > m(Z).

Moreover, on $\boldsymbol{U}$ it meets the extra requirement:

(IV). Let Z in $\boldsymbol{Z}$, with dom(Z) = [0,T] and Z(T) = 0. If now W > T, then for $Y_Z$ defined as

$Y_Z(x) = \begin{cases} Z(x), \text{for } x \in [0,T] \\ 0, \quad \text{for } x \in [T,W] \end{cases}$

m($Y_Z$) ≤ m(Z).

Similarly, we defined measures $m_T$ which are only defined on subsets $\boldsymbol{Z_T}$ of $\boldsymbol{U_T}$. Such measures are required to meet (I), (II), (III) on a given domain [0,T].

We note that if a measure m is an impact measure on a set $\boldsymbol{Z}$ then it is, trivially, also an impact measure on any subset of $\boldsymbol{Z}$. Similarly if a measure is an impact measure on a set $\boldsymbol{Z}$, consisting of functions defined on [0, T], then it is also an impact measure on the set $\boldsymbol{Z^*}$ which consists of the functions in $\boldsymbol{Z}$, restricted to the interval [0,$T_0$], with $T_0$ < T.

This ends our introduction of the definition of an impact measure. In the next part, we will study some examples.

**A study of well-known measures and their impact properties**

Before we consider some well-known measures we introduce another tool for our study.

Given an impact measure m and a set **Z** we known that the set of all numbers a ∈ ]0,T[ (we omit the index Z) for which (III.1) is valid is an interval, open on the right-hand side, ending in T. This interval, to which we now add T, is denoted as $C(m_Z)$. We set $\inf(C(m_Z)) = c(m_Z)$. Similarly, we know that the set of all numbers b ∈ ]0,T[ (again we omit the index Z) for which (III.2) is valid is an interval, open on the right-hand side, ending in T. This interval, to which too we add T, is denoted as $D(m_Z)$. We set $\inf(D(m_Z)) = d(m_Z)$.

We next show that these infima are actually minima.

Before starting the proof we recall that earlier we observed that it is possible to develop our theory for functions Z defined on ]0,T] (and not [0,T]). In that case it is possible that $\inf(C(m_Z)) = 0$, leading to $C(m_Z)$ = ]0,T]. Then the infimum is not a minimum.

Proposition 5.

(i) $C(m_Z) = [c(m_Z), T]$

(ii) $D(m_Z)) = [d(m_Z), T]$

Proof. (i). We already know that

$$]c(m_Z), T] \subset C(m_Z) \subset [c(m_Z), T]$$

If $c(m_Z)$ does not belong to $C(m_Z)$ then there exists Y in **Z**, such that Y >> Z on [0, $c(m_Z)$] and m(Y) ≤ m(Z). As Y and Z are continuous and $c(m_Z)$ < T (by (III.1)) we know that there exists a ∈ $]c(m_Z), T[$ such that Y >> Z on [0,a], with m(Y) ≤ m(Z). This is in contradiction with (III.1), hence $c(m_Z)$ ∈ $C(m_Z)$ and thus $C(m_Z) = [c(m_Z), T]$.

Similarly, part (ii) can be proven.

We include T in these intervals because, if (III.1), resp. (III.2), are valid then Y << Z on [0,T] implies that m(Y) < m(Z).

Next we wonder if (III) implies $c(m_Z) = d(m_Z)$. Surprisingly, this equality does not always hold as shown by the following example.

Example. An example of a measure m defined on a set **Z** such that for all Z in **Z** (III.1) and (III.2) hold, but $c(m_Z) \neq d(m_Z)$.

Remark: we note that this example is not in contradiction with Proposition 3.

For $0 < R < S < T$, T fixed we define the function y = Z(x), see Fig. 3, with

$$\begin{cases} \frac{x}{S} + \frac{y}{S} = 1, & \text{if } x \in [0, R] \\ y = S - R, & \text{if } x \in \,]R, T] \end{cases}$$

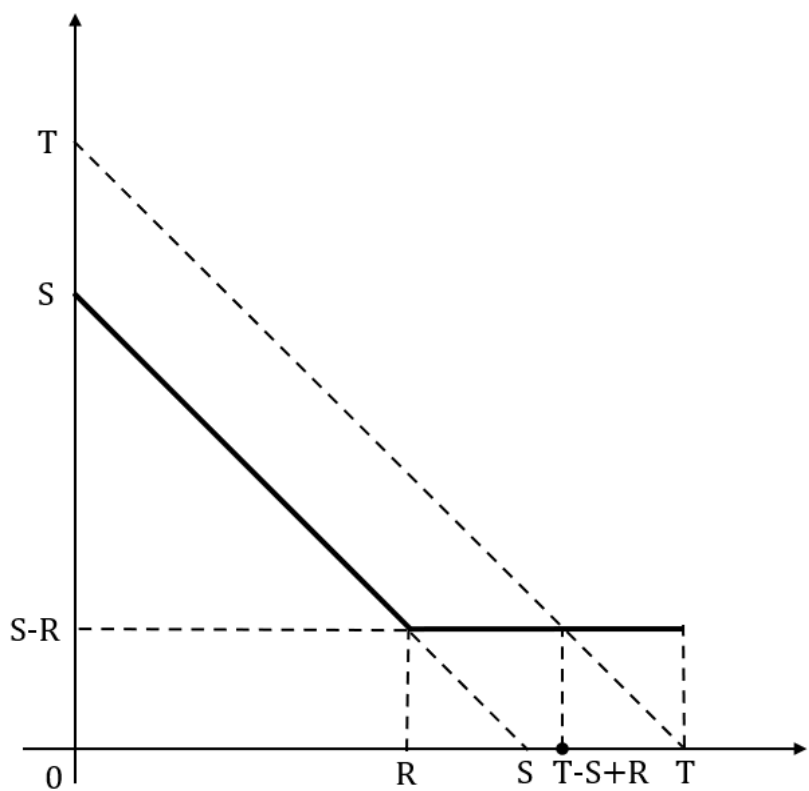

Fig. 3. The function y = Z(x), used in example b

The function Z as defined above, actually depend on R, S, and T, leading to infinity many functions Z. For simplicity we do not mention these parameters but keep in mind that T is fixed, but R and S are variable with $0 < R < S < T$.

Now we set **Z** = {Z; y = Z(x)} and m(Z) = $\int_0^T Z(x)dx$ = (S-R)T + $R^2$/2.

(a) For all Z in **Z**, (III.2) holds with $d(m_Z) \leq R$.

Proof of (a): for all Z in **Z** and Y in **Z** with Y << Z on [0,R], we see that Y<< Z on [0,T] and hence m(Y) < m(Z). This shows that R plays the role of $b_Z$ so that (III.2) holds and clearly $d(m_Z) \leq R$.

(b) for all Z in **Z**, (III.1) holds and $R < c(m_Z) \leq T-S+R$.

Proof of the first inequality in (b). We take one function Z in **Z** (parameters R < S) and take Y in **Z** (with parameters R' and S') with Y >> Z on [0,R] and m(Y) ≤ m(Z) such that (III.1) does not hold for a = R. Note that as Y >> Z on [0,R], Y(0) = S' > Z(0) = S. We write S'= S + δ, where we will determine δ later. Take R' = S < S'. Then m(Y) = (S'- R') T + $(R'^2)/2$ = (S+δ-S)T + $S^2/2$ = δT+ S2/2. We now require that this expression is strictly smaller than $(S-R)T + R^2/2$.

Hence we need: $0 < \delta < \frac{(S-R)T + (R^2 - S^2)/2}{T} = \frac{(S-R)T - (S-R)(S+R)/2}{T} = \frac{(S-R)(2T-S-R)}{2T}$. As $0 < R < S < T$ this expression is strictly positive. So we take $\delta = \frac{(S-R)(2T-S-R)}{4T} > 0$.

Proof that (III.1) holds and of the second inequality in $R < c(m_Z) \leq T-S+R$

Let Z in **Z** with parameters R and S as above and consider any function Y in **Z** (with parameters R' and S') such that Y >> Z on [0, T-S+R]. As S'< T the graph of such a function Y must be horizontal in T-S+R. As now Y(T-S+R) > Z(T-S+R) we see that Y >> Z on [0,T] and thus that m(Y) > m(Z). This shows that (III.1) holds and $c(m_Z) \leq T-S+R$.

From parts (a) and (b) we see that for all Z in **Z**: $c(m_Z) > d(m_Z)$. This completes this first example. Now we give another example where $d(m_Z) > c(m_Z)$.

An example for which for all Z in **Z**: $d(mZ) > c(m_Z)$

For 0 < R < S= (R+T)/2 < T, T fixed we define the function y = Z(x), see Fig. 4, with $y = \begin{cases} S - R, 0 \leq x \leq R \\ S - x, \ R \leq x \leq S \\ \quad 0, \ \ S \leq x \leq T \end{cases}$

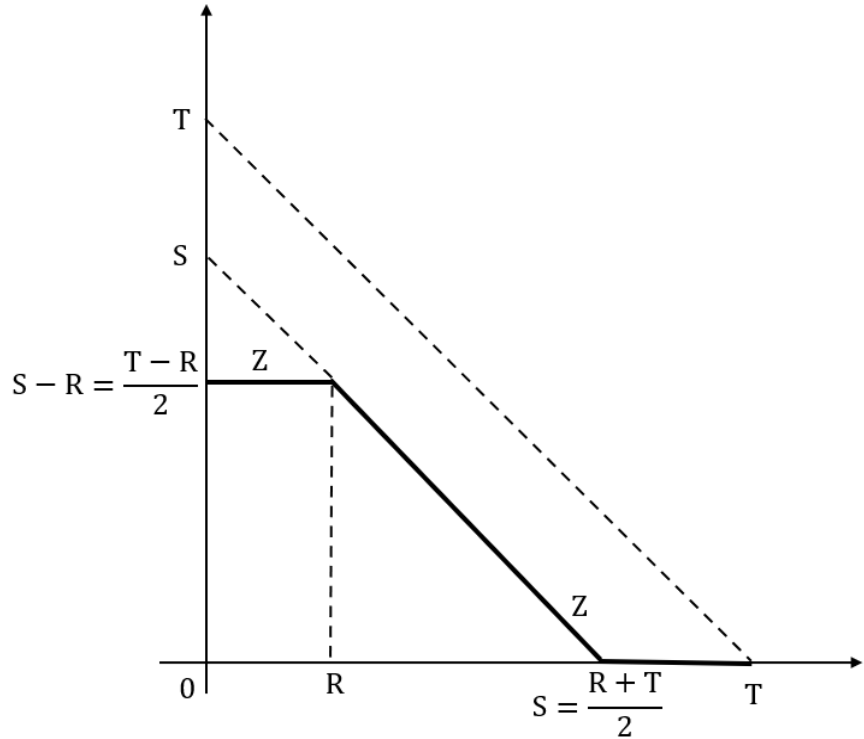

Fig. 4

As for the previous example, we note that the function Z depends on R and T, leading to infinity many functions Z. For simplicity we do not mention these parameters but keep in mind that T is fixed, but R is variable $0 < R < T$ and $S = (R+T)/2$.

Now we set $\boldsymbol{Z} = \{Z; y = Z(x)\}$ and $m(Z) = \int_0^T Z(x)\,dx = R(T-R)/2 + (T-R)^2/8$.

(a). For all Z in $\boldsymbol{Z}$: (III.1) holds with $c(m_Z) \leq R$.

If $Z_{R'}$ is in $\boldsymbol{Z}$ with $Z_{R'} >> Z_R$ on [0,R] then $Z_{R'}(0) = (T-R')/2 > Z_R(0) = (T-R)/2$. Hence $R' < R$. Yet, then also $S' = (R' + T)/2 < (R+T)/2 = S$, which implies that it is impossible that $Z_{R'} >> Z_R$ on [0,R]. This implies that (III.1) holds with $c(m_Z) \leq R$.

(b). For all Z in $\boldsymbol{Z}$: (III.2) does not hold with $b_Z = R$.

For $Z_R$ (Z with parameter R) we choose $Z_{R'}$ such that $Z_{R'} << Z_R$ on [0,R] (their existence is shown further). Then $R' > R$. We denote R' as $R+\delta$ ($\delta > 0$). Then $m(Z_{R'}) = R'(T-R')/2 + (T-R')^2/8 = (R+\delta)(T-R-\delta)/2 + (T-R-\delta)^2/8$. We require that this is strictly larger than $m(Z_R) = R(T-R)/2 + (T-R)^2/8$.

This holds if $-\delta R/2 + \delta(T-R)/2 - \delta^2/2 - \delta(T-R)/4 + \delta^2/8 > 0$

or: -R+T −R −δ -T/2 + R/2 +δ/4 > 0

or: 0 < δ < 2(T-3R)/3.

This means that we have to take R < T/3. In that case $Z_R$ and $Z_R$' exist with $Z_{R'} << Z_R$ on [0,R] and m($Z_{R'}$) > m($Z_R$).

(c). (III.2) holds for $b_Z$ = S ∈ ]0,T[ and hence there exists d($m_Z$) ≤ S such that d($m_Z$) > R (by (a)) ≥ c($m_Z$) (by (b)).

Indeed, for all Z in **Z**, Z(S) = 0. Hence there does not exist Y in **Z** such that Y >> Z on [0,S] so that we do not have to check anything.

This ends the proof of the example for which for all Z in **Z**: d($m_Z$) > c($m_Z$).

If m is a measure on **Z** then we have the following result.

Proposition 6.

Condition (III) ⇔ for all Z in **Z**: $C(m_Z) \cap D(m_Z) \cap ]0,T[ \neq \emptyset$

Proof. a. (⇒) Let Z ∈ **Z,** and let $a_Z$ ∈ ]0,T[ for which (III.1) holds. Let similarly let $b_Z$ ∈ ]0,T[ for which (III.2) holds. Then $c_Z$ = max($a_Z,b_Z$) ∈ $C(m_Z) \cap D(m_Z) \cap ]0,T[$.

b. (⇐) Let Z ∈ **Z,** and let $c_Z$ ∈ $C(m_Z) \cap D(m_Z) \cap ]0,T[$. Then (III.1) holds with $a_Z$ = $c_Z$ and (III.2) holds with $b_Z$ = $c_Z$. Hence (III) holds. □

In the sequel, we will investigate some well-known measures to determine if they meet the requirements for an impact measure. From now on we set **Z** = **U**.

Case (i) The h-index

We recall the definition of the h-index of a continuous, decreasing function Z, denoted as $h_Z$:

$$x = h(Z) \text{ (further denoted as } h_Z) \Leftrightarrow Z(x) = x$$

We note that the h-index of a continuous function Z is well-defined (the definition above is not valid if a function Z would have discontinuities). Moreover, if Z(T) > T, we say that the h-index of Z is not defined. Saying that in this case $h_Z$ = T − as is sometimes done - does not lead to an impact measure in our sense. Yet, a generalized h-index, $h_\theta$, see (Egghe & Rousseau, 2019), is defined for $\theta \geq \left(\frac{Z(T)}{T}\right)$ , the classical h-index being the case $\theta = 1$.

It is clear that the h-index meets the requirements (I) and (IV.1). Consider now requirement (II). Let Y and Z be functions in **Z**, such that $Y \geq Z$. Then $Y(h_Z) \geq Z(h_Z)$ and hence, as Y is decreasing $h_Y \geq h_Z$. This proves (II).

Now, for (III.1): consider Y and Z functions in **Z** , such that $Y >> Z$ on $[0, h_Z]$. Then $Y(h_Z) > Z(h_Z) = h_Z$, and hence, as Y is decreasing, $h_Y > h_Z$. This shows that h meets requirement (III.1) and $c(h_Z) \leq h_Z$. We next show that $c(h_Z) \geq h_Z$ .

We take the number a such that $0 < a < h_Z$ and a number δ such that $0 < δ$. Then define the function Y as follows:

$$Y(x) = \begin{cases} Z + \delta, \;\; x \in [0, a] \\ Z(a) + \delta + \dfrac{\dfrac{h_Z}{2} - Z(a) - \delta}{h_Z - a}(x - a), \quad\;\; x \in [a, h_Z] \\ \dfrac{h_Z}{2}, \quad x \in [h_Z, T] \end{cases}$$

The middle part of the definition of Y(x) is the line connecting the points (a, Z(a)+δ) and $(h_Z, h_Z/2)$. This function Y is decreasing and continuous and hence belongs to **U**. We see that $Y >> Z$ on [0,a]. Now we show that $h_Y \leq h_Z$ which will lead to the conclusion that (III.1) does not hold in a. Assume that $h_Y > h_Z$, then

$$Y(h_Y) \leq Y(h_Z) = Z(h_Z)/2 < Z(h_Z) = h_Z < h_Y.$$

which contradicts the definition of $h_Y$. From this result we see that $c(h_Z) = h_Z$ and $C(h_Z) = [h_Z, T]$.

Now we show that the h-index also meets condition (III.2). Let Y, Z in **Z** with $Y << Z$ on $[0,h_Z]$. Then $Y(h_Z) < Z(h_Z) = h_Z$ and hence $h_Y < h_Z$. This shows that h meets requirement (III.2) and hence taken together with (III.1) h meets requirement  (III).  We know that $d(h_Z) \leq h_Z$. We next consider a function Z which is strictly decreasing, take a number a such that $0 < a < h_Z$ and define Y as follows:

$$Y(x) = \begin{cases} Z(x) - \dfrac{Z(a) - h_Z}{2} \; ; \; x \in [0, a] \\ \dfrac{Z(a) + h_Z}{2} \; ; \quad x \in [a, T] \end{cases} \quad (*)$$

Now, Y belongs to **U**, because it is decreasing, continuous, and positive. We further see that Y << Z on [0,a] as a < $h_Z$ and Z is strictly decreasing (so that Z(a) > $h_Z$). We show that $h_Y \geq h_Z$ and hence (III.2) does not hold for a. Assume that $h_Y < h_Z$ then $Y(h_Y) \geq Y(h_Z) = (Z(a)+h_Z)/2 > h_Z > h_Y$, which is a contradiction. Hence we conclude that $d(h_Z) = h_Z = c(h_Z)$ and $D(h_Z) = [h_Z, T]$.

Case (ii). Pointwise defined measures (Egghe, 2021).

We recall the definition (Egghe, 2021).

A measure m defined on **Z** is pointwise defined if there exists a function y = f(θ,x), θ > 0 such that x = $m_θ$(Z), denoted as $m_{θ,Z} \Leftrightarrow Z(x) = f(θ,x)$.

Taking, with θ=1, f(θ,x) = x we obtain the h-index as studied above. Taking f(θ,x) = θx we obtain the generalized h-index, $h_θ$, see (Egghe & Rousseau, 2019) and taking f(θ,x) = θ$x^p$ (p > 0) we find the generalized Kosmulski-indices, with for θ = 1, the original Kosmulski-indices (Kosmulski, 2006).

For the pointwise defined measures m we have the following theorem.

Theorem 3. Let m be a pointwise defined measure with f(θ,x) increasing in x meets all the requirements to be an impact measure (I), (II), (III) and (IV) and $c(m_{θ,Z}) = m_θ(Z) \geq d(m_{θ,Z})$, with equality if Z is strictly decreasing.

The proof follows the lines of the proof for the h-index with y= f(θ,x) replacing f(x) = x.

Remark

We have shown that for a pointwise defined measure m, and Z strictly decreasing, c($m_Z$) = d($m_Z$), while for Z decreasing, c($m_Z$) ≥ d($m_Z$). Strict inequality for e.g., m = h occurs if, for all x in [a, $h_Z$], Z(x) = $h_Z$, with 0 < a < $h_Z$. If a is the minimum value with this property then c($h_Z$) = $h_Z$ > d($h_Z$) = a, as follows from the previous proof where we replace $h_Z$ by a and a by b, with 0 < b < a in the definition of the function Y (*).

Case (iii). The g-index

We recall the definition of the g-index for the continuous case. For $Z \in \mathbf{Z}$:

$$x = g(Z) \text{ (denoted as } g_Z) \Leftrightarrow \int_0^x Z(s)ds = x^2$$

Theorem 4. The g-index is an impact measure, with $c(g_Z) = g_Z \geq d(g_Z)$, hence $C(g_Z) = [g_Z, T]$ and, for $Z$ strictly decreasing $d(g_Z) = g_Z$, and hence $D(g_Z) = [g_Z, T]$.

Proof. It is obvious that the g-index meets conditions (I) and (IV.1). Now we consider requirement (II). Let $Y, Z \in \mathbf{Z}$, with $Y \geq Z$. Then :

$$\int_0^{g_Z} Y(s)ds \geq \int_0^{g_Y} Z(s)ds = (g_Z)^2$$

This means that $g_Y \geq g_Z$, showing that g meets requirement (II). Next, we consider (III.1). Let $Y, Z \in \mathbf{Z}$, with $Y \gg Z$ on $[0, g_Z]$. Then, using continuity, we have:

$$\int_0^{g_Z} Y(s)ds > \int_0^{g_Y} Z(s)ds = (g_Z)^2$$

and hence, by the definition of the g-index, $g_Y > g_Z$. Hence, we also have $c(g_Z) \leq g_Z$. The g-index also meets the requirement (III.2), hence (III) and $d(g_Z) \leq g_Z$. Indeed, let $Y, Z \in \mathbf{Z}$, with $Y \ll Z$ on $[0, g_Z]$. Again using continuity, we have:

$$\int_0^{g_Z} Y(s)ds < \int_0^{g_Y} Z(s)ds = (g_Z)^2$$

and thus $g_Y < g_Z$. This also leads to $d(g_Z) \leq g_Z$.

The proofs that $c(g_Z) = g_Z$ and that for strictly decreasing $Z$, $d(g_Z) = g_Z$ follows in a similar way as for h. The latter will be shown now.

Let $0 < a < g_Z$ and define for $0 < \delta < Z(a)$ ($\delta$ will be determined further on):

$$Y(x) = \begin{cases} Z(x) - \delta, & x \in [0, a] \\ Z(a) - \delta, & x \in [a, T] \end{cases}$$

Then Y∈ **U** and Y << Z on [0,a]. We will show that there exists δ such that $g_Y > g_Z$, hence $d(g_Z) \geq g_Z$ (leading to equality). As Z is strictly decreasing we have $h_Z < g_Z$, we may obviously assume that $a > h_Z$. Now, for x > a, we have:

$$\int_0^x Y(s)ds = \int_0^a Y(s)ds + \int_a^x Y(s)ds$$

$$= \int_0^a (Z(x) - \delta)ds + \int_a^x (Z(a) - \delta)ds$$

$$= \int_0^a Z(s)ds + xZ(a) - aZ(a) - \delta x.$$

Then $x = g_Y \Leftrightarrow \int_0^a Z(s)ds + xZ(a) - aZ(a) - \delta x = x^2$

$$\Leftrightarrow x^2 + x(\delta - Z(a)) + aZ(a) - \int_0^a Z(s)ds = 0$$

Solving this quadratic equation yields:

$$g_Y = x = \frac{(Z(a) - \delta) \pm \sqrt{(Z(a) - \delta)^2 - 4(aZ(a) - \int_0^a Z(s)ds)}}{2}$$

Z is strictly decreasing and continuous, hence $\int_0^a Z(s)ds > a\,Z(a)$, so that only the plus-sign is meaningful ($g_Y > 0$). The problem is now reduced to finding δ > 0 such that

$$g_Y = x = \frac{(Z(a) - \delta) + \sqrt{(Z(a) - \delta)^2 - 4(aZ(a) - \int_0^a Z(s)ds)}}{2} > g_Z$$

This inequality is equivalent with:

$$\sqrt{4\left(\int_0^a Z(s)ds - aZ(a)\right) + (Z(a) - \delta)^2} > 2g_Z + \delta - Z(a)$$

We first show that the right-hand side of this inequality is positive. We know that $Z(a) < h_Z < a < g_Z$. Hence $2g_Z + \delta - Z(a)$ is certainly positive. This allows us to square the above inequality, leading to:

$$4 \left( \int_0^a Z(s)ds - aZ(a) \right) + (Z(a) - \delta)^2 > (2g_Z + \delta - Z(a))^2$$

$$\Leftrightarrow \int_0^a Z(s)ds - aZ(a) > (g_Z)^2 + \delta g_Z - g_Z Z(a)$$

From this inequality we see that we can find the required δ > 0 if the following inequality holds:

$$\int_0^a Z(s)ds - aZ(a) > (g_Z)^2 - g_Z Z(a) = \int_0^{g_Z} Z(s)ds - g_Z Z(a)$$

As $a < g_Z$, this inequality holds if:

$$\int_a^{g_Z} Z(s)ds < (g_Z - a)Z(a)$$

This inequality follows from the fact that Z is continuous and strictly decreasing. So, with such a δ we have $g_Y > g_Z$ which contradicts (III.2) for $a = g_Z$ and hence $d(g_Z) \geq g_Z$. We conclude that $d(g_Z) = g_Z$. □

A similar proof shows that the generalized g-index $g_\theta$ (Egghe & Rousseau, 2019) is a valid impact measure.

(iv) The R-index

For Z in **U**, the continuous R-index, R > 0, is defined as (Jin et al., 2007; Egghe & Rousseau, 2008):

$$R^2(Z) = (R_Z)^2 = \int_0^{h_Z} Z(s)ds$$

where $h_Z$ is the h-index of Z.

Theorem 5. The R-index is an impact measure and $c(R_Z) = h_Z \geq d(h_Z)$; hence $C(R_Z) = [h_Z, T]$. For Z strictly decreasing we have $d(R_Z) = h_Z$ and hence $D(R_Z) = [h_Z, T]$. This is the first case of a measure m for which $c(m_Z) \neq m_Z$.

Proof. The R-index clearly meets requirements (I) and (IV.1). To prove (II0 we consider Y, Z in **Z**, with Y ≥ Z and hence $h_Y ≥ h_Z$, using the fact that h is an impact measure. Hence:

$$R_Y^2 = \int_0^{h_Y} Y(s)ds \geq \int_0^{h_Z} Z(s)ds = R_Z^2$$

For (III.1) we again take Y,Z in **Z**, now with Y > Z on [0,$h_Z$]. As h meets (III) it follows that $h_Y > h_Z$. Hence, using the continuity of Y and Z we have:

$$R_Y^2 = \int_0^{h_Y} Y(s)ds > \int_0^{h_Z} Y(s)ds > \int_0^{h_Z} Z(s)ds = R_Z^2$$

Hence (III.1) holds for R and $c(R_Z) ≤ h_Z$.

To prove (III.2) we take Y, Z in **Z**, and Y < Z on [0, $h_Z$]. As h satisfies requirement (III) we obtain $h_Y < h_Z$ and hence, using the continuity of Y and Z:

$$R_Y^2 = \int_0^{h_Y} Y(s)ds < \int_0^{h_Z} Y(s)ds < \int_0^{h_Z} Z(s)ds = R_Z^2$$

This shows that (III.2) holds and $d(R_Z) ≤ h_Z$. This proves (III). To finish the proof Theorem 5 we still have to show that $c(R_Z) = h_Z$ and, for strictly decreasing functions Z, that $d(R_Z) = h_Z$. This can be done using the same functions as those used in the proof for h and will not be repeated here. □

Similar results hold for the generalized R-index $R_\theta$ defined as:

$$R_\theta^2(Z) = \int_0^{h_{\theta,Z}} Z(s)ds$$

This is left to the reader.

Next, we study the truncated average.

(V). The truncated average $\mu_\theta$

Given Z in **Z**, we define the θ-truncated average, with 0 < θ ≤ T, as $\mu_\theta(Z)$, denoted as $\mu_{\theta,Z} = \frac{1}{\theta}\int_0^\theta Z(s)ds$. If θ =T then we have the usual average on the interval [0,T].

Theorem 6. The truncated average is an impact measure if θ < T. If θ =T (the usual average) it only meets requirement (I), (II) and (IV.2). For θ < T we have c($\mu_{\theta,Z}$) = θ ≥ d($\mu_{\theta,Z}$) and hence C($\mu_{\theta,Z}$) = [θ,T], while C($\mu_Z$) = {T}. If θ < T and Z is strictly decreasing then d($\mu_{\theta,Z}$) = θ and hence D($\mu_{\theta,Z}$) = [θ,T]. If θ = T, then D($\mu_Z$) = {T}.

Before starting the proof we like to point out that all c- and d-values are independent of the function Z.

Proof. It is clear that all $\mu_\theta$, 0 < θ ≤ T meet requirements (I) and (IV.2). Also (II) is satisfied: for Y,Z in **Z**, Y ≥ Z we have:

$$\mu_{\theta,Y} = \frac{1}{\theta}\int_0^\theta Y(s)ds \geq \frac{1}{\theta}\int_0^\theta Z(s)ds = \mu_{\theta,Z}$$

For the proof that the truncated average meets requirement (III.1) we take Y,Z in **Z**, with Y >> Z on [0, θ]. Then, by the continuity of Y and Z we have that

$$\mu_{\theta,Y} = \frac{1}{\theta}\int_0^\theta Y(s)ds > \frac{1}{\theta}\int_0^\theta Z(s)ds = \mu_{\theta,Z}$$

showing that (III.1) holds for 0 < θ < T and hence c($\mu_{\theta,Z}$) ≤ θT < T. If θ = T we only have $\mu_Y > \mu_Z$ for $a_Z$ = T, which is not sufficient to prove (III.1). To prove (III.2) we take Y,Z in **Z**, with Y << Z on [0, θ]. Then, again using continuity we have

$$\mu_{\theta,Y} = \frac{1}{\theta}\int_0^\theta Y(s)ds < \frac{1}{\theta}\int_0^\theta Z(s)ds = \mu_{\theta,Z}$$

Showing that (III.2) holds for 0 < θ < T and d($\mu_{\theta,Z}$) ≤ θT. Again, if θ = T we only have $\mu_Y < \mu_Z$ for $b_Z$ = T, which is not what is needed for (III.2).

In a similar way as for the other measures we can show that c($\mu_{\theta,Z}$) = θ, for 0 ≤ θ ≤ T, leading to C($\mu_{\theta,Z}$) = [θ,T] for 0 < θ < T and C($\mu_Z$) = {T}. Moreover, for Z strictly decreasing, d($\mu_{\theta,Z}$) = θ (a proof follows), hence D($\mu_{\theta,Z}$) = [θ,T] for 0 < θ < T and D($\mu_Z$) = {T}.

Proof that for Z strictly decreasing, d(μ$_{\theta,Z}$) = θ.

Assume that $0 < a < \theta$. Then we define, for all Z in **Z**, Z strictly decreasing a function Y as follows:

$$Y = \begin{cases} Z - \delta, & on\ [0,a] \\ Z(a) - \delta, & on\ [a,T] \end{cases}$$

with 0 < δ < Z(a) to be determined further on. We see that Y ∈ **Z** and Y << Z on [0,a]. Then

$$\mu_{\theta,Y} = \frac{1}{\theta}\int_0^\theta Y(s)ds = \frac{1}{\theta}\left(\int_0^a Y(s)ds + \int_a^\theta Y(s)ds\right)$$

$$= \frac{1}{\theta}\left(\int_0^a (Z(s) - \delta)ds + \int_a^\theta (Z(a) - \delta)ds\right)$$

$$= \frac{1}{\theta}\left(\left(\int_0^a Z(s)ds\right) - \delta a + (\theta - a)Z(a) - (\theta - a)\delta\right)$$

We require that this expression is strictly larger than $\mu_{\theta,Z} = \frac{1}{\theta}\int_0^\theta Z(s)ds$.

This holds if $\int_a^\theta Z(s)ds < (\theta - a)Z(a) - \delta\theta$. We can find such a δ > 0, actually infinitely many, because Z is strictly decreasing and continuous and hence

$$\int_a^\theta Z(s)ds < (\theta - a)Z(a)$$

This shows that d(μ$_{\theta,Z}$) ≥ θ, and conclude that d(μ$_{\theta,Z}$) = θ. □

We have shown that μ$_\theta$, with 0 < θ < T is an impact measure, while this is not true for the classical average μ. This average does not even meet requirement (III') as c(μ$_Z$) = d(μ$_Z$) = T.

We realize that this is a rather surprising result. Hence we provide a simple discrete example.

Let X = (10,10,10, 5) and Y = (11,11,11,1) then Y >> X on the index set {1,2,3}. The role of T is here played by the number 4. Yet μ$_Y$ = 34/4 < 35/4 = μ$_X$. Such an example cannot be given for μ$_\theta$, 0 < θ < T, θ a natural number, because for such θ we always have μ$_{\theta,Y}$ = 11 > 10 = μ$_{\theta,Z}$.

We recall that in our view, the notion of impact always relates to what happens on the left-hand side and never on the complete interval [0,T].

Case (vii). The truncated total number of items.

The truncated total number of items is defined, for $0 < θ < T$ and $Z$ in **$Z$** as:

$$I_{θ,Z} = \int_0^θ Z(s)ds$$

This is the total number of items in the θ most important sources. For $θ = T$ we have the total number of items, denoted as $I$.

Theorem 7

If $0 < θ < T$, then $I_{θ,Z}$ is an impact measure. If $θ = T$ then it only meets requirements (I), (II) and (IV.1). For $θ < T$ we have: $c(I_{θ,Z}) = θ ≥ d(I_{θ,Z})$ and hence $D(I_{θ,Z}) = [θ, T]$. Yet, $C(I_Z) = \{T\}$ so that $I_Z$ is not an impact measure. If $Z$ is strictly decreasing and $0 < θ < T$, then $d(I_{θ,Z}) = θ$ and $D(I_{θ,Z}) = [θ, T]$, while for $θ = T$, $D(I_Z) = \{T\}$.

The proof follows completely the proof of Theorem 6 and hence is omitted.

The example after Theorem 6 can also be used here. We showed that while $Y >> X$ on $\{1,2,3\}$ and with $\{1,2,3,4\}$ in the role of [0,T] : $I_Y = 34 < 35 = I_Z$, yet with $θ = 1$ we have $I_{θ,Y} = 11 > 10 = I_{θ,Z}$, with $θ = 2$ we have $I_{θ,Y} = 22 > 20 = I_{θ,Z}$, and for $θ = 3$ we have $I_{θ,Y} = 33 > 30 = I_{θ,Z}$.

Case (vii): Percentiles

Definition. Let $Z$ be in **$Z$** defined on [0,T], then a 100 θ% percentile ($0 < θ < 1$) for $Z$ is defined as:

$$P_{θ,Z} = Z(θT)$$

Theorem 8

The 100 θ% percentile $P_{θ,Z}$ is an impact measure and $c(P_{θ,Z}) = θT ≥ d(P_{θ,Z})$ and hence $C(P_{θ,Z}) = [θT, T]$. If $Z$ is strictly decreasing then also $d(P_{θ,Z}) = θT$ and $D(P_{θ,Z}) = [θT, T]$.

Proof. Clearly $P_{\theta,Z}$ always meets requirements (I), (II), and (IV.2). To show (III.1) we consider Y, Z in **Z** with Y >> Z on [0, θT]. Then Y(θT) = $P_{\theta,Y} > P_{\theta,Z} = Z(\theta T)$, which proves (III.1) with $c(P_{\theta,Z}) \leq \theta T$. For (III.2) we take Y,Z in **Z** with Y << Z on [0, θT]. Then, clearly, Y(θT) = $P_{\theta,Y} < P_{\theta,Z} = Z(\theta T)$, which proves (III.2) with $d(P_{\theta,Z}) \leq \theta T$. Similarly, as for the previous measures, we have that $c(P_{\theta,Z}) = \theta T$ and, if Z is strictly decreasing, $d(P_{\theta,Z}) = \theta T$. This equality is shown now. Let Z be given and let $0 < a < \theta T$. Then we define for $0 < \delta < Z(a)$:

$$Y(x) = \begin{cases} Z(x) - \delta, & x \in [0, a] \\ Z(a) - \delta, & x \in [a, T] \end{cases}$$

As $a < \theta T$ and Z is strictly decreasing we have $Z(a) > Z(\theta T)$. Next, we choose δ such that

$$Z(a) > \delta = \frac{Z(a) - Z(\theta T)}{2} > 0$$

Then, as $\theta T > a$: $Y(\theta T) = Z(a) - \delta = (Z(a) + Z(\theta T))/2 > Z(\theta T)$, and hence $P_{\theta,Y} > P_{\theta,Z}$. This proves that (III.2) does not hold for a. Consequently $d(P_{\theta,Z}) \geq \theta T$, leading to the required result that $d(P_{\theta,Z}) = \theta T$.□

(viii) The A-index

The A-index

For Z in **Z**, the A-index (Jin, 2006; Egghe & Rousseau, 2008), denoted as $A_Z$ is in the continuous case defined as:

$$A_Z = \frac{1}{h_Z} \int_0^{h_Z} Z(s) ds$$

It is known (Egghe & Rousseau, 2008) that in the discrete case this is not a good index because it is possible that for the arrays $X_1$ and $X_2$ $h(X_1) > h(X_2)$, while $A(X_1) < A(X_2)$. A simple example given in (Egghe & Rousseau, 2008) consists of taken $X_1 = (10,2)$ and $X_2 = (10,1)$. We will next show that its continuous version is not a good measure either, i.e., is not an impact measure in our sense.

Theorem 9. For all strictly decreasing Z in **Z** , C(A_Z) = D(A_Z) = ∅, hence c(A_Z) and d(A_Z) do not exist. Hence the A-index does not meet requirements (II) and (III) and cannot be considered an impact measure in our sense.

Proof. It suffices to show that T ∉ C(A_Z) and T ∉ D(A_Z) Indeed, consider Z in **Z** , strictly decreasing, and $0 < δ < h_Z$ (where the exact value of δ is determined further on). Then we define Y as follows:

$$Y(x) = \begin{cases} Z(x) - δ, & x \in [0, h_z] \\ \max(Z(x) - δ, 0), & x \in [h_Z, T] \end{cases}$$

We see that Y << Z on [0, h_Z] and hence $h_Y < h_Z$. Now $A_Y = \frac{1}{h_Y} \int_0^{h_Y} Y(s) ds$ $= \frac{1}{h_Y} \int_0^{h_Y} (Z(s) - δ) ds = \frac{1}{h_Y} \int_0^{h_Y} Z(s) ds - δ$.

Now we note that the function J: $x \to \frac{1}{x} \int_0^x Z(s) ds$ (Z > 0 fixed) is continuous and strictly decreasing. This implies that the expression $\frac{1}{h_Y} \int_0^{h_Y} Z(s) ds - \frac{1}{h_z} \int_0^{h_z} Z(s) ds$ is strictly positive. Now we define δ as $min\left(\frac{1}{2}\left(\frac{1}{h_Y} \int_0^{h_Y} Z(s) ds - \frac{1}{h_z} \int_0^{h_z} Z(s) ds\right), h_Z\right)$. Then we have:

$$A_Y \geq \frac{1}{h_Y} \int_0^{h_Y} Z(s) ds - \frac{1}{2}\left(\frac{1}{h_Y} \int_0^{h_Y} Z(s) ds - \frac{1}{h_Z} \int_0^{h_Z} Z(s) ds\right)$$

$= \frac{1}{2}\left(\frac{1}{h_Y} \int_0^{h_Y} Z(s) ds + \frac{1}{h_Z} \int_0^{h_Z} Z(s) ds\right) > \frac{1}{h_Z} \int_0^{h_Z} Z(s) ds = A_Z$ , where we have again used the fact that the function J is strictly decreasing. Thus $T \notin (C(A_Z) \cup D(A_Z))$ and hence C(A_Z) = D(A_Z) = ∅. This proves that the A-index does not meet the requirement (III).

The A-index does not even meet requirement (II). Indeed, we know that Y ≤ Z on [0,T] because, for x ∈ ]h_Z,T] max(Z(x) − δ,0) is equal to Z(x) − δ < Z(x) if Z(x) > δ, and if Z(x) ≤ δ, then Y(x) = 0 < Z(x), unless possibly for x = T, as Z is strictly decreasing and then Y(x) = Z(x). In any case Y ≤ Z while A_Y > A_Z, contradicting (II). □

**Conclusions and suggestions for further research**

In this article, we defined natural properties for a measure m to be considered an impact measure. Essentially these requirements are:

(a) m must have distinguishing power on the left-hand side of the graph of the functions Z in **Z** (the sources with the most impact) for which it is assumed to measure impact. In this we keep the domain [0,T] fixed.

(b) increasing the domain for functions by empty sources may never lead to an increase in impact.

We showed that most well-known functions used to measure impact are also impact measures in our sense, be it with some restrictions. For the average, only a truncated version meets our requirements and a similar remark holds for the number of items.

Our investigations lead to two classifications of impact measures. The first classification makes use of the property (III). Let m and n be two impact measures in our sense and assume that for a certain function Z in **Z**, $c(m_Z) \neq c(n_Z)$. We assume that $c(m_Z) < c(n_Z)$ then for a in $]c(m_Z), c(n_Z)[$ there exists Y >> Z on [0,a] with $m(Y) > m(Z)$ and $n(Y) \leq n(Z)$. In this case, m and n cannot be considered to be "equivalent" (in the sense of not acting in the same way on all functions). We suggest as a topic for further research an investigation of the equivalence (or not) of well-known measures such as the h, g, and the R index.

The second classification uses (IV). Property (IV.1) refers to those measures which stay invariant when adding empty sources, while property (IV.2) refers to measures that may become smaller when adding empty sources. This property too may be investigated further.

We are convinced that our approach solves a problem that was not fully recognized before. Yet, as this is the first time an attempt is made to define the meaning of an impact measure, we admit that it is always possible to propose another set of axioms.

**Acknowledgements.**

The author thanks Ronald Rousseau for helpful discussions, and Li Li (National Science Library, CAS) for making the figures.

# Rank-frequency data and impact in a continuous model: Introducing impact sheaves

Leo Egghe, Hasselt University, Belgium

leo.egghe@uhasselt.be

ORCID: 0000-0001-8419-2932

Ronald Rousseau, KU Leuven, Belgium

ronald.rousseau@kuleuven.be &

University of Antwerp, Belgium

ronald.rousseau@uantwerpen.be

ORCID: 0000-0002-3252-2538

ABSTRACT

First, we introduce the notion of a strong impact measure. This notion is based on the corresponding notion of a strong impact curve. Next, this notion is generalized through the notion of impact sheaves. Intuitively, they can be seen as measurements obtained by a generalized rotating beam. Superficially the corresponding theory bears some resemblance and was inspired by the theory of inequality (concentration, diversity) but it differs in two fundamental aspects. First, absolute numbers are taken into account, and second, there is a special emphasis on the high producers. Several examples of sheaves of impact measures and of strong impact measures are provided. In our opinion, measuring publication-citation impact in science means accentuating highly cited sources.

Keywords: impact, sheaves, impact measures, generalized h-index, generalized g-index, non-normalized Lorenz curve

# 1. Introduction

In this article, we continue Egghe's investigations on the nature and measurement of impact (Egghe, 2021b).

Let Z be a positive, continuous, decreasing function defined on the interval [0, T], T > 0. The set $\boldsymbol{U_T} = \{Z \parallel Z : [0,T] \to \mathbf{R}^+$, continuous and decreasing}, where $\mathbf{R}^+$ denotes the positive real numbers, contains all such functions. The endpoint T will be kept fixed in our article. Then we will simply write $\boldsymbol{U}$ for $\boldsymbol{U_T}$. Sometimes we will use a subset $\boldsymbol{V}$ ($\boldsymbol{V_T}$) of $\boldsymbol{U}$ ($\boldsymbol{U_T}$). Note that a function Z does not have to be strictly decreasing and hence $\boldsymbol{U}$ contains all constant functions, including the zero function $\boldsymbol{0}$. Subsets $\boldsymbol{V}$ will always include $\boldsymbol{0}$ but will be strictly larger than $\{\boldsymbol{0}\}$. The functions in $\boldsymbol{U}$ are continuous models for general rank-frequency functions such as authors and their articles (ranked in decreasing order of their numbers of publications); articles co-authored by one scientist and the received number of citations; one journal, its publications during one year and their received number of citations. Examples outside traditional informetrics (Rousseau et al. 2018) include web pages and their number of inlinks, the distribution of firm sizes, and city sizes in terms of population.

The impact theory we will expound on in this article is very general. There are no further restrictions on the functions Z: they do not have to be power functions, i.e., scale-free functions (Egghe, 2005), or exponential functions.

## 2. Strong impact measures

First, we define the averaging function M (capital Greek letter mu) as:

$$\mathrm{M} : \boldsymbol{U} \to \boldsymbol{U} : Z \to \mathrm{M}(Z) = \mu_Z$$

with

$$\mu_Z : [0,T] \to R^+ : x \to \frac{1}{x} \int_0^x Z(s)ds \text{ and } \mu_Z(0) = Z(0)$$

The graph of the averaging function M will be referred to as the strong impact curve. We consider this graph as the main tool to determine impact. We already note that the strong impact curve can be characterized as a non-normalized Lorenz curve, as known in concentration (inequality) theory (Lorenz, 1905; Rousseau et al., 2018, p. 88).

Definition. A strong impact measure m is a function from **U** to **R⁺**, meeting the following four axioms.

(ax.1). m(Z) = 0 if and only if Z = **0**.

(ax.2). For all Y, Z ∈ **U**:  Z ≤ Y ⇒ m(Z) ≤ m(Y).

(ax.3). For all $Y, Z \in U$: $M(Z) < M(Y)$ on $[0, T[ \Rightarrow m(Z) < m(Y)$

(ax.4). For every X in **U**, there exists $a_X$ in ]0,T[ such that Y=Z on $[0, \min(a_Y, a_Z)]$, implies m(Y) = m(Z).

The main requirement in the context of impact is (ax.3). It states that if the strong impact curve of Y is strictly situated above the strong impact curve of Y on [0,T[, then any strong impact measure of Y is strictly larger than this measure calculated for Z. Further, (ax.4) is a natural condition in our context. It points to the fact that we concentrate on the left-hand side of rank-frequency curves. It is easy to see that well-known measures such as the h-index and the g-index, meet this requirement. Moreover, there exist measures that meet (ax.1), (ax.2), and (ax.3), but not (ax.4). An example is given when considering **V**={**0**, $X_0$} ∪ {$Y_n$, n=2,3,…}, where $X_0(x)$ = T-x+1 on [0,T] and $Y_n$ = $X_0$ on the interval [0,T-T/n] and equal to the constant function (T+n)/n on ]T-T/n, T]. The measure $m_T$ is defined as $m_T(X)$ = X(T) for all X in **V**. Axioms (ax.1), (ax.2), and (ax.3) are trivially satisfied, while $m_T$ does not meet (ax.4), as all values X(T), X in **V**, are different.

We recall that in (Egghe, 2021b) an impact measure m on **V** ⊂ **U** is defined as a function: **V** ⊂ **U**, ≤ → **R⁺**, ≤ : Z → m(Z)  such that:

(I) m(Z) = 0 if and only if Z = **0**, or in words, m maps the zero function to the number zero and this is the only function that is mapped to zero.

(II) For all Z, Y in $\boldsymbol{V} \subset \boldsymbol{U}$ : Z ≤ Y ⇒ m(Z) ≤ m(Y).

(III) $\forall X \in \boldsymbol{V} \subset \boldsymbol{U}$, $\exists a_X \in \ ]0,T[$ such that: for all Y, Z in $\boldsymbol{V}$, Z < Y on [0, min(a$_Y$,a$_Z$)] implies that m(Z) < m(Y).

Recall that in (Egghe, 2021b) it is shown that requirement (III) is equivalent with the conjunction of requirements (III.1) and (III.2), where (III.1) and (III.2) are defined as follows:

(III.1) $\forall Z \in \boldsymbol{V}$, $\exists a_Z \in \ ]0,T[$ such that (Y ∈ $\boldsymbol{V}$ and Y > Z on [0, a$_Z$] implies that m(Y) > m(Z)) .

(III.2) $\forall Z \in \boldsymbol{V}$, $\exists b_Z \in \ ]0,T[$ such that (Y ∈ $\boldsymbol{V}$ and Y < Z on [0, b$_Z$] implies that m(Y) < m(Z)) .

Now, we immediately have the following proposition showing that strong impact measures on $\boldsymbol{U}$ are a subset of the impact measures in the sense of (Egghe, 2021b).

Proposition 1. If m is a strong impact measure on $\boldsymbol{U}$, then m is an impact measure in the sense of Egghe.

Proof. Requirements (I) and (II) are the same as (ax.1) and (ax.2) and hence m complies with (I) and (II). We will next show that also requirement (III) is satisfied. This means that we have to show that if $\forall X \in \boldsymbol{U}$, $\exists a_X \in \ ]0,T[$ such that: for all Y, Z in $\boldsymbol{U}$, Z < Y on [0, min(a$_Y$,a$_Z$)] then m(Z) < m(Y). For every X in $\boldsymbol{U}$ we take a$_X$ equal to the value guaranteed by (ax.4). Setting min(a$_Y$,a$_Z$) = a, we see that Z < Y on [0,a]. Now we construct the functions Z$_a$ and Y$_a$ as follows. Z$_a$(x) = Z(x) on [0, a], while Z$_a$ further connects the points (a,Z(a)) and (T,0) linearly. The function Y$_a$(x) is defined similarly. Then Z$_a$ and Y$_a$ both belong to $\boldsymbol{U}$ and $M(Z_a) < M(Y_a)$, and thus, by (ax.3) m(Z$_a$) < m(Y$_a$). Using (ax.4) and the fact that on [0,a] Z = Z$_a$ and Y=Y$_a$. We obtain m(Z) < m(Y).

Remarks

Remark 1. The opposite of Proposition 1 does not hold. We will see that the h-index, which is shown in (Egghe, 2021b) to be an impact measure, is not a strong impact measure. The h-index is also an example of a measure that meets (ax.4), but not (ax.3).

Remark 2. In the previous proposition, we used (ax.4) to prove that a strong impact measure is an impact measure in the sense of Egghe. Yet, we still need an example that (ax.4) together with requirements (I) and (II), does not imply this result. We first note that if for all Z in **V**: m(Z) = C (a non-negative constant) then this measure trivially meets (ax.4), but it never meets requirement (III). Yet, it does not even meet requirement (I).

We next provide a proper example.

The measure m defined as $m(X) = X(x_0)$, with $x_0$ a fixed number on $]0,T[$. Clearly, this measure meets (ax.4) with $a_X = x_0$. It also meets requirements (I) and (II). Consider the functions $Z_1$ and $Z_2$, defined on $[0,4]$, hence $T = 4$, with $Z_1$ the function that linearly connects the points $(0,10)$ and $(3,2)$ and continues as a constant function from $(3,2)$ to $(4,2)$. The function $Z_2$ linearly connects the points $(0,4)$ and $(3,3)$ and continues linearly to the point $(4,2)$. In this case, $Z_2(x)$ is strictly larger than $Z_1(x)$ for any x in the interval $]18/7,4[$, see Figure 1. So, we take, for instance, $x_0 = 3$. Then, $Z_2 < Z_1$ on $[0,18/7]$ but this does not imply that $Z_2(3) < Z_1(3)$.

This same example is a case of a function m that meets (ax.1), (ax.2), and (ax.4), but does not meet (ax.3). Indeed: $M(Z_2) <$ $M(Z_1)$ on $[0, T]$ as $\mu_{Z_1}(x)$ decreases from 10 to 6 on $[0,3]$ and then further to 5 in the point $x = 4$, while $\mu_{Z_2}(x)$ starts at the value 4 in 0 and hence must stay strictly smaller than $\mu_{Z_1}(x)$ over the entire interval $[0,4]$ (no calculation is necessary).

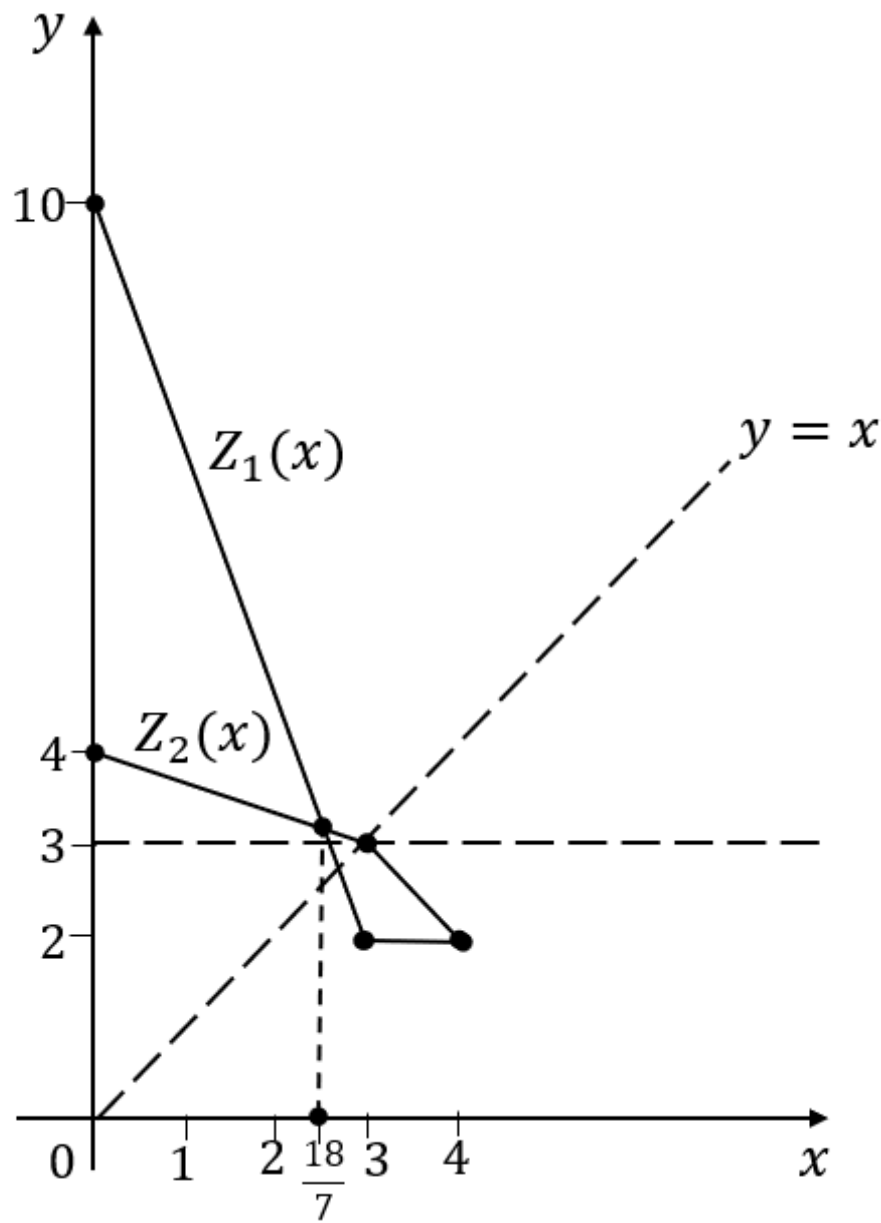

Fig. 1. Counterexample used in Remark 2

Although (ax.4) does not always imply requirement (III), we will next show that requirements (II) and (III) imply (ax.4) for pointwise continuous measures.

Definition. Let S be a set of functions defined on a closed interval [0,T]. We say that a function F defined on S is pointwise continuous iff $\forall x \in [0, T], f_n(x) \to f(x) \Rightarrow F(f_n) \longrightarrow F(f)$, where the function $f$ and all functions $f_n$ belong to S.

Proposition 2. If m is a pointwise continuous impact measure on **U** in the sense of Egghe then it meets (ax.4).

Proof. By (III) we know that $\forall X \in \boldsymbol{U}$, $\exists a_X \in \,]0,T[$ such that: for all Y, Z in **U**, Z < Y on [0, min($a_Y$,$a_Z$)] implies that m(Z) < m(Y). We assume that Y = Z on [0, min($a_Y$,$a_Z$)] and define for all n > 0:

$$Y_n^{\#} = \frac{n+1}{n}\, Y \ and\ Y_n^{\$} = \frac{n}{n+1} Y$$

Then: $Y_n^{\$} < Y < Y_n^{\#}$ and hence, by (II), $m(Y_n^{\$}) \leq m(Y) \leq m(Y_n^{\#})$ on [0,T]. On [0, min($a_Y$,$a_Z$)] we have: $Y_n^{\$} < Z < Y_n^{\#}$ and thus, by (III), we have for all n: $m(Y_n^{\$}) < m(Z) < m(Y_n^{\#})$. As $Y_n^{\#} \to Y$, $Y_n^{\$} \to Y$ and m is a pointwise continuous measure, we have $m(Y) \leq m(Z) \leq m(Y)$ or m(Y)=m(Z), which proves that m meets (ax.4).

We recall from (Egghe, 2021a) that a measure m is called a (classical) PED measure on ]0,T[ if there is a function f such that for all Z in **U**, x=m(Z) iff Z(x)=f(x).

Then we have the following result.

Proposition 3. If m is a PED measure on ]0,T[ then m meets (ax.4).

Proof. Given Z, we take $a_Z$=m(Z) $\in$ ]0,T[. If now Y = Z on [0,$a_Z$], then Y(m(Z)) = Z(m(Z)), which is equal to f(m(Z)) by the definition of a PED measure. Then, by the definition of m(Y) we have m(Y)=m(Z).

A strictly increasing continuous function f, with f(0) = 0, of a strong impact measure is also a strong impact measure. In particular, if m is a strong impact measure, then also $m^c$ (c > 0) is a strong impact function.

## 3. Examples of strong impact measures

In this section, we provide some examples of strong impact measures. In the proofs, we focus on (ax.3) as the other axioms can easily be verified.

1. Partial averages and (generalized) partial sums

All $\mu_\theta(Z) = \frac{1}{\theta}\int_0^\theta Z(s)ds$ and $I_\theta(Z) = \int_0^\theta Z(s)ds$ with $\theta$ in $]0,T[$ are strong impact measures. This also holds for $\mu_0(Z) = Z(0)$.

Proof. If, $M(Z) < M(Y)$ then we have that for every $\theta$ (fixed) in $[0,T[$, $\mu_\theta(Z) < \mu_\theta(Y)$. Then also, $I_\theta(Z) < I_\theta(Y)$.

2. All generalized g-indices, $g_\theta$, $\theta > 0$, are strong impact measures.

Proof. Assume that $M(Z) < M(Y)$, on $[0,T[$. Then, for all x in $]0,T[$: $\int_0^x Z(s)ds < \int_0^x Y(s)ds$ . Given $\theta$, we know that for continuous decreasing functions Z: $0 \neq x = g_\theta(Z)$ iff $\int_0^x Z(s)ds = \theta x^2$. From $\int_0^x Z(s)ds < \int_0^x Y(s)ds$ we derive that $\theta x^2 < \int_0^x Y(s)ds$ . By the definition of $g_\theta(Y)$, we see that $g_\theta(Z) < g_\theta(Y)$, proving that for each $\theta$, $g_\theta$ meets (ax.3).

3. Averages of averages

If, $M(Z) < M(Y)$ then we have that for every $\theta$ (fixed) in $[0,T[$, $\mu_\theta(Z) < \mu_\theta(Y)$. Hence: $\overline{\mu_Z} = \frac{1}{T}\int_0^T \mu_\theta(Z)d\theta < \frac{1}{T}\int_0^T \mu_\theta(Y)d\theta = \overline{\mu_Y}$ . This proves that $\bar\mu$ is a strong impact measure. This is rather remarkable as $\mu_T(Z) = \frac{1}{T}\int_0^T Z(s)ds$ is not even an impact measure in the sense of Egghe.

## 4. Examples of measures in the sense of Egghe that are not strong impact measures

1. The h-index is not a strong impact measure

We return to Figure 1 and the example that (ax.4) does not imply (III). Here $M(Z_2) < M(Z_1)$, but $h(Z_2) = 3 > h(Z_1) = 30/11 \approx 2.73$ (see the line y=x in Figure 1).

2. Generalized h-indices are not strong impact measures.

Proof. Consider a strict positive, fixed value of θ. For Z≠**0** in **U** (with Z(T) = 0), we denote by $x_θ$ in ]0,T] the abscissa of the intersection of y = θx and y = Z(x). Now choose Y in **U**, such that Y(0) > Z(0), Y($x_θ$) = Z($x_θ$), Y > Z on [0, $x_θ$ [ and Y=Z on [$x_θ$,T]. It is clear that now for all x in [0,T]: $\int_0^x Z(s)ds < \int_0^x Y(s)ds$ and hence M($Z$) < M(Y), but $h_θ$(Z) = $h_θ$(Y).

3. Percentiles are not strong impact measures.

Proof. Consider a fixed value of θ in ]0,T]. Given Z in **U**, we consider Y such that Y > Z on [0,θ[, and Y = Z on [θ, T]. As in the previous example, we have $\int_0^x Z(s)ds < \int_0^x Y(s)ds$ and hence M($Z$) < M(Y), but $P_θ$(Z) = Z(θ) = Y(θ) = $P_θ$(Y).

4. The $R^2$ index (and hence also the R-index) is not a strong impact measure

For all X in **V** ⊂ **U**, we assume that the h-index exists (with the h-index of the function **0** being set equal to 0. For Z ∈ **V** ⊂ **U**, the continuous $R^2$-index is defined as (Jin et al., 2007; Egghe & Rousseau, 2008):

$$R^2(Z) = (R_Z)^2 = \int_0^{h_Z} Z(s)ds$$

where $h_Z$ is the h-index of Z.

We consider the functions $Z_3$ and $Z_4$ on the interval [0,4]. Let $Z_3$(x) = $-3x + 8$ on the interval [0,2] and equal to 2 on the interval [2,4]. The function $Z_4$(x) = 3 on the whole interval [0,4]; then h($Z_4$) = 3 and h($Z_3$) = 2 and min(h($Z_3$),h($Z_4$)) = 2.

Now M($Z_4$) < M($Z_3$) on [0,4] as $\frac{1}{x}\int_0^x 3\,ds = 3$ and $\frac{1}{x}\int_0^x(-3s+8)ds = 8 - \frac{3x}{2}$ , leading to a value of 5 in the point 2, and while further $\frac{1}{4}\int_0^4 Z_3(x)dx = 14/4 > 3$. Now $R^2(Z_4) = \int_0^2(-3x+8)dx = 10$ and $R^2(Z_3) = \int_0^3 3dx = 9$.

Hence also $R^2$ and R are not strong impact measures.

## 5. Extending the theory of strong impact measures: using sheaves

We are convinced that the impact curve can play an essential role in the definition of impact. Yet, we also think that indicators such as the h-index reflect some form of impact as studied e.g., within the framework of (Egghe, 2021b). Yet, we propose to extend the theory proposed above. This extended theory uses not just one indicator, but a parameter set, a sheaf, of indicators.

Let $\boldsymbol{V} \subset \boldsymbol{U} = \{Z \; ; \; Z : [0, T] \rightarrow \boldsymbol{R}^+$, continuous and decreasing$\}$. Then a parametric set of applications is a function

$$m: \boldsymbol{V} \times Q_{m,\boldsymbol{v}} \rightarrow \boldsymbol{R}^+ : (Z,\theta) \rightarrow m(Z,\theta) = m_\theta(Z)$$

Here $Q_{m,\boldsymbol{v}} \subset (\boldsymbol{R}^+) \, \boldsymbol{U} \, \{\infty\}$ is a parameter set depending in general on the functions m and Z in $\boldsymbol{V}$. Hence, in each concrete case we will have a set $Q_{m,Z}$. Moreover, for every m and every Z in $\boldsymbol{V}$, we associate a fixed, continuous injection $\psi_{Z,m} : [0,T] \rightarrow Q_{m,Z} : x \rightarrow \psi_{Z,m}(x) = \theta$. As it will always be clear which m is meant we will simply write $\psi_Z(x)$ instead of $\psi_{Z,m}(x)$.

A parametric set of applications together with their corresponding continuous injections is called a sheaf of applications.

We recall the following proposition (Hairer & Wanner, 2008, p. 208).

Proposition 4. A continuous injective real function defined on an interval is strictly monotone.

Hence, functions $\psi_Z(x)$ introduced above, are always strictly monotone.

We next give some examples of sheaves of applications.

a) Areas under a function. Let $m_\theta(Z) = I_\theta(Z) = \int_0^\theta Z(s)ds$, with $\psi_Z(x) = x$ (the identity function). In this case, the same strictly increasing function $\psi$ and the same parameter set $Q_m$ is used for all $Z$; $Q_m = \,]0,T]$.

b) Averages of functions. Let $m_\theta(Z) = \mu_\theta(Z) = \frac{1}{\theta}\int_0^\theta Z(s)ds$, again with $\psi_Z(x) = x$. The function $m_0$ is defined as a limit, leading to $m_0(Z) = Z(0)$. Here too, the same strictly increasing function $\psi$ and the same parameter set $Q_m$ are used for all $Z$; $Q_m = [0,T]$.

c) Generalized h-indices. We recall (Egghe & Rousseau, 2019) that if $Z(T) \leq \theta\, T$, then there exists a unique point $h_\theta$ such that $Z(h_\theta) = \theta h_\theta$. The values $h_\theta$ are called generalized h-indices. This case is illustrated in Figure 2, which also nicely illustrates the use of the term sheaf.

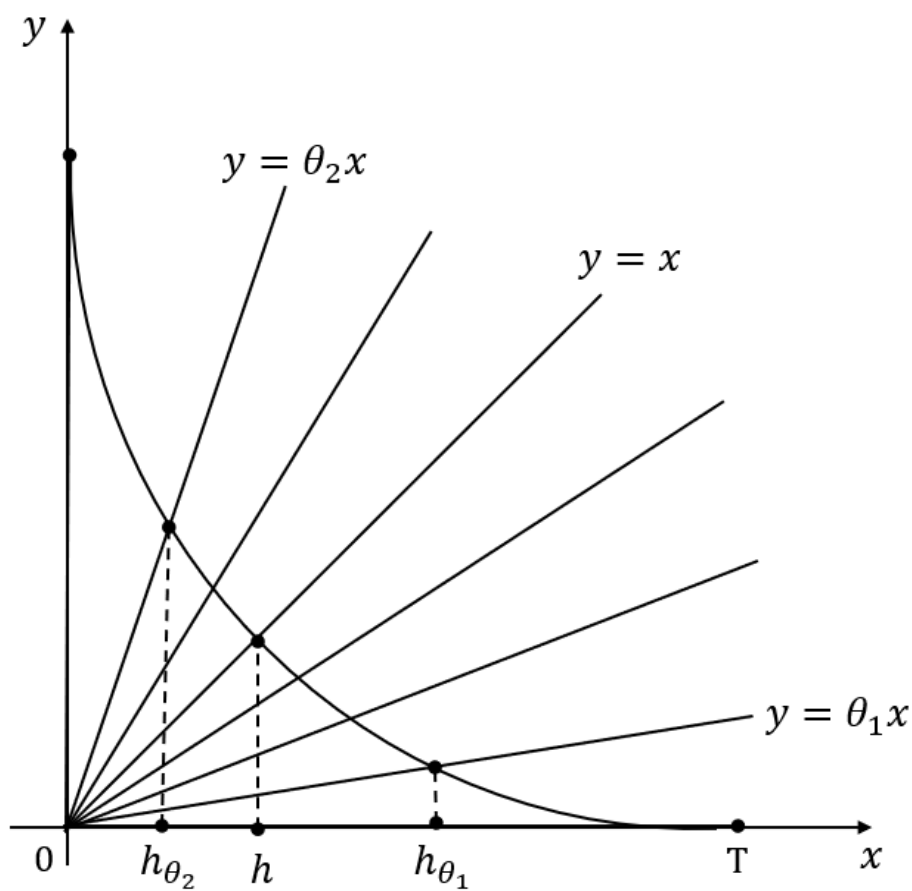

Fig.2 Generalized h-indices.

Let $m_\theta(Z) = h_\theta(Z)$ with $\theta = \psi_Z(x) = \frac{Z(x)}{x}$. Here $\psi$ depends on Z and is always strictly decreasing. Given Z then $Q_m = [Z(T)/T, +\infty]$. For simplicity, we will assume that, when studying generalized h-indices, $Z(T) = 0$ for all $Z \in \boldsymbol{V}$. Under this assumption $Q_m$ does not depend on Z and is equal to $]0, +\infty]$.

d) Generalized g-indices. We recall (van Eck & Waltman, 2008; Egghe & Rousseau, 2019) that if $Y(T) \leq \theta\, T^2$, then there exists a unique point $g_\theta$ in $[0,T]$ such that $Y(g_\theta) = \theta\,(g_\theta)^2$. Here $Y(x)$ is defined as $Y(x) = \int_0^x Z(s)ds$, with $x \in [0,T]$. The $g_\theta$ values are called generalized g-indices.

Let $m_\theta(Z) = g_\theta(Z)$ with $\theta = \psi_Z(x) = \frac{\int_0^x Z(s)ds}{x^2}$. Here too, $\psi$ and $Q_m$, depend on Z; $\psi$ is strictly decreasing. Assuming that for all Z, $\int_0^T Z(s)ds \leq \theta T^2$, $Q_m(Z) = [\frac{\int_0^T Z(s)ds}{T^2}, +\infty[$.

e) Percentiles. Let $m_\theta(Z) = P_\theta(Z) = Z(\theta)$. This is a percentile with $Q_m = [0,T[$ and $\psi_Z(x) = x$. This is the same function $\psi$ as in examples a) and b), be it with slightly different $Q_m$. We admit that usually percentiles are defined as $P_\theta(Z) = Z(\theta T)$, with $0 < \theta < 1$, but for the correspondence with the other cases mentioned above we take $0 < \theta < T$.

**Definition 1**. An impact sheaf

An impact sheaf on a set $\boldsymbol{V} \subset \boldsymbol{U}$ of positive, continuous, decreasing functions defined on [0, T] is a sheaf of applications $m_\theta$ satisfying the following three axioms (AX.1), (AX.2) and (AX.3).

(AX. 1). For all θ in $Q_m$: $m_\theta(\boldsymbol{0}) = 0$.

We will later make an important remark concerning (AX.1), see Theorem 3.

(AX.2). For all Y, Z ∈ **V**, and all θ ∈ $Q_{m,Y} \cap Q_{m,Z}$: Y ≥ Z ⇒ $m_\theta$(Y) ≥ $m_\theta$(Z).

We see that if m meets (AX.2) then for each θ ∈ $Q_{m,Y} \cap Q_{m,Z}$ $m_\theta$ meets the requirement (II) and conversely.

We next come to the axiom (AX.3) corresponding to requirement (III) (or (III.1 and III.2)). For this, we first formulate axioms 3.1 and 3.2.

A sheaf m of applications meets axioms (AX.3.1) and (AX.3.2) if:

(AX. 3.1)  For all a in ]0, T[, and for all Y, Z in **V**: Y > Z on [0, a] ⇒ $m_{\psi_{Z(a)}}(Y) > m_{\psi_{Z(a)}}(Z)$

(AX.3.2) For all a in ]0, T[, and for all Y, Z in **V**: Y < Z on [0, a] ⇒ $m_{\psi_{Z(a)}}(Y) < m_{\psi_{Z(a)}}(Z)$

Note that if we would replace $\psi_Z(a)$ in AX.3.1 by $\psi_Y(a)$ then we would obtain: For all a in ]0, T[, and for all Y, Z in **V**: Y > Z on [0, a] ⇒ $m_{\psi_{Y(a)}}(Y) > m_{\psi_{Y(a)}}(Z)$. This is: For all a in ]0, T[, and for all Y, Z in **V**: Z < Y on [0, a] ⇒ $m_{\psi_{Y(a)}}(Z) < m_{\psi_{Y(a)}}(Y)$. This is just AX.3.2 with the symbols Y and Z interchanged.

Remark. We may in (AX.3.1) replace the requirement " $m_{\psi_{Z(a)}}(Y) > m_{\psi_{Z(a)}}(Z)$ " by "for all x in ]0, a]: $m_{\psi_{Z(a)}}(Y) > m_{\psi_{Z(x)}}(Z)$". Indeed, Y > Z on [0, a] implies that for all x in [0, a] Y > Z on [0,x], and as the axiom holds for all a in ]0, T[, hence also for x in ]0, a[ , we may conclude that $m_{\psi_{Z(x)}}(Y) > m_{\psi_{Z(x)}}(Z)$.

Similarly, we may in (AX.3.2) replace " $m_{\psi_{Z(a)}}(Y) < m_{\psi_{Z(a)}}(Z)$" by "for all x in ]0, a]: $m_{\psi_{Z(x)}}(Y) < m_{\psi_{Z(x)}}(Z)$".

Proposition 5.

A sheaf m meets (AX.3.1) if and only if ∀ θ ∈ $Q_m$: $m_\theta$ meets the requirement (III.1) for impact measures, i.e., for all Z in **V**: ∃ $a_{Z,\theta}$ ∈ ]0,T[ such that for Y in **V**, Y > Z on [0, $a_{Z,\theta}$] implies $m_\theta$(Y) > $m_\theta$(Z), with θ = $\psi_Z(a_{Z,\theta})$.

Similarly, a sheaf m meets (AX.3.2) if and only if $\forall \theta \in \mathbb{Q}$: $m_\theta$ meets the requirement (III.2) for impact measures, i.e., for all Z in $\boldsymbol{V}$: $\exists\, b_{Z,\theta} \in\, ]0,T[$ such that for Y in $\boldsymbol{V}$, Y < Z on $[0, b_{Z,\theta}]$ implies $m_\theta(Y) < m_\theta(Z)$, with $\theta = \psi_Z(b_{Z,\theta})$.

Proof. We only show the proof related to (AX.3.1)

a) A sheaf m meets (AX.3.1) $\Rightarrow$ $\forall\, \theta \in Q_m$: $m_\theta$ meets the requirement (III.1).

$\forall\, \theta \in Q_m$, $\forall\, Z \in \boldsymbol{V}$ we set $a_{Z,\theta} = \psi_Z^{-1}(\theta) \in\, ]0,\, T[$. Then for Y>Z on $[0,\, a_{Z,\theta}]$, we have, by (AX.3.1) that $m_{\psi_Z(a_{Z,\theta})}(Y) > m_{\psi_Z(a_{Z,\theta})}(Z)$ and hence $m_\theta(Y) > m_\theta(Z)$ with $\theta = \psi_Z(a_{Z,\theta})$.

(b) $\forall\, \theta \in Q_m$: $m_\theta$ meets requirement (III.1) $\Rightarrow$ m meets (AX.3.1).

We have that for all a in $]0,\, T[$ and all Y, Z in $\boldsymbol{V}$ with Y > Z on $[0,\, a]$. We can take a = $a_{Z,\theta}$, because $\theta = \psi_Z(a_{Z,\theta})$ (given) and hence $\psi_Z(a_{Z,\theta}) = \psi_Z(a)$. As $\psi_Z$ is an injection we have a = $a_{Z,\theta}$. We know moreover that $m_\theta(Y) > m_\theta(Z)$, and hence, we also have $m_{\psi_Z(a_{Z,\theta})}(Y) > m_{\psi_Z(a_{Z,\theta})}(Z)$ proving (AX.3.1).

We finally formulate (AX.3), making a distinction between $\psi_Z$ and $\psi_Y$ strictly decreasing; and $\psi_Z$ and $\psi_Y$ strictly increasing.

**Definition 2:** The axiom (AX.3)

If $\psi_Z$ and $\psi_Y$ are strictly decreasing, then a sheaf m meets axiom (AX.3) iff for all a in $]0,\, T[$, and for all Y, Z in $\boldsymbol{V}$: Y > Z on $[0,\, a]$ $\Rightarrow m_\theta(Y) > m_\theta(Z)$ for all θ in $[\min(\psi_Z(a),\, \psi_Y(a)),\, +\infty$ $[\,\cap\, Q_m$.

If $\psi_Z$ and $\psi_Y$ are strictly increasing, then a sheaf m meets (AX.3) iff for all a in $]0,\, T[$, and for all Y, Z in $\boldsymbol{V}$: Y > Z on $[0,\, a]$ $\Rightarrow m_\theta(Y) > m_\theta(Z)$ for all θ in $[0,\, \max(\psi_Z(a),\, \psi_Y(a)]\, \cap\, Q_m$.

Theorem 1.

(AX.3) is equivalent with (AX.3.1) ∧ (AX.3.2).

Proof. Part 1. (AX.3) implies (AX.3.1) ∧ (AX.3.2).

We know that if $\psi_Z$ is strictly decreasing, then for all a in ]0, T[, and for all Y, Z in $\boldsymbol{V}$ with Y > Z on [0, a]: $m_\theta(Y) > m_\theta(Z)$, with θ in $[\min(\psi_Z(a), \psi_Y(a)), + \infty$ [ ∩ $Q_m$. Hence, we have:

$$m_{\psi_Z(a)}(Y) > m_{\psi_Z(a)}(Z) \text{ and } m_{\psi_Y(a)}(Y) > m_{\psi_Y(a)}(Z)$$

Using the note following the introduction of (AX.3.1) and (AX.3.2), this shows that (AX.3.1) and (AX.3.2) hold. A similar proof can be given if $\psi_Z$ is strictly increasing.

Part 2. (AX.3.1) ∧ (AX.3.2) implies (AX.3).

Again, using this note and the remark following the introduction of (AX.3.1) and (AX.3.2) we know that for all a in ]0, T[ and for all Y, Z in $\boldsymbol{V}$ with Y > Z on [0, a] :

for all θ in $[\psi_Z(a), + \infty[$ ∩ $Q_m$: $m_\theta(Y) > m_\theta(Z)$

and

for all θ in $[\psi_Y(a), + \infty[$ ∩ $Q_m$: $m_\theta(Y) > m_\theta(Z)$

where we have assumed that $\psi_Z$ and $\psi_Y$ are strictly decreasing.

As $\min(\psi_Z(a), \psi_Y(a))$ is either $\psi_Z$(a) or $\psi_Y$(a) we see that $m_\theta(Y) > m_\theta(Z)$ for all θ in $[\min(\psi_Z(a), \psi_Y(a)), + \infty$ [∩ $Q_m$.

A similar proof can be given if $\psi_Z$ and $\psi_Y$ are strictly increasing.□

Corollary.

A sheaf m of impact measures m meets (AX.3) if and only if for all θ in $Q_m$, $m_\theta$ meets requirement (III).

Notation. Given $\boldsymbol{U}$ as before, we denote by m($\boldsymbol{U}$) the set of all functions m(Z), Z in $\boldsymbol{U}$, where m(Z) is the function $Q_{m,Z} \to \boldsymbol{R}^+$: θ→ $m_\theta(Z)$.

Binary relations on $\boldsymbol{U}$

Given a in ]0, T[ we define the strict partial order relation $<_a$ on $\boldsymbol{U}$ (Roberts, 1979, p.15) as follows:

For Y, Z in **U** we have Z $\prec_a$ Y iff Z < Y on [0, a].

Given a sheaf of impact measures m, the same set **U** as before and a number a in ]0, T] we define the binary relation $<_a$ on m(**U**) as follows:

$$m(Z) <_a m(Y) \text{ iff } m_\theta(Z) < m_\theta(Y) , \theta \in \psi_Z([0,a]) \cup \psi_Y([0,a])$$

Theorem 2

The sheaf m of impact measures meets AX.3 iff, for all a in ]0, T[ m is a strictly increasing function from **V**, $\prec_a$ to m(**V**), $<_a$.

Proof.

The statement that m is a strictly increasing function from **V**, $\prec_a$ to m(**V**), $<_a$ means that if for all x in [0,a]: Z(x) < Y(x), then $m_\theta(Z) < m_\theta(Y)$ for θ on $\psi_Z([0,a]) \cup \psi_Y([0,a])$.

If $\psi_Z$ and $\psi_Y$ are strictly decreasing then, for all a in ]0, T[ we have that $[\min(\psi_Z(a), \psi_Y(a)), +\infty[ \cap Q_m = \psi_Z([0,a]) \cup \psi_Y([0,a])$, while when $\psi_Z$ and $\psi_Y$ are strictly increasing, and for all a in ]0, T[ we have that $[0, \max(\psi_Z(a), \psi_Z(a))] \cap Q_m = \psi_Z([0,a]) \cup \psi_Y([0,a])$. By the definition of (AX.3), this proves Theorem 2.

To the set of axioms studied above, we add the following axiom (AX.4) which will be used further on to prove the main theorem for impact sheaves.

(AX.4) For all a ∈ ]0,T[, and for all Y,Z ∈ **V**, such that Y=Z on [0,a]:

$$\psi_Z|_{[0,a]} = \psi_Y|_{[0,a]}$$

$$\text{and } m_\theta(Z) = m_\theta(Y), \text{ for } \theta \in \psi_Z([0,a])$$

(AX.4) is rather technical but will be essential in the proof of the main theorem. It does not exclude well-known measures such as areas under a function, $I_\theta(Z) = \int_0^\theta Z(s)ds$, averages of functions, $\mu_\theta(Z) = \frac{1}{\theta}\int_0^\theta Z(s)ds$ , the h-index, the g-index and their generalizations (Hirsch, 2005; Egghe, 2006a,b; van Eck and Waltman, 2008), and percentiles, $P_\theta(Z) = Z(\theta)$. For example, for the generalized h-index (assuming that it exists), we have $\psi_Z(x) = \frac{Z(x)}{x}$, hence if Y=Z on [0,a], $\psi_Z(x) = \frac{Z(x)}{x} = \frac{Y(x)}{x} = \psi_Y(x)$ and on [0,a], Z(x) = θx is equivalent with Y(x) = θx. This shows that $h_\theta(Z) = h_\theta(Y)$ on $\psi_Z([0, a])$.

This ends the definition of an impact sheaf for functions defined on an interval [0, T].

We recapitulate: An impact sheaf m on a set **V** of positive, continuous, decreasing functions defined on [0, T] is a sheaf of applications $m_\theta$ satisfying the following four axioms (AX.1), (AX.2), (AX.3) and (AX.4).

(AX. 1). For all θ in $Q_m$: $m_\theta(\mathbf{0}) = 0$.

(AX.2). For all Y, Z ∈ **V** and all θ ∈ $Q_{m,Y} \cap Q_{m,Z}$: Y ≥ Z ⇒ $m_\theta(Y) \geq m_\theta(Z)$.

(AX.3) If $\psi_Z$ and $\psi_Y$ are strictly decreasing, then a sheaf m meets axiom (AX.3) iff for all a in ]0, T[, and for all Y, Z in **V**: Y > Z on [0, a] ⇒ $m_\theta(Y) > m_\theta(Z)$ for all θ in [min($\psi_Z(a), \psi_Y(a)$), + ∞ [ ∩ $Q_m$.

If $\psi_Z$ and $\psi_Y$ are strictly increasing, then a sheaf m meets (AX.3) iff for all a in ]0, T[, and for all Y, Z in **V**: Y > Z on [0, a] ⇒ $m_\theta(Y) > m_\theta(Z)$ for all θ in [0, max($\psi_Z(a), \psi_Y(a)$)] ∩ $Q_m$.

(AX.4) For all a ∈ ]0,T[, and for all Y,Z ∈ **V**, such that Y=Z on [0,a]:

$$\psi_Z|_{[0,a]} = \psi_Y|_{[0,a]}$$

and $m_\theta(Z) = m_\theta(Y)$, for $\theta \in \psi_Z([0, a])$

Remark. If for all Z in **V**, $\psi_Z(x) = x$ then we have: for all a in ]0, T[, and all Y,Z in **V** : Y > Z on [0, a] implies m(Y) > m(Z) on [0, a]. This is the case for: $m_\theta(Z) = \int_0^\theta Z(s)ds$, $m_\theta(Z) = \frac{1}{\theta}\int_0^\theta Z(s)ds$ and $m_\theta(Z) = Z(\theta)$.

We next formulate an important theorem related to (AX.1). Assume that m is an impact sheaf on **V**.

Theorem 3

If m is an impact sheaf, then (AX.1) is equivalent to the expression $\left(\text{if for all } \theta \in Q_{m,Z}: m_\theta(Z) = 0 \Rightarrow Z = \mathbf{0}\right)$

Proof. Assume that $m_\theta(Z) = 0$, for all θ and Z ≠ 0. As Z is continuous and decreasing, we know that there exists a > 0 such that Z > 0 on [0, a]. Consider then the function Y = Z/2 (assumed to belong to **V**), then Z > Y on [0, a]. Then, by (AX.2) 0 = m(Z) ≥ m(Y) ≥ 0. Consequently m(Z) = m(Y) = 0 and thus, for all θ in $Q_{m,Z}$ : $m_\theta(Z) = m_\theta(Y)$, which is in contradiction with (AX.3). □

We note that this result does not hold for 'normal' measures. Indeed, define for all Z in **V**:  m(Z) = Z(a) with a fixed in ]0, T[. Then m(Z) = 0 for all Z in **V** for which Z(a) = 0. Yet, m meets requirements (I), (II), and (III), as 0 < a < T. This shows that m is an impact measure meeting (AX.1) and not Theorem 3. This is the reason why we required that "m(X) = 0 if and only if X=**0**" in (Egghe, 2021b). This observation illustrates the power of using impact sheaves.

Notation. We set for all Y, Z in **V**: [Y, Z [ = {x ∈ [0, T]: Y(y) > Z(y) on [0,x]}. This set can also be described as $[0,x_0[$ where $x_0$ is the smallest element in [0, T] such that $Z(x_0) = Y(x_0)$, at least if such a point $x_0$ exists.

Theorem 4. (AX. 3) is equivalent with the expression: ∀ Y, Z in **V**: m(Y) > m(Z) on $\psi_Z([Y > Z[) \cup \psi_Y([Y > Z[)$.

Proof. We only show this theorem for the case that $\psi_Z$ and $\psi_Y$ are strictly decreasing. We first observe that then $\psi_Z([Y > Z[) \cup$

$\psi_Z([Y > Z[) = \psi_Z([0, x_0[) \cup \psi_Y([0, x_0[) = ]\min(\psi_Z(x_0), \psi_Y(x_0)),$
$+ \infty] \cap Q_m.$

We first show the implication from left to right ( $\Rightarrow$).

$\forall a \in [0,x_0[$ we have that $Y > Z$ on $[0, a] \Rightarrow$ (by (AX.3)) $m(Y) > m(Z)$ on $[\min(\psi_Z(a), \psi_Y(a)), + \infty] \cap Q_m$. Hence $m(Y) > m(Z)$ on $\left(\cup_{a \in [0,x_0[}[\min(\psi_Z(a),\psi_Y(a)),+\infty]\right) \cap Q_m \supset ]\min(\psi_Z(x_0), \psi_Y(x_0)), + \infty] \cap Q_m$. This is so, because if b belongs to $]\min(\psi_Z(x_0), \psi_Y(x_0))]$ then $b > \psi_Z(x_0)$ or $b > \psi_Z(x_0)$, there exists $c < x_0$ such that $b = \psi_Z(c)$, or there exists $c < x_0$ such that $b = \psi_Y(c)$. Hence, $b \in [\min(\psi_Z(c), \psi_Y(c)), + \infty] \cap Q_m$. As c belongs to $[0,x_0[$, this set is a subset of $\left(\cup_{a \in [0,x_0[}[\min(\psi_Z(a),\psi_Y(a)),+\infty]\right) \cap Q_m$. We conclude that $m(Y) > m(Z)$ on $\psi_Z([0, x_0[) \cup \psi_Y([0, x_0[)$.

We next prove the other implication ($\Leftarrow$)

For all a in $]0, T[$, and for all Y, Z in $\boldsymbol{V}$ with $Y > Z$ on $[0, a]$ we have that $a < x_0$ (by definition). Hence, if $\psi_Z$ and $\psi_Y$ are strictly decreasing, then $[\min(\psi_Z(a), \psi_Y(a)), + \infty] \cap Q_m \subset ]\min(\psi_Z(x_0),\psi_Y(x_0)), + \infty] \cap Q_m$. From what is given we may conclude that $m(Y) > m(Z)$ on $[\min(\psi_Z(a), \psi_Y(a)), + \infty] \cap Q$ which shows (AX.3).

A similar proof can be given if $\psi_Z$ and $\psi_Y$ are strictly increasing. $\square$

Corollary

If for all x in $]0, T[$, $\psi_Z(x) = x$ then meeting (AX.3) is equivalent with: $\forall$ Y, Z in $\boldsymbol{V}$: $m(Y) > m(Z)$ on $[Y,Z[$

We next show some implications between the expressions "$m(Z) < m(Y)$" and "$Z < Y$" or "$Z \leq Y$" for some special measures illustrating differences in impact. These measures m are: h, P and μ, defined as follows: $h_\theta$ is the generalized h-index with parameter θ, $0 < \theta \leq +\infty$; $P_\theta(Z)$ is the $100\,\theta\%$ percentile ($0 <$ θ < 1) of Z, defined as: $P_\theta(Z) = Z(\theta T)$ and $\mu_\theta(Z) = \frac{\int_0^\theta Z(s)ds}{\theta}$ with 0 < θ < T and $\mu_0(Z) = \lim_{\theta \to 0} \mu_\theta(Z) = Z(0)$. In all cases, we assume that the functions Z belong to a given set **U**.

Theorem 5

For Z and Y in **U**, strictly decreasing, with Z(0) ≠ Y(0) we have:

$(\forall \theta \epsilon \ ]0, +\infty]: h_\theta(Z) < h_\theta(Y) \ ) \Leftrightarrow (\forall \theta \epsilon \ [0,1[: P_\theta(Z) < P_\theta(Y) \ ) \Leftrightarrow (Z < Y \ on \ [0, T[) $ .

Moreover: $[Z(0) < Y(0)] \Leftrightarrow \exists a \ \in ]0, T[$ such that $Z < Y$ on $[0, a]$

$\Leftrightarrow$ if $x_0$ is the smallest point in $]0, T]$ such that $Z(x_0) = Y(x_0)$

then $Z < Y$ on $[0, x_0[$

These two lines of equivalent expressions are connected through: $(Z < Y \ on \ [0, T[) \Rightarrow (\forall \theta \ \epsilon \ [0, T]: \mu_\theta(Z) < \ \mu_\theta(Y) \ ) \Rightarrow [Z(0) <$ Y(0)] where, in each case, the opposite implication is not valid.

Proof. First, we note (Figure 3) that the equivalences $(\forall \theta \epsilon \ ]0, +\infty]: h_\theta(Z) < h_\theta(Y) \ ) \Leftrightarrow (\forall \theta \epsilon \ [0,1[: P_\theta(Z) < P_\theta(Y) \ ) \Leftrightarrow (Z < Y \ on \ [0, T[)$ , are obvious, also when Z and/or Y are not strictly decreasing.

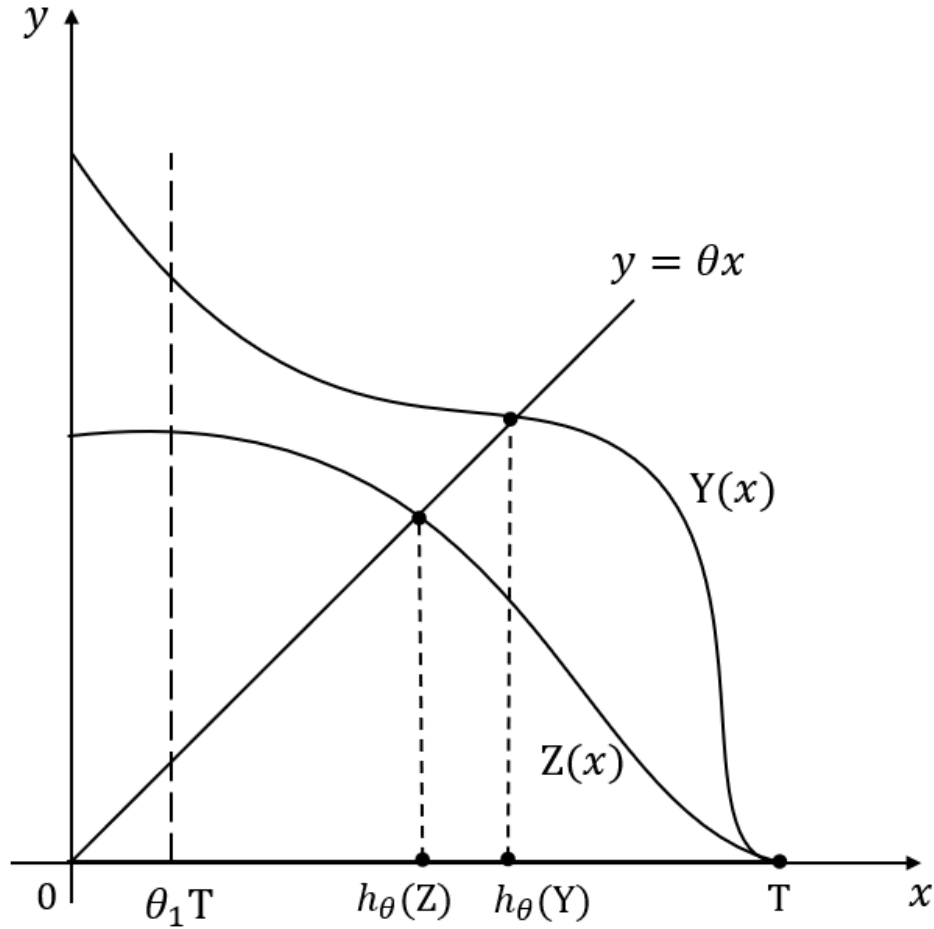

Fig.3. Curves used in Theorem 5

The implication $(Z < Y \text{ on } [0, T[) \implies (\forall \theta \in [0, T]: \mu_\theta(Z) < \mu_\theta(Y))$ is obvious for continuous functions, as is the inequality $Z(0) < Y(0)$, because this is the special case that $\theta = 0$. Finally, the equivalences $[Z(0) < Y(0)] \Leftrightarrow \exists a \in ]0, T[$ such that $Z < Y$ on $[0, a]$

$\Leftrightarrow$ if $x_0$ is the smallest point in $]0, T]$ such that $Z(x_0) = Y(x_0)$

then $Z < Y$ on $[0, x_0[$

too are trivial for continuous functions.

We still have to show that the two opposite implications in the connecting part between the two lines of equivalences do not hold. For this, we construct counterexamples. We take Z and Y decreasing with $Z(0) < Y(0)$, i.e., $\mu_0(Z) < \mu_0(Y)$. First, we want to find functions Z and Y such that $\forall \theta \in [0, T]: \mu_\theta(Z) < \mu_\theta(Y)$, but there is at least one point $x_0$ in $[0,T[$ such that $Z(x_0) \geq Y(x_0)$. We take linear functions Y and Z defined on $[0,T]$ as $Z\colon \frac{x}{T} + \frac{y}{T} = 1$ while Y is the piecewise linear function connecting the point $(0,3T)$ with $(T/4, 5T/8)$, continuing to the point $(T,0)$. Clearly, $Z(0) < Y(0)$, $Z(T/4) = 3T/4 > Y(T/4) = 5T/8$, showing that Y and Z intersect. Finally, as averages of decreasing functions decrease, as $\mu_0(Z) < \mu_0(Y)$ and as the two functions intersect just once, we only have to calculate the average in the point $(T,0)$. For the function Z, this is $T/2$; for the function Y this is $\frac{1}{T}\left(\frac{T^2}{16} + \frac{T}{8}(3T - \frac{5T}{8}) + \frac{5T}{16} \cdot \frac{3T}{4}\right) = \frac{38}{64}T$ , showing that $\forall \theta \, \epsilon \, [0,T]\colon \mu_\theta(Z) < \mu_\theta(Y)$.

For the other implication we take the same function Z. For the function Y we take the piecewise linear function connecting the points $(0,T+1)$ and $(T/4,T/4)$ followed by the line segment connecting $(T/4,T/4)$ with $(T,0)$. As $Z(0) < Y(0)$ we will calculate $\mu_\theta(Z)$ and $\mu_\theta(Y)$ for $\theta = 0.25$ and show that for this value of $\theta$: $\mu_\theta(Z) > \mu_\theta(Y)$.

$$\frac{4}{T}\int_0^{T/4} Z(x)dx = \frac{4}{T}\int_0^{T/4}(T-x)\,dx = \frac{7T}{8}$$

while $\frac{4}{T}\int_0^{T/4} Y(x)dx = \frac{4}{T}\int_0^{T/4}(-\left(\frac{4}{T}+3\right)x + (T+1))dx = \frac{5T+4}{8}$.

If now $T > 2$, then $\frac{7T}{8} > \frac{5T+4}{8}$  or $\mu_{0.25}(Z) > \mu_{0.25}(Y)$.

## 6. Impact-related Lorenz-type curves

In concentration theory, as used e.g., in economics, one has the well-known relation between a concentration measure $m_c$ and the classical Lorenz curve L (Lorenz, 1905):

$L(Z) \leq L(Y)$, with $(L(Z) \neq L(Y))$, implies that $m_c(Z) < m_c(Y)$.  (A)

We do not go deeper on this but note that it is clear from Theorem 5 that such a result cannot hold for impact measures. Indeed, consider the generalized h-indices $h_\theta$. Theorem 5 states that these measures can only meet requirement (A) for all θ if Z < Y on [0,T[ which is too strict to be a reasonable "Lorenz" condition.

Yet, we have the following result for any impact sheaf m:

$\forall \theta \in [0, T[: \mu_\theta(Z) < \mu_\theta(Y) \Rightarrow \exists a \in ]0, T[$ such that $m(Z) <_a m(Y)$ (B)

This follows from (AX.3) and Theorems 2 and 5. Note that Z(0) < Y(0) implies that $\exists a \in ]0, T[$ such that, by continuity, $Z < Y$ on $[0, a]$. Moreover, a stronger result than (B) can be proved, see the *Main Theorem for Impact Sheaves* further on. This observation confirms that the non-normalized Lorenz-type curve $\mu_Z: \theta \rightarrow \mu_\theta(Z)$, referred to as the strong impact curve, is a key element in the study of impact sheaves.

We next explain why we do not use the function I(Z): θ → $I_\theta(Z)$ = $\int_0^\theta Z(s)ds$.

To study impact, we consider intervals of the form [0, a], a < T. Of course, we want to include the case a=0 and hence [0, a] = {0}. Yet, $\lim_{\theta \to 0} I_\theta(Z) = \lim_{\theta \to 0} \int_0^\theta Z(s)ds = 0$ , while $\lim_{\theta \to 0} \mu_\theta(Z) = \lim_{\theta \to 0} \frac{\int_0^\theta Z(s)ds}{\theta} = \lim_{\theta \to 0} Z(\theta) = Z(0)$.

Moreover, we note that for $\theta > 0 : I_\theta(Z) < I_\theta(Y) \Leftrightarrow \int_0^\theta Z(s)ds < \int_0^\theta Y(s)ds \Leftrightarrow \frac{\int_0^\theta Z(s)ds}{\theta} < \frac{\int_0^\theta Y(s)ds}{\theta} \Leftrightarrow \mu_\theta(Z) < \mu_\theta(Y)$

*The Main Theorem for Impact Sheaves*

Let m be an impact sheaf defined on *V* = *U*. Then the following expressions are equivalent:

(i) m meets requirement (AX.3), i.e., if $\psi_Z$ and $\psi_Y$ are strictly decreasing, for all a in ]0, T[, and for all Y, Z in *U*: Y > Z on [0, a] ⇒ $m_\theta(Y) > m_\theta(Z)$ for all θ in [min($\psi_Z$(a), $\psi_Y$(a)), + ∞ [ ∩ Q.

If $\psi_Z$ and $\psi_Y$ are strictly increasing, then the measure m meets (AX.3) iff for all a in ]0, T[, and for all Y, Z in **U**: Y > Z on [0, a] $\Rightarrow m_\theta(Y) > m_\theta(Z)$ for all θ in [0, $\max(\psi_Z(a), \psi_Y(a))] \cap$ Q.

(ii) For all a in ]0, T[, m is strictly increasing from **U**, $<_a \to$ m(**U**), $<_a$

(iii) For all Z,Y in **U**, let $x_0$ be the smallest real number in ]0, T] such that $Z(x_0) = Y(x_0)$. Then $Z(0) < Y(0)$ implies that for all a in ]0,$x_0$[ : m(Z) $<_a$ m(Y). If $x_0$ does not exist then this property holds for all a in ]0,T[.

(iv) For all Z,Y in **U**, let $x_0$ be the smallest real number in ]0, T] such that $Z(x_0) = Y(x_0)$. Then μ(Z) < μ(Y) on [0, T[ implies that for all a in ]0,$x_0$[ : m(Z) $<_a$ m(Y). Also here, if $x_0$ does not exist then the property holds for all a in ]0,T[.

(v) For all Z,Y in **U**, Z < Y on [0, T[ implies that for all a in ]0, T[ : m(Z) $<_a$ m(Y).

Proof. (i) is equivalent with (ii). This is Theorem 2.

(ii) implies (iii) If $Z(0) < Y(0)$ then, by continuity, there is a number a in ]0, T] such that Z < Y on [0, a]. Hence, for all a in ]0,$x_0$[ we have: Z < Y on [0, a] is equivalent with Z $<_a$Y implies (by (ii)) m(Z) $<_a$ m(Y).

(iii) implies (iv) implies (v) is trivial because Z < Y on [0, T[ implies μ(Z) < μ(Y) on [0, T[, which implies that Z(0) < Y(0).

(v) implies (ii) For all a in ]0, T[ and for all Z,Y in **U** such that Z $<_a$ Y we know that Z < Y on [0, a] by the definition of $<_a$. Now we use the functions $Z_a$ and $Y_a$ again. Recall that $Z_a(x) = Z(x)$ on [0, a] continuing by connecting (a,Z(a)) and (T,0) linearly. The function $Y_a(x)$ is defined similarly. The functions $Z_a$ and $Y_a$ both belong to **V** = **U**.

Now $Z_a < Y_a$ on [0, T[. As a < T we know by (v) that m($Z_a$) $<_a$ m($Y_a$). As now $Z_a = Z$ and $Y_a = Y$ on [0, a], we have, by (AX.4), $\psi_Z|_{[0,a]} = \psi_{Z_a}|_{[0,a]}, \psi_Y|_{[0,a]} = \psi_{Y_a}|_{[0,a]}$ and hence $m_\theta(Z) = m_\theta(Z_a)$ for $\theta \in \psi_Z([0,a])$ and $m_\theta(Y) = m_\theta(Y_a)$ for $\theta \in \psi_Y([0,a])$. From this, we obtain that m(Z) <ₐ m(Y), proving (ii).

The previous main theorem clearly shows the importance of concentrating attention on the left-hand sides of rank-frequency functions when studying the notion of impact.

Moreover, μ(Z) < μ(Y) on [0, T[ in point (iv) means that the strong impact curve of Y is strictly situated above the strong impact curve of Z.

## 8. Numbers considered as impact sheaves: a number-theoretic example

In this section, we show that the framework presented above can be applied in other contexts too. Concretely, we present an example in the context of (c+1)-ary numbers, such as binary numbers (c=1), octal numbers (c=7), classical decimal numbers (c=9), and so on.

Recall that **N** = {0,1, 2, ... }, while **N₀** = **N** \ {0}. For a given c in **N₀**, we set C = {0,1,..,c}. We define a number function $Z_C$ as $Z_C$: $C\setminus\{0\} \to C \subset$ **N**: i → $a_i$ = Z(i). The null function **0** is the function mapping each i to the number zero. It exists for each C. Similar to the theory expanded before, we denote by **Z_C** the set of all number functions and note that these functions are generally not decreasing.

For θ in Q = C \ {0}, (note that #Q = c) we define:

$$m_\theta(Z_C) = \sum_{j=1}^{\theta} \frac{Z_C(j)}{(c+1)^{j-1}} = \sum_{j=1}^{\theta} \frac{a_j}{(c+1)^{j-1}} \in [0, c+1[$$

The corresponding function $\psi_Z(i) = i$. It is the same for all $Z_C$ in **Z_C** and hence we will not explicitly use the function $\psi_Z$ in this section. We further note that the set Q is the same for all $Z_C$ in **Z_C**.

In this context we say that m is an impact sheaf m on the set $\boldsymbol{Z_C}$, to be compared with the impact sheaves in the continuous context, because it meets the following four axioms (AX.1), (AX.2), (AX.3) and (AX.4).

(AX.1). For all θ in Q: $m_\theta(Z_C) = 0$ iff $Z_C = 0$.

(AX.2). For all Y, Z $\in \boldsymbol{Z_C}$ and all $\theta \in Q$: $Y \geq Z \Rightarrow m_\theta(Y) \geq m_\theta(Z)$.

(AX.3) For all θ in Q, and for all Y, Z in $\boldsymbol{Z_C}$: Y > Z on {1, …, θ}, i.e., $b_j > a_j$, 1≤j≤θ, with Y: $\boldsymbol{N_0} \to C \subset \boldsymbol{N}$: $i \to b_i = Y(i)$, clearly implies $\forall j, 1 \leq j \leq \theta: m_j(Y) > m_j(Z)$.

(AX.4) For all i in Q, and for all Y, Z $\in \boldsymbol{Z_C}$, such that Y=Z on {1,…,i}:

$$m_\theta(Z) = m_\theta(Y), \text{ for } \theta \in \{1, …, i\}$$

We further have the following stronger result, that for each fixed θ in Q, $m_\theta$ is a strong impact measure. We first write down the requirements for $m_\theta$.

(ax.1). $m_\theta(Z) = 0$ if and only if Z = $\boldsymbol{0}$.

(ax.2). For all Y, Z $\in \boldsymbol{Z_C}$: $Z \leq Y \Rightarrow m_\theta(Z) \leq m_\theta(Y)$.

(ax.3). For all $Y, Z \in \boldsymbol{U}$: $M(Z) < M(Y)$ on $C \backslash \{0\} \Rightarrow m_\theta(Z) < m_\theta(Y)$

(ax.4). For every X in $\boldsymbol{Z_C}$, there exists $i_X$ in {1, …, θ} such that Y=Z on {1, …,i}, implies $m_\theta(Y) = m_\theta(Z)$.

Here, the function M is defined as

$$M: \boldsymbol{Z_C} \to \boldsymbol{Z_C}: Z \to M(Z) = \mu_Z$$

with

$$\mu_Z: C \backslash \{0\} \to R^+: i \to \frac{1}{i} \sum_{j=1}^{i} Z(j)$$

Proposition 6. For each fixed θ in Q, $m_\theta$ is a strong impact measure.

Proof. The measure $m_\theta$ obviously meets (ax.1),(ax.2), and (ax.4). For (ax.3) the point is that for these functions $M(Z) < M(Y)$ is equivalent with $Z(1) < Y(1)$.

## 9. Discussion

In an earlier publication (Egghe, 2021b) we introduced a set of axioms required for impact measures. In that theory, the h-index and the R-index met the requirements for being an impact measure. To make a distinction with the theory introduced now, we have referred to the 'new' impact measures as strong impact measures.

In the current article, we used the term 'impact' in a generic way. In the standard case of a scientist, journal, or edited book publishing articles (or chapters) which receive citations (from sources in a certain database), we may use the term citation impact. Yet, considering the historical Zipf curve of cities in a country and their inhabitants (Auerbach, 1913) the associated 'impact' could be described as population-visibility impact (a notion related to economic and political power), where – at least in this case – military or human development aspects are not explicitly taken into account. Note that rightly, for two countries with about the same number of cities and about the same total population, the one that has several megacities (if the other has few), is considered to have the highest population-visibility impact.

Also, visibility on the Internet as aspired by 'influencers' is a form of impact that could be quantified by our approach (studying the total set of influencers and their followers). Similarly for streaming data of singers, or bestsellers list, or the richest persons on earth or per country.

For the readers' information, we would like to tell that Leo Egghe got the idea about the notion and results on impact sheaves before strong impact measures were considered. This was a natural way of thinking since impact sheaves are natural extensions of impact measures as developed in Egghe (2021b) in the sense that the same measures appear in impact sheaves, while not all impact measures are strong. Yet we chose to introduce strong impact measures first for didactical reasons, since the results on impact sheaves are somewhat more intricate.

As the notion of impact is an important element in research evaluations on all levels, we think that the theory proposed in this article may contribute to such exercises.

## 10. Conclusion

We introduced and defined the notion of strong impact measures in connection with strong impact curves. Generalizing these notions, we presented the idea of impact sheaves. We have shown that when a single function cannot solve a particular problem, the idea of a sheaf may yield the solution. Superficially the corresponding theory bears some resemblance and was inspired by the theory of inequality (concentration, diversity) but it differs in two fundamental aspects. First, absolute numbers are taken into account, and second, there is a special emphasis on the high producers. Several examples of impact sheaves and strong impact measures were provided. These include the sheaves of areas under a function, partial averages, the h- and the g-index and their generalizations, and well as percentiles. Considered on their own, many of these are strong impact measures, but the h-index and their generalizations, as well as percentiles are not. We finally present a didactical example within the theory of $(c+1)$-ary numbers. In concluding this article, we like to mention that we see multiple possibilities for further work.

Acknowledgments. The authors thank Ad Meskens for help with the references and Li Li for providing excellent figures.

# Global impact measures

Leo Egghe[1] and Ronald Rousseau[2,3] *

[1] Hasselt University, 3500 Hasselt, Belgium

leo.egghe@uhasselt.be

ORCID: 0000-0001-8419-2932

[2] KU Leuven, MSI, Facultair Onderzoekscentrum ECOOM,

Naamsestraat 61, 3000 Leuven, Belgium

ronald.rousseau@kuleuven.be  &

[3] University of Antwerp, Faculty of Social Sciences,

Middelheimlaan 1, 2020 Antwerp, Belgium

ronald.rousseau@uantwerpen.be

ORCID: 0000-0002-3252-2538

**Abstract**

We present a continuous theory of global impact measures. Such measures combine inequality (like the Lorenz theory) with productivity, leading to the notion of global impact and its measurement.

**Keywords**: Lorenz curve; inequality; global impact measures

* Corresponding author

## 1. Introduction

In this work, our investigations will lead to the introduction of a new type of measures. These measures are called global impact measures and take concentration as well as production into account. They can be considered as elements in a theory on impact as outlined in other publications (Egghe, 2021; Egghe & Rousseau, 2022a; Egghe & Rousseau, 2022b).

Many distributions studied in informetrics, such as authors and their publications (Lotka, 1926), websites and inlinks (Rousseau, 1997; Faloutsos et al., 1999), or topics and journals dealing with them (Bradford, 1934) can be described by power law relations (Egghe, 2005) or similar long-tailed distributions (Laherrère & Sornette, 1998). Moreover, these power laws have many applications in other fields such as demography (cities and their inhabitants), linguistics (words and their uses), economics (incomes in a market economy), ecology (fragmentation of forests), astronomy (initial mass functions) and many more, see (Pareto, 1895; Auerbach, 1913; Zipf, 1941, 1949; Salpeter, 1955; Newman, 2005; Saravia et al., 2018). One common characteristic of these distributions is the high concentration of items among a few sources. As such the study of concentration or inequality, with its social implications, is one of the main topics studied in our field (Rousseau et al., 2018, Section 9.5).

In the next section we introduce a dominance order in the case of a non-normalized Lorenz curve and prove an impact-concentration theorem. This then leads in the following section to the definition of global impact measures, followed by a number of practical examples. We conclude by pointing out new opportunities for further studies in the science of science.

## 2. Continuous dominance and the impact-concentration theorem

Let T > 0, let $\boldsymbol{U}$ be the set $\{Z: [0, T] \to \mathbb{R}^+, Z$ continuous and decreasing$\}$, $\boldsymbol{U_0}$ = $\{Z \in \boldsymbol{U}; Z > 0 \text{ on } [0, T[ \}$ and $\boldsymbol{Z}$ = $\{Z \in \boldsymbol{U}; Z \text{ strictly decreasing}\}$. Then $\boldsymbol{Z} \subset \boldsymbol{U_0} \subset \boldsymbol{U}$.

Definition. The continuous Lorenz curve L(x).

Given Z in $\boldsymbol{U}$, we define the continuous Lorenz curve of Z as the graph of the function

$$[0,1] \to [0,1]: x \to \frac{\int_0^{xT} Z(s)ds}{\int_0^T Z(s)ds}$$

Definition. The classical Lorenz dominance order.

The dominance order on $\boldsymbol{U}$ (Marshall-Olkin-Arnold, 2011), is defined as:

if Z, Y ∈ **U**, then $Z - <_L Y$ iff $\forall x \in [0, T]$: $\frac{\int_0^x Z(s)ds}{\int_0^T Z(s)ds} \leq \frac{\int_0^x Y(s)ds}{\int_0^T Y(s)ds}$

or equivalently: $\forall x \in [0,1]$: $\frac{\int_0^{xT} Z(s)ds}{\int_0^T Z(s)ds} \leq \frac{\int_0^{xT} Y(s)ds}{\int_0^T Y(s)ds}$.

The relation $Z - <_L Y$ means that the continuous Lorenz curve of Y is situated above the continuous Lorenz curve of Z, providing an argument in favor of using the concave, and not the convex, form of the Lorenz curve.

Definition. A continuous concentration measure

A function m from **U** to the positive real numbers is a concentration measure if it is an order morphism from **U**, -<$_L$ to the positive real numbers. This means that X-<$_L$ Y implies that m(X) ≤ m(Y), with equality only if X and Y have the same Lorenz curve. In practice one often requires that m(**0**) = 0, with **0** the zero function on [0,T].

Definition. The non-normalized Lorenz curve

Given Z in **U**, we define the function $I_Z : [0, T] \to R^+ : x \to \int_0^x Z(s)ds$. This function is concavely increasing. Its graph will be said to be the non-normalized Lorenz curve of Z.

Clearly the graph of $I_Z$ is the graph of a cumulative function. Yet, because we use it here within a generalization of the classical Lorenz curve (Lorenz, 1905) we refer to it as a non-normalized Lorenz curve.

Definition: The non-normalized dominance order on **U**

If Z, Y ∈ **U**, then $Z - < Y$ iff $\forall x \in [0, T]$: $I_Z(x) = \int_0^x Z(s)ds \leq I_Y(x) = \int_0^x Y(s)ds$

Clearly, -< on **U** is reflexive, antisymmetric, and transitive. Hence it is a partial order (Roberts, 1979).

From the definition of the Lorenz curve of a function Z in **U** it follows that it is the image of its non-normalized Lorenz curve, through a linear mapping with matrix $\begin{pmatrix} 1/T & 0 \\ 0 & 1/(I_Z(T)) \end{pmatrix}$ which is a composition of a horizontal and a vertical contraction (T > 1).

Related to this observation we formulate two remarks.

Remark 1. If Z, Y ∈ **U**, and $I_Z(T) = I_Y(T)$ then Z -< Y implies $Z - <_L Y$.

Remark 2. If Z, Y ∈ **U** and Z ≤ Y then Z -< Y, but the relation $Z -<_L Y$ may or may not hold.

Notation

We denote the average of Z ∈ **U** by $\mu_Z = \frac{1}{T}\int_0^T Z(s)ds = \frac{I_Z(T)}{T}$.

Theorem. The impact-concentration theorem.

$\forall Z, Y \in \textbf{U}_\textbf{0}$, with Z ≠ Y the following expressions are equivalent.

(i) Z -< Y ;

(ii) $\exists Y^* \neq Z \in \textbf{U}_\textbf{0}$, such that $\mu_Z = \mu_{Y^*}$, Z -< Y* ≤ Y and $Z -<_L Y^*$;

(iii) $\exists Y^* \neq Z \in \textbf{U}_\textbf{0}$, such that $\mu_Z = \mu_{Y^*}$, Z -< Y* -< Y and $Z -<_L Y^*$;

Proof. The implications (ii) ⇒ (iii) ⇒ (i) trivially follow from the facts that -< is transitive and that Z ≤ Y implies Z -< Y.

We next prove that (i) implies (ii).

If $I_Z(T) = I_Y(T)$ we may set Y* = Y, see Remark 1. We next assume that $I_Z(T) < I_Y(T)$, and hence $\mu_Z < \mu_Y$.

Define I*(x) on [0,T] as min($I_Y(x)$, $I_Z(T)$). Then Z -< Y implies that $\forall x \in [0,T]$: $I_Z(x) \leq I_Y(x)$ and hence, $\forall x \in [0,T]$: $I_Z(x) \leq I^*(x) \leq I_Y(x)$. The following construction is illustrated in Fig.1.

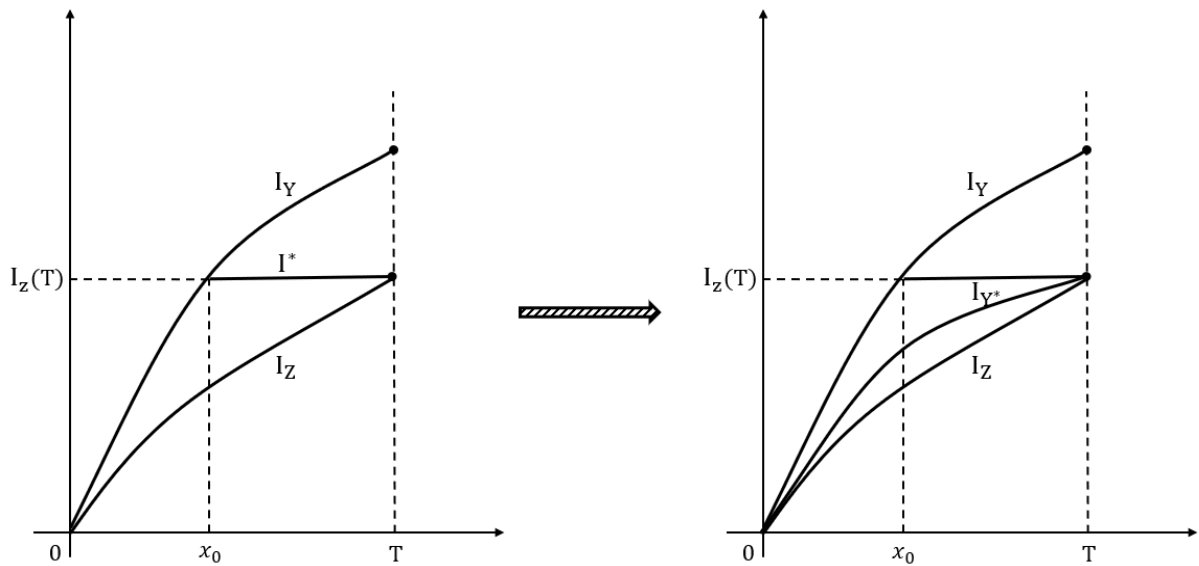

Fig. 1. Illustration of the construction of the function $I_{Y^*}$

As Z -< Y (Z,Y different) there exists, by continuity, a point $x_0 \in ]0,T[$ such that $I_Z(x_0) < I_Y(x_0)$. As Z is strictly decreasing on [0,T] it is not zero on an interval of finite length. Hence, we see that $I_Z(x_0) < I_Z(T)$ and $I_Z(x_0) < I^*(x_0)$.

The function $I^*$ has a most one kink (recall that the integral of a continuous function is differentiable). Hence, we can work around this kink by replacing $I^*$ locally by a smooth, concavely increasing curve, still denoted as $I^*$ such that $I_Z(x) < I^*(x) \leq I_Y(x)$ on $[0,T]$ and such that $I^*$ is in no point increasing faster than $I_Y$. Now we define $Y^*$ as $(I^*)'$ leading to: $I_{Y^*} = I^*$. From this, we see that $Z \prec Y^* \prec Y$, actually by its construction, we even have $Y^* \leq Y$.

Now $I_Z(T) = I_{Y^*}(T)$ and hence $\mu_Z = \mu_{Y^*}$. By Remark 1 we then have that $L_Z < L_{Y^*}$ on $]0,1[$ or $Z \prec_L Y$.

## 3. Global impact measures

Let $\boldsymbol{U_\mu} = \{Z \in \boldsymbol{U_0}: \mu_Z(T) = \mu\}$ and let m be a function $\boldsymbol{U_0} \to \mathbb{R}^+$. Then we have the following theorem, leading to the definition of measures of a new type.

Theorem

The following three expressions are equivalent:

(i) $\forall Z,Y \in \boldsymbol{U_0}$, $Z \neq Y$: $Z \prec Y \Rightarrow m(Z) < m(Y)$

(ii) $\forall Z,Y \in \boldsymbol{U_0}$ : $Z \prec Y \Rightarrow m(Z) \leq m(Y)$ and, for all $\mu > 0$, if $Z, Y \in \boldsymbol{U_\mu}$: ($Z \prec_L Y$ and $Z \neq Y$, $\Rightarrow m(Z) < m(Y)$)

(iii) $\forall Z,Y \in \boldsymbol{U_0}$ : $Z \leq Y \Rightarrow m(Z) \leq m(Y)$ and, for all $\mu > 0$, if $Z, Y \in \boldsymbol{U_\mu}$ ($Z \neq Y$), we have: ($Z \prec_L Y \Rightarrow m(Z) < m(Y)$)

Proof.

(i) $\Rightarrow$ (ii). Let $Z \prec Y$, and $Z \neq Y$, then we know by (i) that $m(Z) < m(Y)$. Of course, if $Z = Y$, then $m(Y)=m(Z)$ so that always $m(Z) \leq m(Y)$. If now, for $\mu > 0$, $Z, Y \in \boldsymbol{U_s}$: $Z \prec_L Y$, with $Z \neq Y$, then $I_Z(T) = I_Y(T)$ and by Remark 1, $Z \prec Y$, and thus $m(Z) < m(Y)$. This proves this implication.

(ii) $\Rightarrow$ (iii) is trivial as $Z \leq Y$ implies $Z \prec Y$.

(iii) $\Rightarrow$ (i). Let $Z \neq Y$ and $Z \prec Y$. By the "impact-concentration theorem", we know that there exists $Y^* \neq Z$ in $\boldsymbol{U_0}$ such that $Z \prec Y^* \prec Y$ and $Z \prec_L Y^*$ with $\mu_Z = \mu_{Y^*}$ simply denoted as $\mu$. Hence $Z$ and $Y^*$ belong to $\boldsymbol{U_\mu}$. It follows from (iii) that $m(Z) < m(Y^*)$ and $m(Y^*) \leq m(Y)$, which proves (i).

Definition

We say that a measure m as defined above is increasing on $\boldsymbol{U_0}, \prec$ if $Z \prec Y \Rightarrow m(Z) \leq m(Y)$, and strictly increasing if $Z \neq Y$: $Z \prec Y \Rightarrow m(Z) < m(Y)$.

Definition. Global impact measures

A function $\boldsymbol{U_0} \to \mathbb{R}^+$ such that $\forall\, Z, Y \in \boldsymbol{U_0}$, $Z \neq Y$: $Z \prec Y \Rightarrow m(Z) < m(Y)$ is called a global impact measure.

Corollary

The following expressions are equivalent

(i) m is a global impact measure on $\boldsymbol{U_0}$

(ii) m is an increasing function on $\boldsymbol{U_0}, \prec$ and for every $\mu > 0$ (fixed) m is a concentration measure on $\boldsymbol{U_{0,\mu}}$

(iii) m is an increasing function on $\boldsymbol{U_0}, \leq$ and for every $\mu > 0$ (fixed) m is a concentration measure on $\boldsymbol{U_{0,\mu}}$

## 4. Examples of global impact measures

In this section we provide some examples of global impact measures.

4.1. The generalized Gini-index

Obviously, the area under the non-normalized Lorenz curve is a global impact measure. As we work on $\boldsymbol{U_0}$ there is no function with Gini-value equal to zero, but zero is the infimum (largest lower bound).

4.2. The length of the non-normalized Lorenz curve of Z in $\boldsymbol{U_0}$ minus T. This measure is denoted as $\mathcal{L}(Z)$.

Now $\mathcal{L}(Z) = \int_0^T \sqrt{1 + \left(\frac{d}{ds}(I_Z(s))\right)^2}\, ds - T = \int_0^T \sqrt{1 + (Z(s))^2}\, ds - T$

The length of the non-normalized Lorenz curve (or this length minus T) is an increasing function on $\boldsymbol{U_0}$, $\leq$ and for every $\mu > 0$ (fixed) m is a concentration measure on $\boldsymbol{U_{0,\mu}}$ as the length of the standard Lorenz curve (or this length minus $\sqrt{2}$ ) is a bona fide concentration curve (Dagum, 1980).

Alternatively, we know that if f is a strictly convex, continuous, and increasing function, then $Z \prec Y$, $Z \neq Y$, implies that $\int_0^T f(Z(s))\,ds < \int_0^T f(Y(s))\,ds$ . Taking now $f(s) = \sqrt{1 + s^2}$ , shows that $\mathcal{L}$ is a global impact measure. We subtract T because we require that when Z tends to zero (pointwise), $\mathcal{L}(Z)$ also tends to zero.

4.3. $m(Z) = \int_0^T (Z(s))^p\,ds$, $p > 1$

This measure m is a global impact measure because $f(s) = s^p$ ($p > 1$) is a strictly convex, continuous, and increasing function.

Considering now the case p = 2, we define $(\sigma_Z)^2$ as $\frac{1}{T}\int_0^T (Z(s) - \mu_Z)^2 ds$, and see that $\int_0^T (Z(s))^2 ds = (\sigma_Z)^2 + (\mu_Z)^2$. Taking $\mu_Z$ fixed, we find that the squared variance $\left(\frac{\sigma_Z}{\mu_Z}\right)^2 = 1 + \int_0^T \left(\frac{Z(s)}{\mu_Z}\right)^2 ds$ is a, well-known, concentration measure on $\boldsymbol{U_{0,\mu}}$.

### 4.4. The Theil measure.

We define the (generalized) Theil measure for Z in $\boldsymbol{U_0}$ as

$$Th_g(Z) = \int_0^T Z(s)\, ln(Z(s))ds$$

Clearly, $Th_g$ is increasing in Z. Next, we observe that $Th(Z)$, the analytical Theil concentration measure, defined as

$$Th(Z) = \frac{1}{T}\int_0^T \frac{Z(s)}{\mu_Z} ln\left(\frac{Z(s)}{\mu_Z}\right) ds$$

is equal to

$\frac{1}{T\,\mu_Z}\int_0^T Z(s)ln\left(\frac{Z(s)}{\mu_Z}\right)ds = \frac{1}{T\,\mu_Z}\left(\int_0^T Z(s)ln(Z(s))\,ds - \int_0^T Z(s)ln(\mu_Z)\,ds\right)$

$= \frac{1}{T\,\mu_Z}\left(\int_0^T Z(s)ln(Z(s))\,ds - T\,\mu_Z\,ln(\mu_Z)\right)$

$= \frac{1}{T\,\mu_Z}\int_0^T Z(s)ln(Z(s))\,ds - ln(\mu_Z) = \frac{Th_g(Z(s))}{T\,\mu_Z} - ln(\mu_Z).$

Consequently: $Th_g(Z) = T.\mu_Z. (Th(Z) + ln(\mu_Z))$. Hence for $\mu = \mu_Z$ constant, $Th_g$ is a strictly increasing function of the concentration measure Th. Using the corollary above, we conclude that $Th_g$ is a global impact measure.

Alternatively, $Th_g$ is a global impact measure because the function $s \to s\,ln(s)$ is, for s > 0, strictly convex, continuous, and increasing.

## 5. Conclusion

We investigated a continuous theory of domination, leading to so-called global impact measures. In this context, we consider the notion of impact, as e.g., shown by a citation curve (articles ranked according to the number of received citations) as a combination of the notions of inequality (concentration) and productivity. In this way, our article extends earlier approaches, in which we mainly focused on the high producers (Egghe, 2021; Egghe & Rousseau, 2022a). This article belongs to a series of investigations studying the notion of impact. The main ideas of which were summarized in (Egghe & Rousseau, 2022b).

Equality or evenness, the opposite of inequality or concentration, plays a key role in studies of interdisciplinarity or, more generally, diversity studies (Wagner et al., 2011; Rousseau et al., 2019). Of course, concentration and diversity are also essential notions in other fields such as economics and ecology, and have widespread implications. For this reason, we expect that global impact measures, will lead to new opportunities for studies in informetrics and, more generally, the science of science.

*Acknowledgment.* The authors thank Li Li (National Science Library, CAS) for providing excellent figures.

*Conflict of interest.* Leo Egghe and Ronald Rousseau belong to the distinguished reviewers board of the journal *Scientometrics* as Price Medal awardees.

*Funding.* No funding has been received for this investigation

*Author contributions:*

Leo Egghe: conceptualization, formal analysis, investigation, methodology, writing-original draft, writing-review and editing.
Ronald Rousseau: investigation, validation, writing-review and editing.

Global informetric impact: a description and definition

Leo Egghe[1] and Ronald Rousseau[2,3] *

[1] Hasselt University, 3500 Hasselt, Belgium

leo.egghe@uhasselt.be

ORCID: 0000-0001-8419-2932

[2] KU Leuven, MSI, Facultair Onderzoekscentrum ECOOM,

Naamsestraat 61, 3000 Leuven, Belgium

ronald.rousseau@kuleuven.be   &

[3] University of Antwerp, Faculty of Social Sciences,

Middelheimlaan 1, 2020 Antwerp, Belgium

ronald.rousseau@uantwerpen.be

ORCID: 0000-0002-3252-2538

Abstract

Inspired by the Lorenz curve for evenness or concentration, and the corresponding axioms, we construct a theory leading to the notion of impact. In this theory we construct a relation which plays the role of the Lorenz dominance order for evenness or concentration theory. The notion of impact we obtain is such that well-known impact bundles such as percentiles, the cumulative number of items produced by the x most productive sources, the average production of the x most productive sources, the generalized h- and g-indices, and the highest number of citations indeed measure impact in our sense of the word.

Keywords: global impact; global impact bundle; non-normalized Lorenz curve; information production process (IPP)

* Corresponding author

# 1. Introduction

## *1.1 Evenness*

As our approach to studying impact is to some extent inspired by the notion of concentration or its opposite evenness, when both are based on the notion of a Lorenz curve, we first recall how we see evenness. Similar  to the notion of 'impact' also the notion of 'evenness' can be defined in plain words. E.g., in studying biodiversity, species evenness refers to how close in numbers each species in an environment is. Yet, plain words can never lead to a scientifically exact notion. The resulting problem of defining evenness, more precisely, to be able to say when situation A is more even than situation B, has been solved by using the Lorenz curve (Lorenz, 1909; Nijssen et al., 1998). Note though that one has to admit that some situations are not intrinsically comparable, namely in the case that their Lorenz curves intersect. Mathematically speaking, Lorenz curves lead to a partial order, not to a complete order relation, a fact already pointed out by (Sen, 1973) in the more general context of measuring concentration and diversity. Once a Lorenz curve is accepted as the proper representation of evenness or concentration, one can define measures that respect the partial order induced by the Lorenz curve, such as the Gini index, the repeat rate (Simpson's measure or the Herfindahl-Hirschman measure), the entropy measure (Shannon's or Theil's), and many more. They differ in the sensitivity of their response to changes in the Lorenz curve and hence, depending on the application, one is more suitable than the other.

Evenness and concentration are interesting properties on their own. Yet, evenness is often studied in combination with the number of species, or more generally, the number of cells in the context of nominal data. This then leads to the notions of diversity (in ecology), or inequality (as a general term) and functions suitable to measure them (Allison, 1978; Rousseau & Van Hecke, 1999). Finding the 'best' among generally accepted measures for diversity or inequality is still an open problem (in our opinion). Moreover, Stirling (2007) has forcefully argued that, in the context of diversity, besides species richness and evenness, the disparity between cells must be taken into account, making the problem of a precise definition and associated measurement, even harder, see also (Jost, 2009; Leinster & Cobbold, 2012). In the field of informetrics, diversity measures have been used to study interdisciplinarity, another notion that is hard to formulate in words. As this article does not intend to study interdisciplinarity we refer the reader for recent progress and discussions related to measuring interdisciplinarity to (Wagner et al., 2011; Rousseau et al., 2019; Leydesdorf et al., 2019; Rousseau, 2020) among others. How the Lorenz curve has led us to the notion of the – in informetrics - fundamental notion of impact will be shown in this article. We will introduce requirements that are to some extent related to the notion of impact. Then we will investigate how these requirements are related, and finally come to the appropriate (mathematical) notion of impact. Of course, we could have (bluntly) presented our final result, but then the reader had no clue why this was the best and why other notions could not be used. So, please follow us in our investigations, or, if you are not interested in the underlying 'why?' go the Definition 7. A semi-intuitive presentation of our whole impact framework is published in (Egghe & Rousseau, 2022a).

*1.2 Impact*

In this article, we continue our impact-related investigations (Egghe, 2022; Egghe & Rousseau, 2021, 2022b,c) with the final aim to come to an acceptable definition of the notion of 'impact' and a corresponding method of measuring it. As mentioned above, we are inspired by the developments related to the notions of evenness and concentration. We like to point out here that we use the notion of impact in an academic/informetric sense. Although rather general, i.e., not only restricted to a publication and citation context, it should nevertheless not be confused with 'making an impact' in the daily sense, such as when saying that the Prime Minister has a huge impact on her country's citizenry. We always consider continuous models for quantitative data, represented by non-negative numbers, that hence can be ranked from high to low.

*1.3 Notations and terminology: bundles and impact bundles*

We next recall the used notations and terminology.

Let T > 0, let **U** be the set $\{Z: [0, T] \to \mathbf{R}^+, Z$ continuous and decreasing}. As we use a continuous model for rank-frequency functions (Zipf-type) (Egghe, 2005, p. 105), the number T is a real number, not necessarily a natural number as in the discrete case. As we will keep T fixed, this means that in the case of articles and their citations the number of publications is kept fixed, while in the case of authors and their publications it is the number of authors that is kept fixed. When applied in the context of a general information production process (IPP) the term 'impact' should be considered in a generalized sense. As in (Egghe & Rousseau, 2022a, 2022b) we focus on the left-hand sides of these curves as impact is related to the high achievers for the studied phenomenon (typically those receiving a lot of citations) and not to those at the end.

On the set $\boldsymbol{U}$ and its subsets, we have the relations $<$ and $\leq$, where $Z < Y$ iff $\forall\, x \in [0, T[: Z(x) < Y(x)$, and hence, by continuity, $Z(T) \leq Y(T)$; and $Z \leq Y$ iff $\forall\, x \in [0, T]: Z(x) \leq Y(x)$. The relation $\leq$ is a partial order relation on $\boldsymbol{U}$, while $<$ is a strict partial order (Roberts, 1979, p.15).

If $Z < Y$ in $\boldsymbol{U}$, then certainly Y "has more impact" than Z, but we will propose a meaningful theory in which the opposite relation does not necessarily hold.

From (Egghe & Rousseau, 2022b) we recall the notions of a bundle and an impact bundle.

## Definition 1: Bundle measures and bundles

Let $\boldsymbol{Z} \subset \boldsymbol{U}$ and let m be a map defined on a subset of ($\boldsymbol{Z} \times [0,+\infty]$),

$$m: \text{dom}(m) \subset (\boldsymbol{Z} \times [0,+\infty]\,) \rightarrow \mathbf{R^+}: (Z,\theta) \rightarrow m_\theta(Z),$$

to which we will refer as a measure, or as we want to stress the bundle context, a bundle measure. For fixed Z in $\boldsymbol{Z}$, we have the function m(Z): dom(m) ∩ ({Z} × [0,+∞]) $\rightarrow \mathbf{R^+}$ : $\theta \rightarrow m_\theta(Z)$. Then, for each Z, the domain dom(m(Z)) is required to be a finite or infinite, open, closed, or half-open interval. It is called the parameter set of m(Z). Then a bundle $\mathbf{B}$ is defined as a pair of maps $\mathbf{B} = (m, \psi)$:

$$m: \text{dom}(m) \subset (\boldsymbol{Z} \times [0,+\infty]) \rightarrow \mathbf{R^+}: (Z,\theta) \rightarrow m_\theta(Z),$$

$$\psi : \boldsymbol{Z} \times [0,T] \rightarrow [0, +\infty]$$

The function $\psi$ represents the rate of 'scanning' the graph of the function Z in $\boldsymbol{Z}$, from high to low values or vice versa. Moreover, $\psi|_{\{Z\}\times[0,T]}: \{Z\} \times [0,T] \rightarrow [0,+\infty]$, is a continuous injection, henceforth abbreviated as $\psi_{\boldsymbol{B},Z}$.

When studying two functions Z and Y at the same time, we will always assume that $\psi_{\boldsymbol{B},Z}$ and $\psi_{\boldsymbol{B},Y}$ are both increasing or both decreasing. When it is clear about which bundle $\mathbf{B}$ we are talking we simply write $\psi_Z$ (without the symbol **B**). When, considering a parameter θ for a single function Z we assume that θ ∈ dom(m(Z)), or when considering two functions we always assume that θ ∈ dom(m(Z)) ∩ dom(m(Y)), and hence that this intersection is non-empty. In all these cases, we will simply write "all admissible θ".

To eliminate uninteresting cases, we require in this article that a bundle always meets the following axioms:

(AX. 1). If **0** (the zero function) belongs to **Z**, then $m_θ(\mathbf{0}) = 0$ for all θ.

(AX.2). The monotonicity requirement. For all Y, Z ∈ **Z,** and all admissible θ: $Z ≤ Y ⇒ m_θ(Z) ≤ m_θ(Y)$.

These axioms were already used in (Egghe, 2022; Egghe & Rousseau, 2022b).

Before coming to the definition of an impact bundle we state the following definitions:

For Z, Y ∈ **Z** ⊂ **U**, a, b ∈ [0,T], a < b we define:

$Z <_a Y ⇔ Z < Y$ on [0,a]; this relation is a strict partial order.

If m is a bundle measure then we define similarly:

$m(Z) \prec_a m(Y) \Leftrightarrow m(Z) < m(Y)$ on $\psi_Z([0,a]) \cup \psi_Y([0,a])$, or stated otherwise: for all $\theta \in \psi_Z([0,a]) \cup \psi_Y([0,a])$: $m_\theta(Z) < m_\theta(Y)$.

We note that if $\psi_Z$ and $\psi_Y$ are decreasing, then $\psi_Z([0,a]) \cup \psi_Y([0,a])$ is equal to the set of admissible $\theta$-values in $[\min(\psi_Z(a),\psi_Y(a)), +\infty]$, and similarly, if $\psi_Z$ and $\psi_Y$ are increasing, then $\psi_Z([0,a]) \cup \psi_Y([0,a])$ is equal to the set of admissible $\theta$-values in $[0, \max(\psi_Z(a),\psi_Y(a))]$.

Definition 2: An impact bundle (Egghe & Rousseau, 2022b)

A bundle $\mathbf{B} = (m, \psi)$ is an impact bundle if, besides (AX.1) and (AX.2) it satisfies the following axiom (AX.3):

(AX.3). For all $a \in ]0,T[$: $Z <_a Y$ implies $m(Z) \prec_a m(Y)$.

If a bundle $\mathbf{B} = (m, \psi)$ is an impact bundle, i.e., it satisfies (AX.1), (AX.2) and (AX.3) we also state this as: $\mathbf{B}$ satisfies the requirements for (IB), or even more concisely: $\mathbf{B}$ is (IB).

Finally, we recall the following definition.

If Z and Y belong to **U** then we say that Z and Y intersect in the point $x_0 \in [0,T]$ if $Z(x_0) = Y(x_0)$. We say that they intersect strictly in the point $x_0 \in ]0,T[$, if there exists moreover an interval $]x_0 - \delta, x_0+\delta[$, $\delta > 0$, such that Z-Y is strictly positive in $]x_0 - \delta, x_0[$ and strictly negative on $]x_0, x_0+\delta[$ or vice versa, with the positive and negative parts interchanged.

## 2. Preliminary investigations about the notion of 'impact'

Roughly speaking (the problem will be made more precise later), we will try to find conditions such that "Z, Y $\in$ **U**, and Z is smaller than Y (in a sense to be made more precise) implies that there exists an admissible θ such that $m_\theta(Z) < $ (or ≤) $m_\theta(Y)$". We then say that Y has more impact (in a generalized sense) than Z.

We note that requiring that for all admissible θ, $m_\theta(Z) < m_\theta(Y)$ is asking too much as suggested by the following case. Let m = h, leading to the generalized h-index family $h_\theta$ (van Eck & Waltman, 2008; Egghe & Rousseau, 2019) and assuming that all X $\in$ **Z** $\subset$ **U** take the same value in T. Then the set of all admissible θ-values is {θ ∈ ]0,+∞]; θ ⩾ (X(T))/T}. The requirement h$_θ$(Z) < h$_θ$(Y) (assuming Z(0) < Y(0)) implies that Z < Y on [0,T[ (Egghe & Rousseau, 2022b; Theorem 5). This example is illustrated in Fig. 1. Recall that we provide this example because we do not want to restrict the notion of "more impact" to the relation < on **U**.

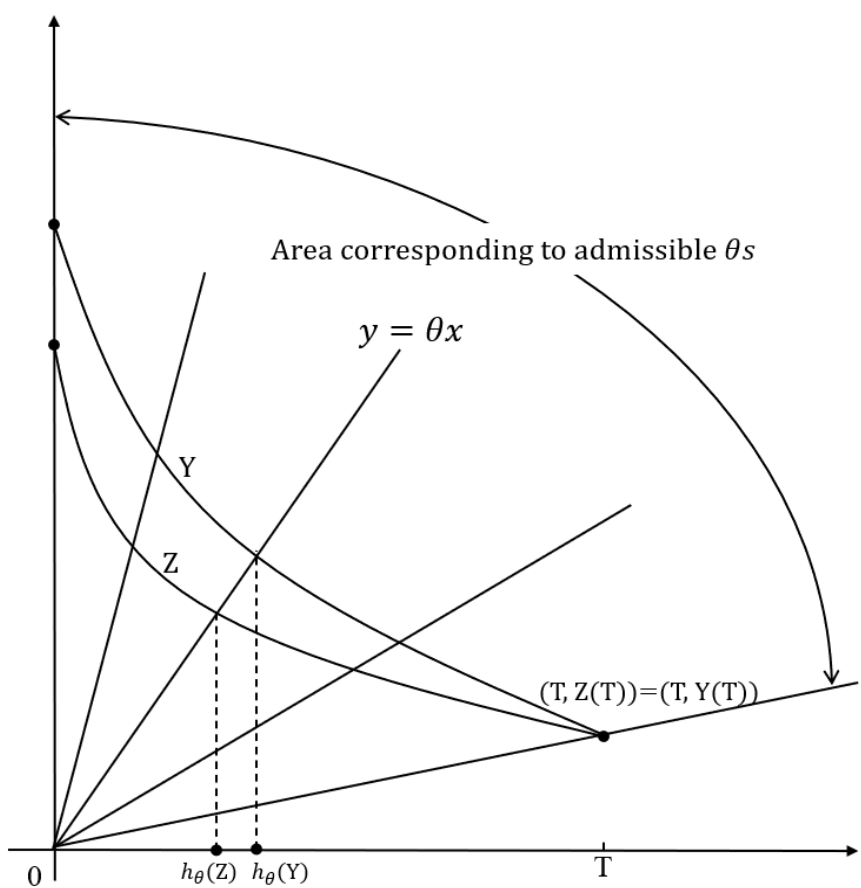

Fig. 1 Illustration of the fact that requiring that for all admissible θ, m$_θ$(Z) < m$_θ$(Y) may lead to Z < Y (on [0,T[].

In (Egghe & Rousseau, 2022b,c) we defined for every $Z \in \boldsymbol{U}$, the area function I(Z) on [0,T] as:

$$I(Z): [0,T] \to \mathbf{R}^+ : \theta \to I_\theta(Z) = \int_0^\theta Z(s)ds$$

The graph of I(Z) is concavely increasing and continuous, connecting the points (0, 0) and (T, $I_T$(Z)). where $I_T$(Z) is the area under the function Z, above the interval [0,T]. If Z is a constant, equal to C, then I(Z) is linearly increasing from (0,0) to (T, CT). The graph of I(Z) will, for obvious reasons, be called the non-normalized Lorenz curve. Indeed, the concave form of the Lorenz curve is obtained in the same way, but normalized such that T =1 and $I_T$(Z) = 1.

An example. If $Z(x)= a(T-x)^2$, a > 0, x $\in$ [0,T], then $I_\theta$(Z) $=\frac{1}{3}a\theta(3T^2 - 3T\theta + \theta^2)$, with θ $\in$ [0,T]. This is illustrated in Fig.2.

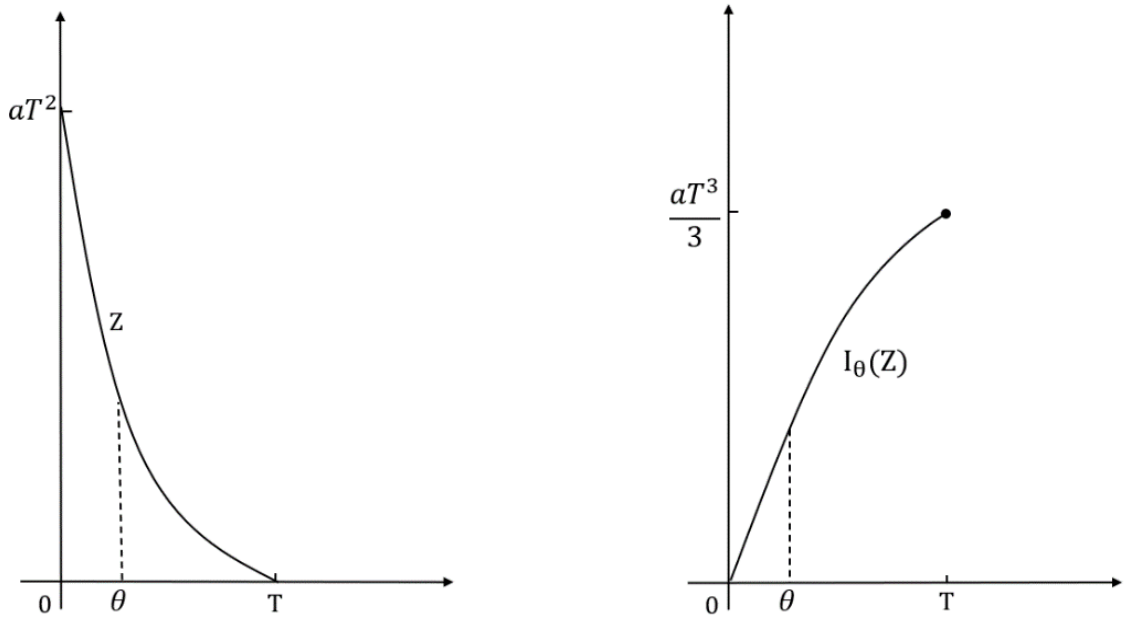

Fig. 2 Function Z and its corresponding area function

We further defined (Egghe & Rousseau, 2022b,c) the relation $-<$ for functions Z and Y in **U**, as $I(Z) \leq I(Y)$. If it is required that Z $\neq$ Y, then we denote the corresponding inequality as $Z -<_{\neq} Y$. This relation is called the non-normalized Lorenz dominance order.

In (Egghe & Rousseau, 2022b) we also defined for every $Z \in$ **U**, the average function $\mu(Z)$ on [0,T] as:

$$\mu(Z): ]0,T] \to \mathbf{R}^+: \theta \to \mu_\theta(Z) = \frac{1}{\theta}\int_0^\theta Z(s)ds$$

$$\text{and } \mu_0(Z) = Z(0) = \lim_{\theta \to 0} \mu_\theta(Z)$$

Yet, the non-normalized Lorenz curve is the more important as requiring that $\mu(Z) < \mu(Y)$ on $[0, T[$ is a stronger requirement than requiring that $Z -<_{\neq} Y$ on [0,T] as the first implies that $Z(0) <$ $Y(0)$, while this is not required by $Z -<_{\neq} Y$. We note the following elementary proposition (without proof).

Proposition 1

$\forall$ Z, Y $\in$ **U**:

$\mu(Z) \leq \mu(Y)$ on $[0, T] \Leftrightarrow Z(0) \leq Y(0)$ and $I(Z) \leq I(Y)$ on $[0, T]$.

Now we first recall some bundle properties investigated in previous work (CES, PED), introduce some other, rather obvious ones, and investigate their relationship. This section can be considered a preparation for the next section, where we will come closer to our aim. Readers not interested in the related mathematical developments may skip the proofs without any lack of understanding.

We recall from previous work (Egghe & Rousseau, 2021) the notion of a complete evaluation system. Given m(X): $\theta \rightarrow m_\theta(X)$, then the condition: for all $\theta$, $m_\theta(Z) = m_\theta(Y)$, implies that $Z = Y$ (for all admissible θ-values). We say, in short, that m(X) is or satisfies (CES). It is shown in (Egghe & Rousseau, 2021) that the generalized h-index, as well as the generalized g-index (van Eck & Waltman, 2008; Egghe & Rousseau, 2019), are examples of bundle measures that satisfy the condition (CES).

We further recall that a measure p is a (classical) Pointwise Explicitly Defined measure on ]0,T[ (in short: we say that p is, or satisfies (PED)) if there is a continuous function f of two variables such that for all Z ∈ **U**, and all admissible θ, $x = p_\theta(Z)$ iff $Z(x) = f(\theta,x)$ (Egghe, 2021). The generalized h-index $h_\theta$ is an example of a (PED) measure with $f(\theta,x) = \theta x$. We will further assume in this article that, in the context of (PED)-measures, for fixed θ, f is increasing in x.

In our search to define the notion of impact and related orders, we next investigate the following three properties and their relations with (PED) and (CES).

A bundle has property ($W_L$) if the bundle measure m satisfies:

Z, Y ∈ **Z** ⊂ **U**, (Z $-<_{\neq}$ Y on [0,T]) ⇒ (∃ θ : $m_\theta(Z) < m_\theta(Y)$ )

A bundle has property ($W_1$) if the bundle measure m satisfies

Z, Y $\in$ $\textbf{\textit{Z}} \subset \textbf{\textit{U}}$, (Z $\leq_{\neq}$ Y on [0,T]) $\Rightarrow$ ($\exists\, \theta: m_\theta(Z) < m_\theta(Y)$ )

A bundle has property ($W_2$) if the bundle measure m satisfies:

Z, Y $\in$ $\textbf{\textit{Z}} \subset \textbf{\textit{U}}$, ($\exists\, x \in$ [0,T], such that Z(x) < Y(X)) $\Rightarrow$ ($\exists\, \theta : m_\theta(Z) < m_\theta(Y)$ ).

Then we have the following Theorem, showing that (PED) is the strongest property, while ($W_1$) is the weakest.

Theorem 1

Given a bundle $\textbf{B}$ = (m, $\psi$), then for all $\textbf{\textit{Z}} \subset \textbf{\textit{U}}$, we have

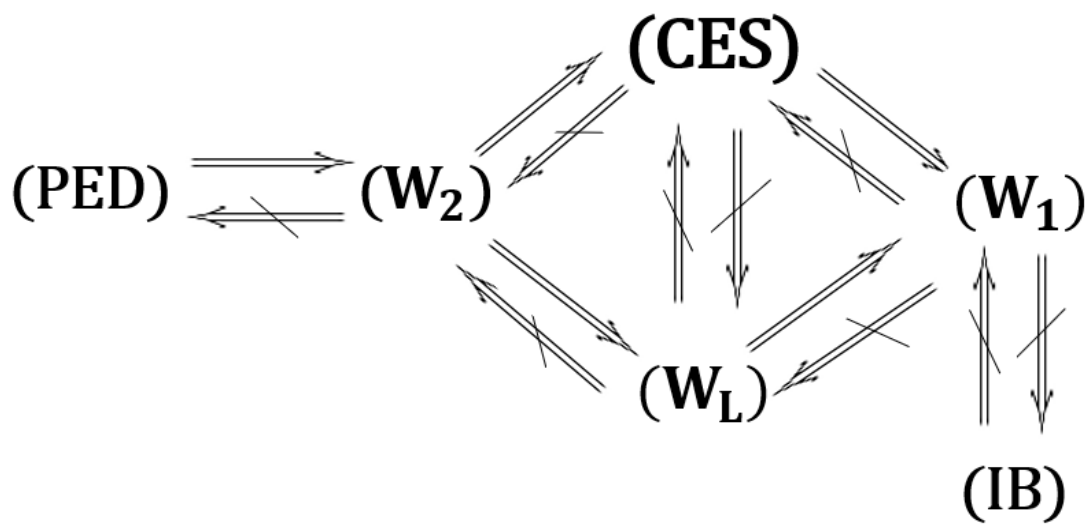

Proof. (PED) $\Rightarrow$ ($W_2$)

Let $x_0 \in$ [0,T] such that Z($x_0$) < Y($x_0$). As m satisfies the requirements for (PED), there exist an admissible θ such that $x_0$

= $m_\theta(Z)$ and hence $Z(x_0) = f(\theta, x_0) < Y(x_0)$, with f the function defining PED. From this we derive that $\exists\, \theta$ such that $m_\theta(Y) > m_\theta(Z)$. Indeed, if $m_\theta(Y) \leq m_\theta(Z)$ then $f(\theta, m_\theta(Y)) \leq f(\theta, m_\theta(Z))$ (where we have used that f is increasing in its second variable). By the (PED) property we then have $Y(m_\theta(Y)) \leq Z(m_\theta(Z))$. As Y is decreasing, we have $Y(x_0) = Y(m_\theta(Z)) \leq Y(m_\theta(Y)) \leq Z(m_\theta(Z))=Z(x_0)$. This is in contradiction with $Z(x_0) < Y(x_0)$. Hence there exists an admissible $\theta$ such that $m_\theta(Z) < m_\theta(Y)$, proving $(W_2)$.

$(W_2) \nRightarrow (PED)$

The point is that if the function $Z \in \boldsymbol{Z}$ is composed with a strictly increasing continuous function $\varphi$ such that $\varphi \circ Z$ is still decreasing then property $(W_2)$ still holds, while this is not necessarily true for (PED).

We consider the following example, already studied in (Egghe, 2022). Let $\boldsymbol{Z} = \left\{ Z_{T,S}; Z_{T,S}(x) = S\left(1 - \frac{x}{T}\right), x \in [0,T] \right\} \subset \boldsymbol{U}$, with S,T > 0. This is the set of straight lines connecting (T,0) with (0,S). It is shown in (Egghe, 2022) that the generalized g-index $g_\theta$ is not (PED) on $\boldsymbol{Z}$ (but it is (PED) on an infinite part of $\boldsymbol{Z}$). We know from (Egghe, 2022) that

$$h_\theta\left(Z_{T,S}\right) = \frac{S}{\frac{S}{T} + \theta}$$

and

$$g_\theta\left(Z_{T,S}\right) = \frac{S}{\frac{S}{2T} + \theta}$$

Now, $\left[\left(g_\theta\left(Z_{T,S}\right) > g_\theta\left(Z_{T,S'}\right)\right) \Leftrightarrow \left(h_\theta\left(Z_{T,S}\right) > h_\theta\left(Z_{T,S'}\right)\right)\right]$

$$\Leftrightarrow \left[\left(\frac{S}{\frac{S}{2T} + \theta} > \frac{S'}{\frac{S'}{2T} + \theta}\right) \Leftrightarrow \left(\frac{S}{\frac{S}{T} + \theta} > \frac{S'}{\frac{S'}{T} + \theta}\right)\right]$$

Simplifying shows that the previous equivalence holds. Now as $h_\theta$ is (PED), it also satisfies (W$_2$), and thus by the previous calculation also $g_\theta$ satisfies (W$_2$). This proves that (W$_2$) does not necessarily imply (PED).

(W$_2$) $\Rightarrow$ (CES)

If for all admissible $\theta$: $m_\theta(Z) \geq m_\theta(Y)$ then we know by (W$_2$) that for all $x \in [0,T]$: $Z(x) \geq Y(x)$. Similarly, if $\forall \theta$: $m_\theta(Y) \geq m_\theta(Z)$ then we have for all $x \in [0,T]$: $Y(x) \geq Z(x)$. Hence, if $\forall \theta$: $m_\theta(Z) = m_\theta(Y)$, we conclude that $Z(x) = Y(x)$. This proves (CES).

(CES) $\not\Rightarrow$ (W$_2$)

We know (Egghe & Rousseau, 2021) that if we take for $m_\theta$ the average $\mu_\theta$ then this bundle measure is (CES) on $\boldsymbol{U}$. Consider now the functions Z and Y in Fig. 3.

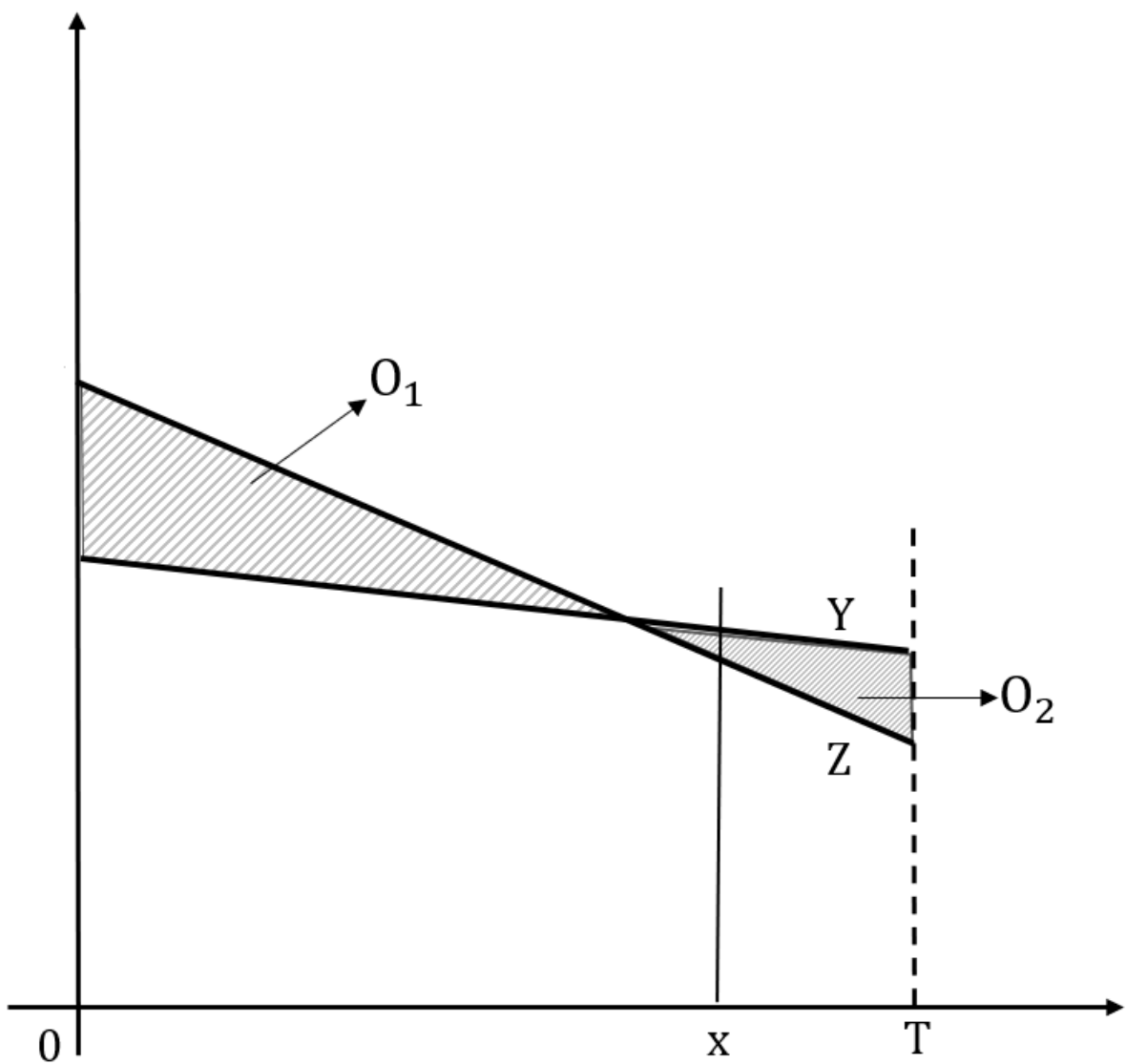

Fig. 3. The example used in the proof of (CES) $\nRightarrow$ (W$_2$)

If we take the areas $O_1$ and O2 such that $O_1 > O_2$, then we have for all θ ∈ [0,T] that $\mu_\theta(Z) > \mu_\theta(Y)$, and yet there exist points x ∈ [0,T] such that Z(x) < Y(x).

(CES) ⇒ ($W_1$)

We prove: NOT($W_1$) implies NOT(CES)

Assume that there exist $Z \leq Y$ and $Z \neq Y$, such that $\forall \theta : m_\theta(Z) \geq m_\theta(Y)$. We also know that Z ≤ Y implies that $\forall \theta : m_\theta(Z) \leq m_\theta(Y)$ (by the monotonicity requirement). Combining these two statements yields: $\forall \theta: m_\theta(Z) = m_\theta(Y)$. As Z ≠ Y, this shows that the (CES) requirement is not satisfied.

($W_2$) ⇒ ($W_L$) ⇒ ($W_1$)

These implications are obvious.

($W_L$) ⇏ (CES)

We take $\boldsymbol{Z} \subset \boldsymbol{U}$ with $\boldsymbol{Z}$ = {all linearly decreasing, positive functions f on [0,T], with f(T/4) = T/2}. Then no two different functions in $\boldsymbol{Z}$ satisfy (Z, Y ∈ $\boldsymbol{U}$, Z −<$_{\neq}$ Y on [0,T] ), i.e., the left-hand side of ($W_L$). Hence all functions in $\boldsymbol{Z}$ satisfy ($W_L$). Yet, if we consider the constant measure $m_\theta(Z) = Z(T/4) = T/2$ then this measure is not (CES).

$(W_1) \nRightarrow$ (CES)

Indeed, if $(W_1) \Rightarrow$ (CES) then also $(W_L) \Rightarrow$ (CES), which is not true.

$(W_1) \nRightarrow (W_L)$

We will use the bundle measure $m_\theta(X) = X(\theta) + X(T)$, $\theta \in [0,T]$, and show that it always satisfies $W_1$, but not always $W_L$. If Z, Y $\in \boldsymbol{U}$, and $Z \leq Y$ (with $Z \neq Y$) on $[0,T]$, then there exist $\theta_0 \in [0,T]$ such that $Z(\theta_0) < Y(\theta_0)$ and hence $m_{\theta_0}(Z) = Z(\theta_0) + Z(T) <$ $m_{\theta_0}(Y) = Y(\theta_0) + Y(T)$. This shows that $m_\theta$ meets the requirement to be $(W_1)$.

Consider now Fig.4. Z, Y $\in \boldsymbol{U}$, with $Z(x) = (1-aT)$, with $a < 0.5$, and $Y(x) = \sqrt{T^2 - x^2}$, $x \in [0,T]$. As $a < 0.5$ we already know that $aT < (1-a)T$. Now we will adapt $a$ such that $O_1 > O_2$ (see Fig.4).

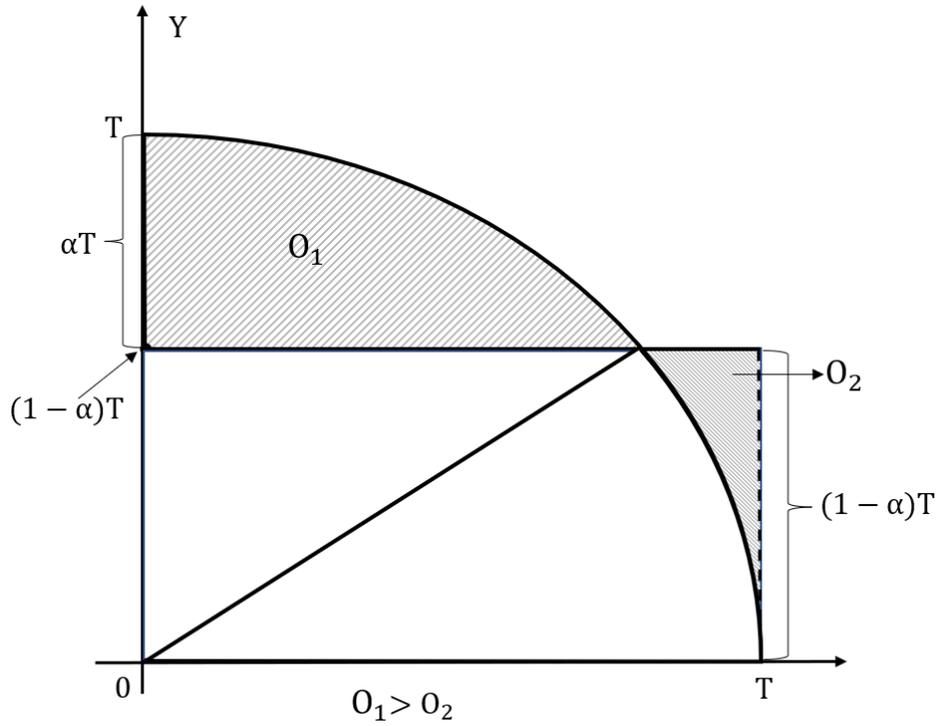

Fig.4. The example used in the proof of (W₁) ⇏ (W_L)

$$O_1 = \int_{(1-\alpha)T}^{T} \sqrt{T^2 - x^2}\, dx > O_2 = (1-\alpha)T^2 - \int_{0}^{(1-\alpha)T} \sqrt{T^2 - x^2}\, dx$$

This is equivalent with: $\int_{0}^{T} \sqrt{T^2 - x^2}\, dx > (1-\alpha)T^2$ or $\frac{\pi T^2}{4} > (1-\alpha)T^2$, leading to the extra requirement $\alpha > 1 - \frac{\pi}{4} \approx 0.2146$. Hence if $1 - \frac{\pi}{4} < \alpha < \frac{1}{2}$ we have $Z -<_{\neq} Y$ on [0,T], but for all $\theta \in [0,T]$, $m_\theta(Z) = Z(\theta) + Z(T) \geq m_\theta(Y) = Y(\theta) + Y(T)$, because $Z(T) - Y(T) \geq Z(\theta) - Y(\theta)$.

(CES) ⇏ (W_L)

We use the same example as for $(W_1) \nRightarrow (W_L)$, but with $\boldsymbol{Z} = \{Z,Y\}$. This example satisfies (CES), as the requirement for all $\theta \in [0,T]$ $m_\theta(Z) = m_\theta(Y)$ is never met, but it does not satisfy $(W_L)$ as shown above.

$(W_L) \nRightarrow (W_2)$

Assume that $(W_L) \Rightarrow (W_2)$, then it would follow from $(W_2) \Rightarrow$ (CES) that $(W_L) \Rightarrow$ (CES), which is a contradiction.

Finally, we show:

(IB) $\nRightarrow (W_1)$ and $(W_1) \nRightarrow$ (IB)

Let $\boldsymbol{Z}$ be the set of functions in $\boldsymbol{U}$, such that $Z(0) = C$ (a fixed constant). Then any bundle measure m meets (IB) as the antecedent of (AX.3) is never satisfied. If we now define for all Z in $\boldsymbol{Z}$: $m(Z) = Z(0) = C$ then $(W_1)$ is not satisfied as there exist Z, Y in Z such that $Z \leq_{\neq} Y$, but $m_\theta(Z) < m_\theta(Y)$ is never true.

For the other case, we let $\boldsymbol{Z}$ be a subset of $\boldsymbol{U}$ for which any two functions intersect strictly in the interval $]0,T[$. Then any bundle satisfies $(W_1)$. Let now $\boldsymbol{B}$ be a bundle such that for any Z: $m(Z) = C$ (a fixed constant). Then this bundle is not (IB), because $Z <_a Y$ holds for some a, Z and Y but $m(Z) \prec_a m(Y)$ never holds.

This ends the proof of Theorem 1. □

Corollary. (IB) does not imply (CES), (PED), ($W_2$), or ($W_L$).

Proposition 2

Given $Z,Y \in \boldsymbol{U}$ then the bundle $\boldsymbol{B} = (m,\psi)$ satisfies property ($W_1$) $\Leftrightarrow$ (if for all θ, $m_\theta(Z) = m_\theta(Y)$ then either $Z = Y$ on $[0,T]$ or there exists a point $q \in ]0,T[$ such that $Z$ and $Y$ intersect in q.

Proof. We first show the implication from left to right.

If ($W_1$) holds, then we know that if for all admissible θ-values $m_\theta(Z) \geq m_\theta(Y)$ either implies that $Z=Y$ on $[0,T]$ or there exists $x_0$ in $[0,T]$ such that $Z(x_0) > Y(x_0)$. Similarly, if for all admissible θ- $m_\theta(Z) \leq m_\theta(Y)$ then either $Z=Y$ on $[0,T]$ or there exists $y_0$ in $[0,T]$ such that $Z(y_0) < Y(y_0)$. Hence if for all admissible θ-values $m_\theta(Z)$ = $m_\theta(Y)$ we know that either $Z=Y$ on $[0,T]$ or (there exists $x_0$ in $[0,T]$ such that $Z(x_0) > Y(x_0)$ and there exists $y_0$ in $[0,T]$ such that $Z(y_0) < Y(y_0)$). By continuity we obtain: if for all admissible θ-values $m_\theta(Z) = m_\theta(Y)$ then, either $Z=Y$ on $[0,T]$ or $Z$ and $Y$ strictly intersect in a point $q \in ]0,T[$.

Proof of the implication from right to left.

We have to show that ($W_1$) holds. Let Z, Y ∈ **U**, with Z ≤ Y, but Z ≠ Y. Then Z = Y is not true, but also the other alternative, namely, Z and Y intersect on [0,1] is not true, because Z ≤ Y. hence, there exists an admissible θ such that $m_θ(Z) ≠ m_θ(Y)$. By the monotonicity requirement, applied to Z ≤ Y, we know that $m_θ(Z) ≤ m_θ(Y)$, and hence there exists an admissible θ such that $m_θ(Z) < m_θ(Y)$, which proves ($W_1$). □

Corollary

Let **Z** be a subset of **U** such that no two functions intersect in ]0,T[. Then: ($W_1$) ⇔ (CES)

Proof. This follows immediately from the previous proposition and Theorem 1.

We add some comments to the previous developments. Given **Z** ⊂ **U**, with Y, Z ∈ **Z**, the requirement "∃ $θ$ such that $m_θ(Z) < m_θ(Y)$" is rather weakly related to the notion of impact, while requiring that there exist fixed $θ_1$ and $θ_2$ such that for all $θ ∈ [θ_1, θ_2]$, $m_θ(Z) < m_θ(Y)$ looks like asking too much when Z $−<_{≠}$ Y.

## 3. Searching for a precise requirement, leading to the notion of impact

Let us first have a better look at the relations Z -< Y and Z $-<_{\neq}$ Y.

Z -< Y $\Leftrightarrow$ for all x $\epsilon$ [0,T]: $I_x(Z) = \int_0^x Z(s)ds \leq I_x(Y) = \int_0^x Y(s)ds$.

Similarly, Z $-<_{\neq}$ Y $\Leftrightarrow$ Z$\neq$Y and for all x $\epsilon$ [0,T]: $I_x(Z) = \int_0^x Z(s)ds \leq I_x(Y) = \int_0^x Y(s)ds$.

Considered as bundle measures, θ-values for I as well as μ lie in the interval [0,T].

Theorem 2

For, Z, Y ∈ **U**, the following three expressions are equivalent

(i) μ(Z) ≤ μ(Y) on [0,T]

(ii) Z(0) ≤ Y(0) and Z -< Y

(iii) Z -< Y

Proof. (i) $\Leftrightarrow$ (ii) $\Rightarrow$ (iii) is obvious.

Now we prove (iii) ⇒ (ii). The relation Z -< Y implies that I(Z) ≤ I(Y) on [0,T].

Hence $\frac{I_x(Z)}{x} = \mu_x(Z) \leq \frac{I_x(Y)}{x} = \mu_x(Y)$ for x ∈ ]0,T]

⇒ $\lim_{x \to 0} \mu_x(Z) \leq \lim_{x \to 0} \mu_x(Y) \Leftrightarrow Z(0) \leq Y(0)$ □

Proposition 3

$Z -<_{\neq} Y \Rightarrow \exists\, x \in [0,T]$ such that $Z(x) < Y(x)$

Proof.

Assume that for all x ∈ [0,T] Z(x) ≥ Y(x) then Y -< Z, which is in contradiction with $Z -<_{\neq} Y$.

Proposition 4

Let Z, Y ∈ **U**, and a ∈ [0,T] such that $\forall\, x \in [0,a],\ Z(x) \geq Y(x)$ then Z -< Y implies that Z=Y on [0,a].

Proof. Assume that there exists $x_0$ ∈ ]0,a] such that $Z(x_0) \neq Y(x_0)$, then we know that $Z(x_0) > Y(x_0)$ and hence, Z ≥ Y on $[0,x_0] \subset$ [0,a], contradicting the fact that Z -< Y. If Z(0) ≠ Y(0), then Z(0) > Y(0) and thus Z > Y on a certain interval [0,c], again contradicting Z -< Y. □

For further use, we introduce the following "finite number of transitions" axiom.

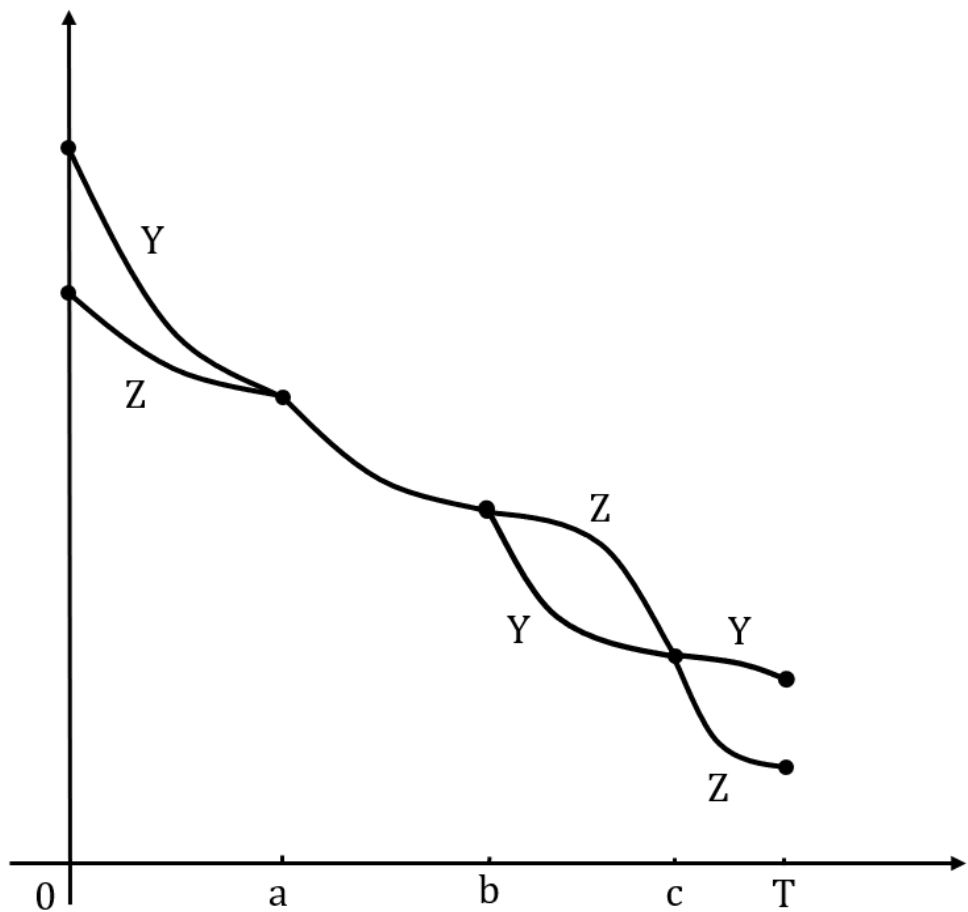

Fig. 5 An example of intersecting functions Y and Z

Definition 3

A couple of functions (Z,Y), Z≠Y, in *U* has a transition in the point t if Z(t)=Y(t) and any of the following cases occurs:

(i) t ∈ ]0,T[and ∃ δ > 0 such that (Z-Y)(x) ≠ 0 if x∈]t-δ, t[ and (Z − Y)(x) ≠ 0 if x∈]t, t+δ[,

(ii) t ∈ ]0,T[ and ∃ δ > 0 such that (Z-Y)(x) ≠ 0 if x∈]t-δ, t[ and (Z − Y)(x) = 0 if x∈]t, t+δ[,

(iii) t ∈ ]0,T[ and ∃ δ > 0 such that (Z-Y)(x) = 0 if x∈]t-δ, t[ and (Z − Y)(x) ≠ 0 if x∈]t, t+δ[,

(iv) The point t = 0 and ∃ δ > 0 such that (Z-Y)(x) ≠ 0 if x∈[0,δ[.

(v) The point t = T if T is the first point on [0,T] such that Z(T) = Y(T)

Example. The functions Y and Z shown in Fig.5 have transition points in a, b and c.

The finite number of transitions axiom or the FNT axiom

A couple of functions (Z,Y), Z≠Y, in **U** satisfies the FNT axiom if (Z,Y) has a finite number of transitions in [0,T].

This axiom is very natural, ruling out mathematical functions that do not occur in practical applications.

Definition 4: the transition point $x_1$ ∈ [0,T]

Given Z and Y, meeting the requirements of (FNT), we denote by $x_{1,Z.Y} \in [0,T]$ (from now on simply denoted as $x_1$ as it will always be clear which Z and Y we are talking about) the following point, depending on the case:

a) if Z and Y are never equal then $x_1 = T$.

b) otherwise, $x_1$ is the first transition point.

Note that, in this article, the notation $x_1$ will always denote the transition point as defined here.

Theorem 3

If $Z,Y \in \boldsymbol{U}$, such that the couple (Z,Y) satisfies the FNT axiom then $Z -<_{\neq} Y$ implies that either, (i) there exist points *a* and $b \in [0,T]$, $a < b \leq T$, such that Z=Y on [0,*a*] and Z < Y on ]*a,b*],

or,(ii) if a point *a* for which Z=Y on [0,*a*] does not exist, then there exists a point $b > 0$ such that Z < Y on [0,*b*].

Proof. If $Z -<_{\neq} Y$ then $Z(0) \leq Y(0)$.  If $Z(0) < Y(0)$ then, by the definition of $x_1$, we may take any *b*, $0 < b < x_1$, proving part (ii) of Theorem 3.

If $Z(0) = Y(0)$, then we may take the number *a* equal to $x_1$, and because $Z -<_{\neq} Y$, and Z-Y has only a finite number of transitions, there exists a number $b > a=x_1$ such that $Z < Y$ on $]a,b]$, proving part (i) of Theorem 3.□

Theorem 3 can be interpreted as stating that if $Z -<_{\neq} Y$ then on the left-hand side of these curves we have an interval where they are equal (possibly only the point 0), followed by an interval where $Z < Y$, or Z and Y start with an interval where $Z < Y$. From this, we conclude that a reasonable requirement for measuring impact is that if $Z -<_{\neq} Y$ then one of the two requirements is met:

either there is a left-hand part for which $\forall \theta \in Q$ $m_\theta(Z) = m_\theta(Y)$ followed by an interval such that $\forall \theta \in Q$ $m_\theta(Z) < m_\theta(Y)$, or there is a left-hand part such that $\forall \theta \in Q$ $m_\theta(Z) < m_\theta(Y)$.

Because of their importance in the following developments, we recall the following notations, and add some new ones.

Notations

For Z, Y $\in$ ***U***, a, b $\in$ [0,T], a < b we define:

$Z <_a Y \Leftrightarrow Z < Y$ on $[0,a]$; this relation is a strict partial order;

$Z <_{a,b} Y \Leftrightarrow (Z = Y$ on $[0,a])$ and $(Z < Y$ on $]a,b])$; this relation too is a strict partial order;

If m is a bundle measure $(\theta, Z) \rightarrow m_\theta(Z)$, then

$m(Z) \prec_a m(Y) \Leftrightarrow m(Z) < m(Y)$ on $\psi_Z([0,a]) \cup \psi_Y([0,a])$, or stated otherwise: for all $\theta \in \psi_Z([0,a]) \cup \psi_Y([0,a])$: $m_\theta(Z) < m_\theta(Y)$;

$m(Z) \prec_{a,b} m(Y) \Leftrightarrow (m(Z) = m(Y)$ on $\psi_Z([0,a]) \cup \psi_Y([0,a]))$ and $(m(Z) < m(Y)$ on $\psi_Z(]a,b]) \cup \psi_Y(]a,b]))$

Note that these relations on m(**Z**) are not transitive in general.

Definition 5. A bundle **B** = (m, ψ) satisfies condition (GIB) if for all a, b, $0 \leq a \leq T$ and $0 \leq a < b \leq T$: the bundle measure m satisfies the following two conditions: for Z, Y $\in$ **U** :

$Z <_a Y \Rightarrow (m(Z) \prec_a m(Y))$

$Z <_{a,b} Y \Rightarrow m(Z) \prec_{a,b} m(Y)$

Note. It will soon become clear why we use the term (GIB).

This definition leads to the following Theorem 4.

Theorem 4

If the bundle **B** = (m, ψ) satisfies condition (GIB) and the couple (Z,Y) satisfies (FNT) then:

$(Z \quad -<_{\neq} \quad Y) \Rightarrow \big( \big[ \exists a \in [0, T[ \text{ and } \exists b \in ]a, T] \text{ such that } m(Z) =_a m(Y) \land m(Z) <<_{a,b} m(Y) \big] \lor [\forall b < x_1 : m(Z) <<_b m(Y)] \big)$

Proof. This follows from condition (GIB) and Theorem 3.

We note that Theorem 4 includes IPPs for which the corresponding curves, Z and Y, begin with an equal part. This case was not yet included in our earlier research. This is the main reason for using the relation $Z -<_{\neq} Y$ in combination with the FNT axiom.

We further observe that most, we would even say all, well-known impact bundles satisfy condition (GIB). We recall that when studying strong impact measures (Egghe & Rousseau, 2022b) we encountered a serious restriction on the number of candidate measures meeting the requirement $Z -<_{\neq} Y$.

Definition 6

A bundle **B** meeting the requirement (GIB) is called a global impact bundle. Like for the case of an impact bundle, we state this, in short, as **B** is (GIB).

Theorem 5

A global impact bundle is an impact bundle, or stated otherwise (GIB) $\Rightarrow$ (IB).

Proof.  We only have to check (AX.3). (GIB) implies that if $\forall\, a \in [0,T]$, $Z <_a Y$, it follows that m(Z) $\prec_a$ m(Y). This shows that m is an impact bundle measure.

Remark. The opposite implication of Theorem 5 does not hold. Indeed, there exist impact bundles that are not global. To prove this remark and for further use, we study the following example.

Example A

Let **Z** = {Z,Y}, with Z(x)  = T − x/2, x $\in$ [0,T]. Now we choose a number $\theta_0$ such that T/2 <  $\theta_0$ < T and a positive number b such that 0 < b < $\theta_0$/2 < T/2.  Then the function Y is defined as follows:

Y = T on an interval $[0, \theta_0]$ and $Y(x) = $ T $-\left(b+\frac{T}{2}\right)(\frac{x-\theta_0}{T-\theta_0})$ on $[\theta_0,$ T], i.e., the linear segment connecting the points $(\theta_0,$ T) and (T, T/2-b). This construction is illustrated in Fig. 6.

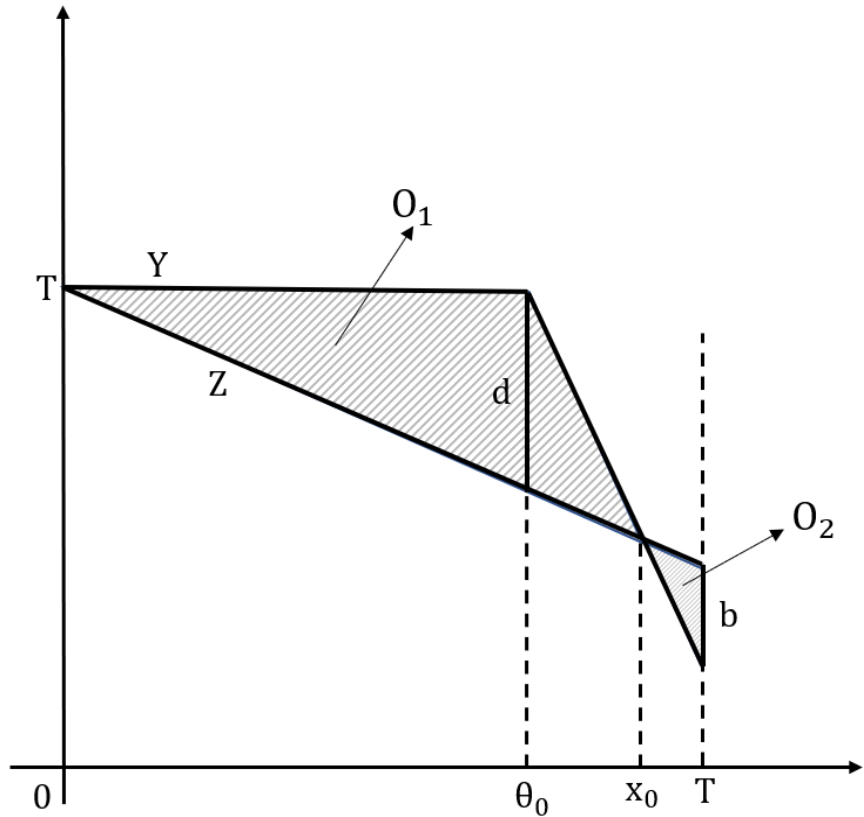

Fig. 6. Example A

If $O_1 > O_2$ then $Z -<_{\neq}$ Y. Hence, we calculate the surface of these two areas. We see that d = $\theta_0/2$ and $x_0$ = ($\theta_0$ (T+2b))/($\theta_0$+2b). Then $O_1 > O_2 \Leftrightarrow \theta_0 d + (x_0-\theta_0)d > (T-x_0)b \Leftrightarrow (d+b)x_0 > Tb \Leftrightarrow \frac{\theta_0(T+2b)}{\theta_0+2b} > \frac{2Tb}{\theta_0+2b} \Leftrightarrow \theta_0(T+2b) > 2Tb$ . As $\theta_0 >$ T/2, we see that $\theta_0(T + 2b) > \frac{T}{2}(T + 2b)$. As T+2b > 4b or T > 2b, this shows the required inequality. From this result, we conclude that $Z -<_{\neq} Y$.

We now consider for θ ∈ [0,T], $m_\theta(Z) = Z(\theta) + Z(T)$.

Then m is an impact bundle because Z(0) = Y(0) and $Z -<_{\neq} Y$ implies that never Z < Y on [0,a], and hence (AX.3) is satisfied.

Yet m is not a global impact bundle because we can find a > 0 such that : $\forall \theta \in [0,a]$: $m_\theta(Z) = Z(\theta) + Z(T) > Y(\theta) + Y(T) = m_\theta(Y)$, or Y(θ) − Z(θ) < Z(T) - Y(T). Taking now a < $\theta_0$, we see that $m_\theta(Z) > m_\theta(Y)$ becomes T − (T − θ/2) < T/2 − (T/2-b). This inequality is correct if we take a < min(2b, $\theta_0$).

Next, we establish the relation between properties (GIB), (PED), (W₂), and (W_L).

Theorem 6. For all bundles **B** = (m, ψ) and all **Z** ⊂ **U** the following relations (shown in the diagram) hold:

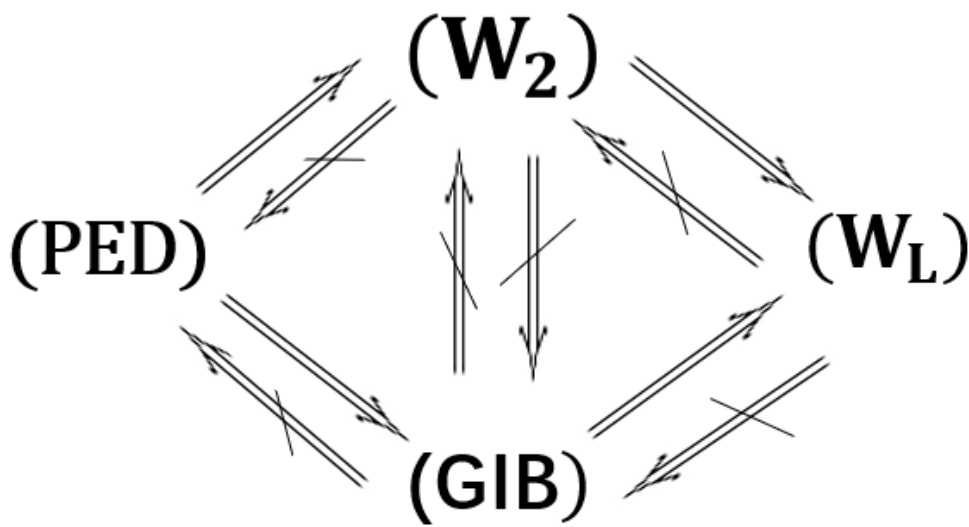

Proof. The upper relations are already shown in Theorem 1.

(i) (PED) $\Rightarrow$ (GIB) $\Rightarrow$ (W$_L$)

(GIB) is an obvious consequence of (PED), whatever the value of $x_1$

(W$_L$) follows from (GIB) by Theorem 4 (under the extra assumption that each couple (Z,Y) satisfies the (FNT) requirement).

(ii) (GIB) $\not\Rightarrow$ (PED)

We take for all $\theta \in ]0,T]$: $m_\theta = \mu_\theta$. Then, by its very definition, this bundle measure satisfies (GIB). We also know that in general $\mu_\theta$ is not (PED), as shown in (Egghe, 2021).

(iii) (W$_2$) $\not\Rightarrow$ (GIB)

We consider example A. We already know that it does not satisfy (GIB). Yet, it does meet the requirement (W$_2$). Indeed, for all x ∈ ]0,x$_0$[, we have Z(x) < Y(x). Now we consider θ = θ$_0$ and show that:

$$m_{\theta_0}(Z) < m_{\theta_0}(Y)$$

$$\Leftrightarrow T - \frac{\theta_0}{2} + \frac{T}{2} < T + \frac{T}{2} - b$$

$$\Leftrightarrow \theta_0 > 2b$$

which is correct. This proves that this example satisfies requirement (W$_2$) but not (GIB).

(iv) (W$_L$) $\nRightarrow$ (GIB)

We know that in general (W$_2$) ⇒ (W$_L$). Hence Example A is a case for which W$_L$ is satisfied, but (GIB) is not.

(v) (GIB) $\nRightarrow$ (W$_L$), assuming that **B** also satisfies (TNT)

We consider the counterexample used to show that (CES) $\nRightarrow$ (W$_2$). We know that this example does not satisfy (W$_2$). Yet, with m$_\theta$ = μ$_\theta$ and using Theorem 3 we see that (GIB) is satisfied.

This proves Theorem 6.

The previous theorems and considerations lead us to the notion of global impact. We consider the formulation of this definition as the culmination of our investigations on impact.

Definition 7. The notion of 'impact'

For Z, Y $\in \boldsymbol{Z} \subset \boldsymbol{U}$, we say that Y has strictly more impact than Z on [0,T] iff Z $-<_{\neq}$ Y.

We emphasize that because of the developments above, this definition is a logical way to define impact in a general sense. We can also use $-<_{\neq}$ to define "strictly more impact" in the discrete case, i.e., in the case of vectors Z, Y (replacing the integral in I(Z) by a finite sum in Section 2), hence also in practical situations. Note that, for vectors Z, Y, the condition FNT in Section 3 is always satisfied so that Z $-<_{\neq}$ Y expresses that "Y has strictly more impact than Z" without any restrictions on $-<_{\neq}$.

Combining now Theorems 4 and 6 we have the following diagram (Fig. 7), where we only show those implications which we have shown to hold, not those that do not hold:

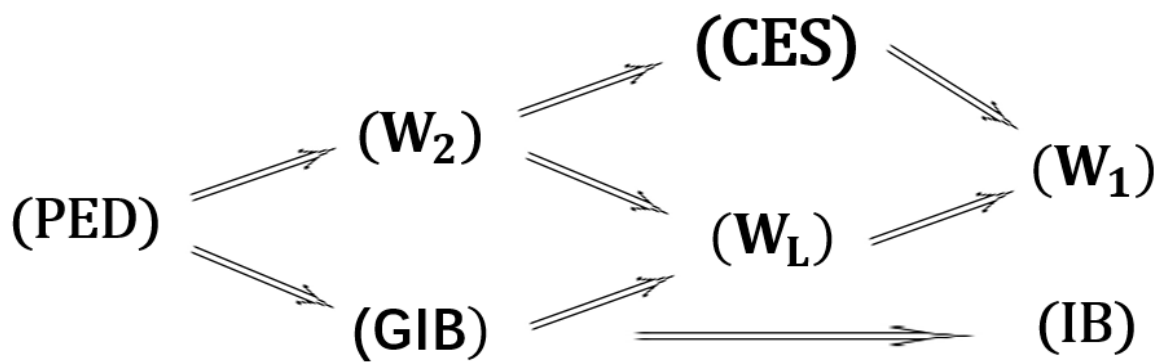

Fig.7. Diagram showing implication relations between seven bundle properties

## Discussion and conclusion

As stated in the introduction, our approach is inspired by the standard Lorenz curve. For impact studies, the role of the standard Lorenz curve is played by the non-normalized Lorenz curve, which was introduced as the graph of the area function on the interval [0,T]. The dominance relation for Lorenz curves is now replaced by the relation $-<_{\neq}$, so that the expression Z is more even than Y, or Y shows a larger concentration than Z, becomes now "Y has strictly more impact than Z".

Acceptable evenness (concentration) measures are those that respect the Lorenz order. Similarly, acceptable global impact bundle measures satisfy condition (GIB). We note here an important difference, namely that in the case of evenness one has measures (functions), whereas in the case of impact one needs bundles, which can roughly be described as parametrized functions.

Similar to the case of evenness (and variety or species richness) which are just aspects of the notion of diversity, we have here kept the length of the interval fixed (fixed T). In the case of citations received by a set of publications, this means that we have kept this number of publications fixed. Comparing situations with variable Ts is left for further investigations.

We like to add the following final comments. We are strongly convinced that in order to study the notion of impact for real data, requiring that the whole set of data must be taken into account is asking too much and is not necessary. Recall that even the best scientists have uncited publications (Egghe et al., 2011) and that even Nobel-Prize winning articles are not always fully appreciated (Hu & Rousseau, 2017). When using a continuous, decreasing function on an interval [0,T] as a model (as we did), this means that impact does not have to be defined in relation with the whole curve (here, on the whole interval [0,T]), although that is, of course, not forbidden either.

Acknowledgment. The authors thank Li Li (National Science Library, CAS) for providing excellent figures.

Conflict of interest. Leo Egghe declares no conflict of interest. Ronald Rousseau is a member of the editorial board of *Quantitative Science Studies*.

Funding. No funding has been received for this investigation

Author contributions:

Leo Egghe: conceptualization, formal analysis, investigation, methodology, writing-original draft, writing-review and editing.

Ronald Rousseau: investigation, validation, writing-review and editing.